\documentclass[aps,prd,twocolumn,groupedaddress,floatfix]{revtex4}

\usepackage{graphicx}
\usepackage{amsmath}

\begin{document}

\newcommand{\aap}{Astron.~Astrophys.}
\newcommand{\apjl}{Astrophys.~J.}
\newcommand{\mnras}{Mon.~Not.~Roy.~Astron.~Soc.}
\newcommand{\physrep}{Phys.~Rep.}
\newcommand{\jcap}{JCAP}

\newcommand{\simgt}{\lower.5ex\hbox{$\; \buildrel > \over \sim \;$}}
\newcommand{\simlt}{\lower.5ex\hbox{$\; \buildrel < \over \sim \;$}}
\newcommand{\bmf}[1]{\mbox{\boldmath$#1$}}
\newcommand{\sang}[1]{\left\langle#1\right\rangle}

\title{
Combining cluster observables and stacked weak lensing
to probe dark energy: Self-calibration of systematic uncertainties
}

\author{Masamune Oguri$^{1}$ and Masahiro Takada$^{2}$}
\affiliation{
$^1$Division of Theoretical Astronomy, National Astronomical
Observatory of Japan, 2-21-1 Osawa, Mitaka, Tokyo 181-8588, Japan\\ 
$^2$Institute for the Physics and Mathematics of the Universe (IPMU),
The University of Tokyo, Chiba 277-8582, Japan 
}

\date{\today}

\begin{abstract}
We develop a new method of combining cluster observables (number
counts and cluster-cluster correlation functions) and stacked weak
lensing signals of background galaxy shapes, both of which are
available in a wide-field optical imaging survey. Assuming that the
clusters have secure redshift estimates, we show that the joint
experiment enables a self-calibration of important systematic errors
inherent in these measurements, including the source redshift
uncertainty and the cluster mass-observable relation, by adopting a
{\it single} population of background source galaxies for the lensing
analysis. The single source galaxy population allows us to use the
relative strengths of the stacked lensing signals at different cluster 
redshifts for calibrating the source redshift uncertainty, which in 
turn leads to accurate measurements of the mean cluster mass in each 
redshift and mass bin. In addition, our formulation of the stacked
lensing signals in Fourier space simplifies the Fisher matrix
calculations, as well as the marginalization over the cluster
off-centering effect which is one of the most significant
uncertainties in the stacked lensing analysis. We show that upcoming
wide-field surveys covering more than a few thousand square degrees 
yield stringent constraints on cosmological parameters including dark  
energy parameters, without any priors on nuisance parameters that
model systematic uncertainties. Specifically, the stacked lensing
information improves the dark energy figure of merit (FoM) by a factor 
of 4, compared to that from the cluster observables alone. The
primordial non-Gaussianity parameter can also be constrained with a
level of  $\sigma(f_{\rm NL})\sim 10$. In this method, the mean source
redshift is well calibrated to an accuracy of 0.1 in redshift, and the
mean cluster mass in each bin to 5--10\% accuracies, which
demonstrates the success of the self-calibration of systematic
uncertainties from the joint experiment.  
\end{abstract}


\maketitle

\section{Introduction}

The abundance and its evolution of clusters of galaxies are thought to 
be one of main probes of cosmology including dark energy (e.g.,
\cite{DETF}). This is because dark matter plays an essential role in the
formation processes, indicating that the abundance and clustering
properties of clusters can be predicted reasonably well by theory. In
addition, the redshift evolutions of the cluster abundance and their
clustering amplitudes are sensitive to the linear growth rate, which
in turn is sensitive to the expansion history of the universe. In fact
recent careful analyses of X-ray \cite{Vikhlininetal:09,Mantzetal:10a}
and optical \cite{Rozoetal:10} clusters have demonstrated that cluster
abundance can provide useful constraints on cosmological
parameters. In addition, clusters serve as powerful tests of
non-standard cosmological scenarios including the modified theory of
gravity \cite{Schmidtetal:09,Reyesetal:10,Rapettietal:10} and
primordial non-Gaussianity
\cite{Dalaletal:08,Oguri:09,Roncarellietal:10,Cunhaetal:10,Sartorisetal:10}.

Perhaps the most significant challenge in cluster cosmology lies in
the determination of masses for individual clusters or for a cluster
sample as a whole. In many cases, we resort to mass-observable scaling  
relations, which have been calibrated by the intensive observations of
(a sub-sample of) the clusters, in order to infer clusters masses.
Popular choices of such observables include X-ray luminosities
\cite{ReiprichBohringer:02,Mantzetal:08,Rykoffetal:08}, X-ray
temperatures
\cite{Finoguenoetal:01,Shimizuetal:03,Sandersonetal:03,Arnaudetal:05},
gas masses \cite{Okabeetal:10}, optical richnesses
\cite{Johnstonetal:07}, the Sunyaev-Zel'dovich (SZ) effect (the
integrated gas pressure along the line-of-sight)
\cite{Bonamenteetal:08,Marroneetal:09} and its X-ray analog
\cite{Kravtsovetal:06,Zhangetal:08,Mantzetal:10b}. However, it is not
an easy task to derive cluster masses observationally for
calibrations, and moreover, we have to take account of its redshift
evolution for robust cosmological studies.    

Weak gravitational lensing provides a very powerful tool for measuring
cluster masses. Gravitational lensing induces coherent tangential
distortions of background galaxy shapes around a cluster, from which we
can directly measure total mass profiles of the cluster (see
\cite{BartelmannSchneider:01} for a review).  While accurate mass
determinations with weak lensing are feasible only for massive clusters,
by stacking signals from many clusters we can measure average masses of
clusters down to less massive halos
\cite{Fischeretal:00,McKayetal:01,Sheldonetal:01,GuzikSeljak:01,Mandelbaumetal:05,Sheldonetal:09,Okabeetal:10}.
This stacked lensing technique has indeed been applied to calibrate the
mass-observable relation to derive interesting constraints on
cosmological parameters (e.g., \cite{Rozoetal:10}). However, the stacked
lensing analysis involves various systematic uncertainties.  For
instance, all weak lensing analysis is subject to the uncertainty in the
redshift distribution of source galaxies used in the lensing
analysis. In cosmic shear analysis, the estimation of source redshifts
sometimes becomes the most significant source of systematic errors that
challenge the use of weak lensing for cosmological studies
\cite{Hutereretal:06,Nishizawaetal:10,Hearinetal:10}. In addition, one
of the most significant uncertainties in stacked lensing analysis would
be cluster centroiding (off-centering) errors which arises from the
limitation of the mass centroid determination from available data
\cite{Johnstonetal:07,Mandelbaumetal:10}.

Clusters of galaxies are biased tracers of the underlying mass
distribution. Thus one can obtain constraints on the growth of
structure from spatial clustering of massive clusters as
well. Furthermore, the bias contains information on the cluster mass,
which offers another opportunity to infer cluster masses in a
statistical way. This opens up a possibility to calibrate the
mass-observable relation from observing data themselves
\cite{MajumdarMohr:04,LimaHu:04,LimaHu:05}.

In this paper, we explore the potential of future optical cluster
surveys, such as the Hyper Suprime-Cam (HSC) \cite{Miyazakietal:06},
Dark Energy Survey (DES) \cite{DES}, and Large Synoptic Survey Telescope
(LSST) \cite{LSST}, for constraining cosmological parameters. In these
surveys, many clusters out to high-redshift ($z>1$) will be identified
in multi-color optical images, or with aid of Sunyaev-Zel'dovich (SZ)
effects observed by Atacama Cosmology Telescope (ACT) 
\cite{ACT:09,ACTPol:10}, South Pole Telescope (SPT) \cite{SPT:10}, or 
the all-sky Planck survey 
\footnote{http://www.sciops.esa.int/index.php?project=PLANCK}.
We combine number counts of these clusters with stacked weak
lensing measurements, i.e., tangential shear profiles around the
clusters taken in the stacking average. At small-scale the shear
profiles constrain mean masses of cluster samples, whereas large-angle
tangential shear profiles serve as a clean probe of halo-mass cross
power spectrum. The cluster number counts and tangential shear
profiles respectively contain important systematics, the
mass-observable relation and the source redshift uncertainty. We show
how the combination of these two  observables can help self-calibrate
these systematics to obtain robust constraints on cosmological
parameters.  

This paper is organized as follows.  In Sec.~\ref{sec:idea}, we sketch
out the basic idea of our technique, focusing on how we can overcome
various difficulties inherent in cluster cosmology using the stacked
weak lensing. We calculate the signals in Sec.~\ref{sec:signal}, and the
covariance between observables in Sec.~\ref{sec:cov}. We show results on
the cosmological parameter forecast in Sec.~\ref{sec:forecase}. We
discuss possible systematics in Sec.~\ref{sec:discuss}, and conclude in
Sec.~\ref{sec:conc}.  We choose the best-fit cosmological parameters
from the seven-year Wilkinson Microwave Anisotropy Probe (WMAP)
observations \cite{Komatsuetal:10} as our fiducial cosmological model,
with the matter density $\Omega_M=0.266$, baryon density
$\Omega_b=0.04479$, the dark energy density $\Omega_{\rm DE}=0.734$ and
its equation of state $w(a)=w_0+(1-a)w_a=-1$, the dimensionless Hubble
constant $h=0.710$, the power spectrum tilt $n_s=0.963$, and the
normalization of the power spectrum $\sigma_8=0.801$.  Throughout the
paper we assume a flat universe, i.e., $\Omega_M+\Omega_{\rm DE}=1$, and
use the unit $c=1$ for the speed of light. 

\section{Basic Ideas}
\label{sec:idea}

\begin{figure*}[t]
\begin{center}
\includegraphics[width=0.9\textwidth]{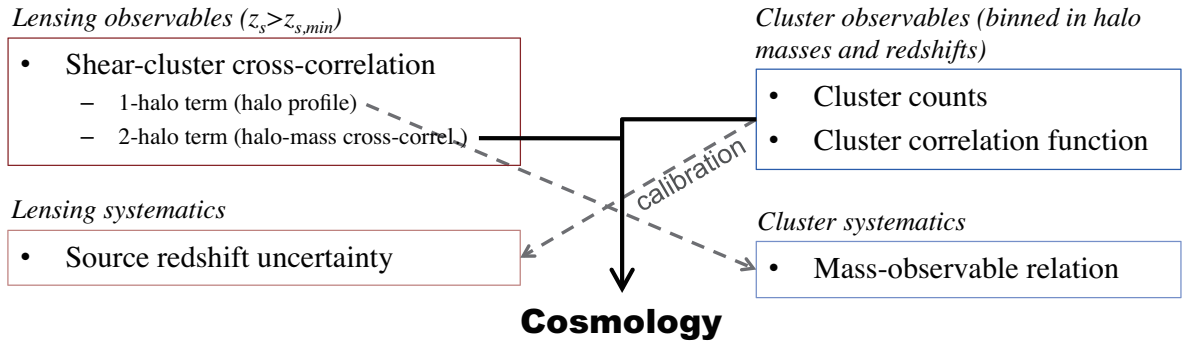}
\caption{A conceptual flow-chart of the method studied in this
  paper. We propose a joint experiment of combining the cluster
  observables and the shear-cluster correlation (or the so-called
  stacked lensing) to constrain cosmological parameters including dark
  energy parameters. The most serious systematic error inherent in
  each observable, the source redshift uncertainty and the
  mass-observable relation, can be calibrated out by combining the two 
  observables (also see text around Eq.~[\ref{eq:idea_stacked}]). The
  cosmological power of this method arises mainly from the large-angle
  signals of the shear-cluster cross-correlation (2-halo term) and
  from the cluster observables. In addition, we consider how other
  systematic errors such as variations in the halo profile and the
  cluster centering offset affect cosmological constraints. } 
\label{fig:idea}
\end{center}
\end{figure*}

The lensing distortion signal for a source galaxy in a particular
direction $\bmf{\theta}$ on the sky and at redshift $z_s$ is expressed
as 
\begin{equation}
\gamma(\bmf{\theta}; z_s)= 4\pi G \int_0^{z_{s}}\!\!dz~ 
a\chi(z) \left[1-\frac{\chi(z)}{\chi(z_s)}\right]
\Delta\Sigma(\chi\bmf{\theta},z),
\end{equation}
where $\chi=\chi(z)$ is the comoving angular diameter distance to
redshift $z$, which is same as the comoving radial distance for a flat
universe, $\Delta\!\Sigma$ denotes the mass distribution field along
the line of sight, and the redshift is related to the scale factor $a$
via $1+z=1/a$.  

One of the most serious systematic errors in weak lensing is the
photometric redshift error, i.e., uncertainty in estimating redshifts  
of source galaxies. Since a spectroscopic survey of all the source
galaxies is observationally too expensive, in most cases redshifts of
source galaxies have to be inferred based on their broadband
photometries, the so-called photo-$z$ information. For cosmic shear
tomography \cite{Hu:99,Huterer:02,TakadaJain:04}, which is potentially
the most powerful cosmological probe, future surveys need to calibrate
mean redshifts of each tomographic bins to about 0.1\% in order for
constraints on cosmological parameters not to be significantly 
degraded \cite{Hutereretal:06}. This level of calibration is indeed
very challenging. 

In this paper, we propose a method which adopts the stacked lensing
technique for the cosmological purpose, yet is insensitive to the 
photo-$z$ errors. The key idea to overcome the source redshift
uncertainty is to use a {\it single} population of background
source galaxies to extract the lensing information, where source
galaxies can be selected based on the photo-$z$ information, in order
to control and to calibrate its uncertainty.  

More specifically, we consider the following situation.  We study
shear signals around a catalog of galaxy clusters, whose redshifts are
assumed to be known either spectroscopically or photometrically.  The
photometric redshift estimate of each cluster is much more robust and
accurate than photo-$z$ estimates of individual galaxies, because 
photo-$z$ estimates are more secure for early-type galaxies and
photo-$z$ uncertainties are statistically reduced by combining many
member galaxies. We assume that redshifts of source galaxies are not
overlapped with the cluster redshifts, i.e., all the source galaxies
are located behind any clusters in the catalog. By cross-correlating
the shapes of source galaxies with the cluster distribution in a
particular redshift slice centered at $z_l$, we can measure the
shear-cluster cross-correlation (or the stacked lensing signal) as a
function of separation angles between the cluster center and the
source galaxies: 
\begin{eqnarray}
&&\hspace{-2em}\sang{\gamma_+}(\theta;z_l)\equiv\left.
\langle n_{\rm cl}(\bmf{\theta}'; z_{l})\gamma_+
(\bmf{\theta}-\bmf{\theta}')\rangle \right|_{z_s>z_{s,{\rm min}}}\nonumber\\
&&\hspace{-2em} = 4\pi G (1+z_l)^{-1}\chi_l
\left[1-\chi_l\left\langle \frac{1}{\chi(z_s)}\right\rangle\right]
\langle\Delta\Sigma(\chi_l\theta,z_l)\rangle,
\label{eq:idea_stacked}
\end{eqnarray}
where $\chi_l\equiv \chi(z_l)$ and 
\begin{equation}
\left\langle\frac{1}{\chi(z_s)}\right\rangle\equiv
\left[\int_{z_{s,{\rm min}}}^{\infty}\!dz_s~\frac{dp}{dz_s}
\frac{1}{\chi(z_s)}\right]\left[\int_{z_{s,{\rm min}}}^{\infty}
\!dz_s~\frac{dp}{dz_s}\right]^{-1}. 
\label{eq:ave_chis}
\end{equation}
The redshift $z_{s, {\rm min}}$ denotes the minimum redshift used to
define a source galaxy sample. The function
$\langle\Delta\Sigma(\chi_l\theta,z_l)\rangle$ denotes the average
mass profile around the lensing clusters at the redshift $z_l$, and
$dp/dz_s$ is the normalized redshift distribution of source 
galaxies defined so as to satisfy $\int_0^{\infty}\!dz_sdp/dz_s=1$. 
Thus the dependence of the stacked lensing on source redshifts is only
via a single quantity, the average $\langle\chi(z_s)^{-1}\rangle$ 
over the source galaxy population, and its effect is just to cause an
offset in the overall amplitude of the stacked shear profile.

Eq.~(\ref{eq:idea_stacked}) indicates that the stacked lensing method
in principle allows one to extract the lensing contribution {\it at}
each cluster redshift from the total shear signals in source galaxy images.
Thus, if cluster samples at different redshift slices are available as
we study below, we can measure the redshift evolution of the stacked
lensing signals, which provides a clue to the tomography of shear
signals, i.e., how the shear signals in source galaxy images are built
up over structures along the line of sight. A key in this step is that
the stacked lensing signals at different redshifts all depend on the
single quantity $\langle \chi(z_s)^{-1}\rangle$. Hence, even if the mean
redshift of source galaxies is completely unknown as an extreme case,
the quantity $\langle \chi(z_s)^{-1}\rangle$ can be self-calibrated by
combining all the stacked lensing signals, as long as the stacked
lensing signals at different cluster redshifts do not carry perfectly
degenerate information. 

\begin{figure}[t]
\begin{center}
\includegraphics[width=0.42\textwidth]{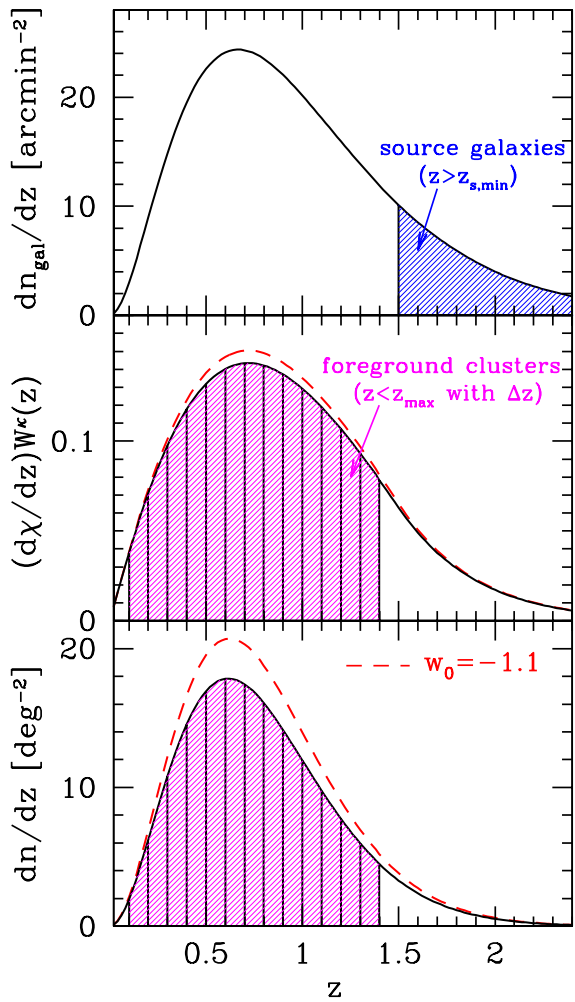}
\caption{The setup assumed in the paper. The population of source
 galaxies for weak lensing analysis is assumed to be located behind
 {\it any} clusters used for cosmological analysis ({\it top}). 
 We study the cluster-shear cross correlation (whose signal is
 proportional to the lensing weight function $W^\kappa(z)$; {\it
 middle}) and the number of clusters as a function of cluster
 redshifts ({\it bottom}). }
 \label{fig:wei}
\end{center}
\end{figure}

On the other hand, the cluster observables (number counts of clusters
and the cluster-cluster power spectrum) also involve several systematic
uncertainties. The most important one is the uncertainty in the
mass-observable relations. While the theoretical predictions for cluster
observables are primarily given as a function of halo masses, the halo
masses need to be inferred from the cluster observables, which causes
uncertainties because of the imperfect relation between cluster observables
and halo masses. Although previous studies have considered
a possibility of calibrating the mass-observable relations by using
clustering information of clusters (e.g., \cite{MajumdarMohr:04}), in
our case cluster masses in each bin can directly be calibrated based on
the stacked lensing signals at small angular scales, which simply
measures the mean mass profile of clusters (the 1-halo term).

Thus the weak lensing signals of distant galaxies and the cluster
catalogs offer a promising synergy. Fig.~\ref{fig:idea} shows a
conceptual flow-chart of the method studied in this paper. The
combination of the two observables, the stacked lensing and the
cluster observables, enables the self-calibration of the two major
systematic errors, the source redshift uncertainty and the
mass-observable relation. The cosmological power arises mainly from
the large-angle signals of the stacked lensing as well as from the
cluster observables, where both the observables are given as a
function of different bins of cluster mass indicators (e.g.,
richnesses) and redshifts. In addition, we will carefully address
other systematic errors that affect the lensing observables, such as
variations in the cluster mass profile, the offset of cluster
centers, and a possible multiplicative uncertainty in estimating the
lensing shear amplitude from galaxy shapes.

\begin{figure}[t]
\begin{center}
\includegraphics[width=0.42\textwidth]{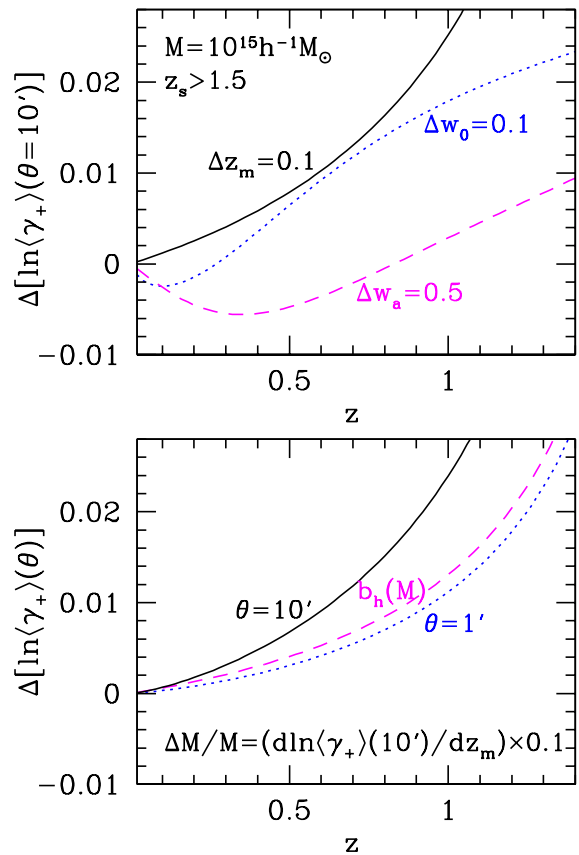}
\caption{{\it Upper panel}: The sensitivity of the tangential shear 
$\langle\gamma_+\rangle(\theta)$ at $\theta=10'$ (roughly
corresponding to the virial radius for a cluster with the mass
$M=10^{15}h^{-1}M_\odot$), as a function of the cluster redshift
$z$. We show the fractional change of $\gamma_+$ by changing the mean
redshift of the source galaxy distribution $z_m$ ({\it solid}) and the
dark energy equation of state $w_0$ ({\it dotted}) and $w_a$ ({\it
  dashed}). See Sec.~\ref{sec:signal} for details of the setup and
calculations. {\it Lower panel}: The change of $\langle\gamma_+\rangle(\theta)$
for different $\theta$, when the mass of the cluster is shifted by 
$(d\ln \langle\gamma_+\rangle(10')/dz_m)\times 0.1$ in
order to mimic the fractional change due to $\Delta z_m=0.1$ (shown 
in the upper panel). The solid curve shows the fractional change of
$\langle\gamma_+\rangle(\theta)$ at $\theta=10'$, which indeed mimics
the slid line in the top panel, wheres the dotted curve indicates the
change at $\theta=1'$. At $\theta\gg10'$ the stacked lensing signal is
dominated by the 2-halo term, whose mass dependence is simply the halo
bias $b_h(M)$; the fractional change of $b_h(M)$ is shown by the
dashed curve.} 
 \label{fig:zdep}
\end{center}
\end{figure}

For a further clarity, the setup of the method is illustrated in
Fig.~\ref{fig:wei}. Unlike the conventional weak lensing tomography,
we consider a single population of source galaxies defined so as not
to be overlapped with clusters. We assume that redshifts of the
clusters can be measured accurately, and clusters are divided into
redshift bins. Since all cluster samples in different redshift bins
share exactly the same source population, the bias in the source
redshift distribution affects the shear signals in different redshift
bins coherently in a way easily predicted by theory, which is the
reason the source redshift uncertainty can be calibrated.  

The self-calibration of the source redshift uncertainty and the
mass-observable relations is feasible only when these two
uncertainties are not completely degenerate.  We check this point in
Fig.~\ref{fig:zdep}. By increasing the source galaxy redshift, the
stacked lensing signal at scales around the virial radius, $10'$
considered here, increases monotonically with increasing cluster
redshift. Although we can shift the mass of the cluster in a way to
mimic this trend, the different mass dependence on the tangential
shear at different radii suggests that we can break the degeneracy
between the source redshift and the mass-observable relation by
observing tangential shear signals over a wide range of radii. 
Furthermore, the trend is quite different from the changes of the
tangential shear due to different dark energy equation of state,
suggesting that these uncertainties are not degenerate with the
uncertainty of cosmological parameters either.

\section{Cosmological Observables}
\label{sec:signal}

\subsection{A Flat Cold Dark Matter Model}

Throughout this paper we work in the context within a spatially flat
universe. The Hubble expansion rate, $H(z)\equiv(da/dt)/a$, is given
by the scale factor $a(t)$ and is given in terms of energy density
parameters as
\begin{equation}
H^2(a)=H_0^2\left[\Omega_{M}a^{-3}+\Omega_{\rm DE}
e^{-3\int^{a}_1da'(1+w(a'))/a'}\right],
\end{equation}
where $H_0=100 h~$km~s$^{-1}$~Mpc$^{-1}$ is the present-day Hubble
parameter. The parameter $w(a)$ specifies the equation of state for
dark energy as $w(a)\equiv p_{\rm DE}(a)/\rho_{\rm DE}(a)$. 
Note that $\Omega_M+\Omega_{\rm DE}=1$ and we have used the convention 
$a(t_0)=1$ today. The comoving angular diameter distance $\chi(z)$ is
given as (note $1+z=1/a$),
\begin{equation}
\chi(z)=\int^z_0dz'\frac{1}{H(z')},
\end{equation}
which is related to the angular diameter distance $D_A(z)$
via $\chi(z)=(1+z)D_{\rm A}(z)$. 

Throughout this paper we adopt a parametrized form of the dark energy
equation of state, $w(a)=w_0+(1-a)w_a$ \cite{ChevallierPolarski:01}. 
In this case, the Hubble expansion rate can be described as  
\begin{equation}
H^2(a)=H_0^2\left[\Omega_Ma^{-3}+\Omega_{\rm DE}
a^{-3(1+w_0+w_a)}e^{-3w_a(1-a)}\right].
\end{equation}

Another important ingredient is the density fluctuation and its
redshift evolution. In linear theory after matter-radiation equality,
all Fourier modes of the mass density perturbation, 
$\delta(\bmf{x})(\equiv \delta \rho_m(\bmf{x})/\bar{\rho}_m)$, grow at
the same rate, the growth rate. For general dark energy equation of
state $w(a)$, the growth rate $D(a)$ can be obtained by solving the
following differential equation for $G(a)\equiv D(a)/a$
\cite{WangSteinhardt98}: 
\begin{eqnarray}
\frac{d^2G}{d\ln a^2}+\left[\frac{5}{2}-\frac{3}{2}w(a)\Omega_{\rm
    DE}(a)\right]\frac{dG}{d\ln
  a}&&\nonumber\\
&&\hspace*{-8em}+\frac{3}{2}\left[1-w(a)\right]\Omega_{\rm DE}(a)G=0,
\label{eq:growth}
\end{eqnarray}
with the boundary condition $G(a)=1$ and $dG/d\ln a=0$ at $a\ll 1$.
The linear density fluctuation is characterized by the linear power
spectrum $P_m^{\rm L}(k;z)$, which is related with the primordial
curvature fluctuation as \cite{Takadaetal:06}
\begin{eqnarray}
\frac{k^3}{2\pi^2}P_m^{\rm L}(k;z)&=&
\delta_\zeta^2\left(\frac{2k^2}{5H_0^2\Omega_M}\right)^2 
\left[T(k)D(a)\right]^2\nonumber\\
&&\times\left(\frac{k}{k_0}\right)^{n_s-1+(1/2)\alpha_s\ln(k/k_0)},
\label{eq:linearps}
\end{eqnarray}
where $T(k)$ is the transfer function of matter perturbations with
baryon oscillations \cite{EisensteinHu:98}, $\delta_\zeta$ is the
normalization of the fluctuation at $k_0=0.002~$Mpc$^{-1}$ following
the convention in \cite{Komatsu:09}, and $\alpha_s$ is the running
index of the primordial power spectrum. We adopt no running, i.e.,
$\alpha_s=0$ as our fiducial choice.

\subsection{Cluster Cosmology}
\label{sec:cluster}

The galaxy cluster observables we will consider in this paper are the 
number counts and the cluster power spectrum which can both be measured
from a given survey region. In this section we briefly review the basics
of the cluster observables. 

\subsubsection{Number Counts}

The number counts of clusters in a given mass and redshift bin is the
most fundamental quantity in cluster cosmology. The cosmological power
of cluster number counts arises from their exponential sensitivity to
the amplitude of the initial density perturbations. However, to
implement this experiment, the total mass of each cluster, which is
dominated by dark matter, has to be inferred from available
observables such as lensing, member galaxies, X-ray and the SZ effect.
We follow the previous work \cite{LimaHu:05} to model the
mass-observable relation including uncertainties in the mass inference
from available data.  The model relation assumes that the probability 
of obtaining the mass inferred from observables, $M_{\rm obs}$, for a
cluster with the true mass of $M$ can be represented by the log-normal
distribution:  
\begin{equation}
p(M_{\rm obs}|M)=\frac{1}{\sqrt{2\pi}\sigma_{\ln M}}
\exp\left[-x^2(M_{\rm obs})\right]\frac{1}{M_{\rm obs}},
\label{eq:mobs}
\end{equation}
where
\begin{equation}
x(M_{\rm obs})\equiv \frac{\ln M_{\rm obs}-\ln M - 
\ln M_{\rm bias}}{\sqrt{2}\sigma_{\ln M}}.
\end{equation}
Unless otherwise specified, we assume $\ln M_{\rm bias}=0$ and
$\sigma_{\ln M}=0.3$, though uncertainties of the mass variance
$\sigma_{\ln M}^2$ and the mass bias $\ln M_{\rm bias}$, including
their mass and redshift dependences, are fully taken into account 
in cosmological parameter forecasts. 

\begin{figure}[t]
\begin{center}
\includegraphics[width=0.42\textwidth]{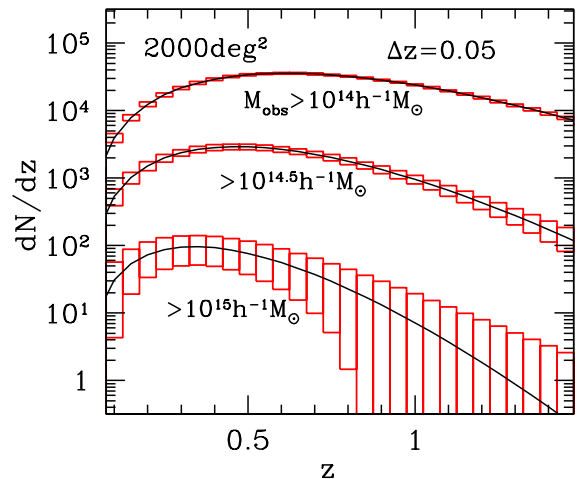}
\caption{The predicted number distributions $dN/dz$ of clusters for
  the survey area of 2000~deg$^2$. Here we consider three mass thresholds,
  $M_{\rm obs}>10^{14}h^{-1}M_\odot$, $10^{14.5}h^{-1}M_\odot$, and 
  $10^{15}h^{-1}M_\odot$, where the observed mass $M_{\rm obs}$ is
  related with the true mass via Eq.~(\ref{eq:mobs}). Boxes indicate
  the binned uncertainties (see Sec.~\ref{subsec:nc}) for the bin size
  of $\Delta z=0.05$.} 
 \label{fig:num}
\end{center}
\end{figure}

The average number density of clusters that lie within the $i$-th
redshift bin $z_{i,{\rm min}}<z<z_{i,{\rm max}}$ and the $b$-th mass
bin defined  in terms of the observed mass, 
$M_{b,{\rm min}}<M_{\rm obs}<M_{b,{\rm max}}$, is given by
\begin{eqnarray}
&&\hspace{-3em}\bar{n}_{i(b)}=\int_{z_{i,{\rm min}}}^{z_{i,{\rm max}}}
\!\!dz\frac{d^2V}{dzd\Omega}\int^{M_{b,{\rm max}}}_{M_{b,{\rm min}}}
\!\!dM_{\rm obs} \nonumber\\
&&\hspace{2em}\times 
\int\!\! dM \frac{dn}{dM}p(M_{\rm obs}|M)\nonumber \\
&&\hspace{-1em}=\int
dz\frac{d^2V}{dzd\Omega}\int dM \frac{dn}{dM}S_{i(b)}(M;z),
\label{eq:numdens}
\end{eqnarray}
where $d^2V/dzd\Omega=\chi^2/H(z)$ denotes the comoving volume element
per unit redshift and per unit steradian, and $S_{i(b)}(M;z)$ is the
selection function in the $i$-th redshift bin and the $b$-th mass bin,
defined by 
\begin{eqnarray}
S_{i(b)}(M;z)&\equiv&\Theta(z-z_{i,{\rm min}})\Theta(z_{i,{\rm max}}-z)\nonumber\\
&&\hspace*{-3.5em}\times
\frac{1}{2}\left[{\rm erfc}\{x(M_{b,
{\rm min}})\}-{\rm erfc}\{x(M_{b,{\rm max}})\}\right],
\label{eq:select_cluster}
\end{eqnarray}
where $\Theta(x)$ is the Heaviside step function and ${\rm erfc}(x)$ is
the complementary error function. The 
redshift dependence of $S_{i(b)}(M)$ may come from a possible
variation of the minimum and maximum mass with redshift, and also from
a possible redshift evolution of $\ln M_{\rm bias}$ and $\sigma_{\ln
  M}$ as considered later. In the following we will use the indices
$i,j,\dots $ and $b,b',\dots$ to denote the redshift bins and the
mass bins, respectively, for notational convenience. 

The function $dn/dM$ in Eq.~(\ref{eq:numdens}) denotes the the halo
mass function, for which we employ a fitting formula presented by
\cite{Bhattacharyaetal:10}. When we include the local-type primordial
non-Gaussianity parametrized by $f_{\rm NL}$, we employ the same
method as adopted in \cite{Oguri:09} to compute the modified mass
function. 

For a given survey area of $\Omega_s$ the mean number counts of clusters
in the redshift and mass bin is given by
\begin{equation}
N_{i(b)}=\Omega_s \bar{n}_{i(b)}.
\label{eq:counts}
\end{equation}
We show in Fig.~\ref{fig:num} how the number counts of clusters scale
with redshift and mass threshold. 

\subsubsection{Cluster-Cluster Power Spectrum}

Another useful cluster observable for cosmology is the cluster-cluster
correlation function. While clusters are biased tracers of the
underlying mass distribution, the cluster bias is, in contrast to the
galaxy bias, well described by the halo bias. As in the case of the
halo mass function, the halo bias is accurately predictable for a
given cosmological model, e.g., based on a suite of $N$-body
simulations, and is sensitive to halo mass. For this reason, combining
the cluster number counts with the cluster correlation function can
provide a promising way for self-calibrating the mass-observable relation  
\cite{MajumdarMohr:04,LimaHu:04,LimaHu:05,Oguri:09,Cunhaetal:10}.
Once the mass-observable relation is sufficiently calibrated, the
cluster correlation function itself carries clean cosmological
information in the sense that the information is mainly from the
linear or weakly nonlinear regimes. Furthermore, if the primordial
non-Gaussianity exists, it induces characteristic scale-dependent
features on the halo bias at large length scales \cite{Dalaletal:08},
which can be efficiently extracted by measuring the cluster
correlation function. 

In this paper we use the Fourier-counterpart of the cluster correlation
function, the cluster power spectrum, for parameter forecasts.
Using the Limber's approximation \cite{Limber:54}, we can compute the
angular power spectrum between pairs of clusters, which are in the $b$-
and $b'$-th mass bins, respectively, and at the same $i$-th redshift
bin:
\begin{equation}
C^{\rm hh}_{i(bb')}(\ell)=\int\!\!d\chi W^{\rm h}_{i(b)}(z)
W^{\rm h}_{i(b')}(z)\chi^{-2}P^{\rm L}_m\!\left(k=\frac{\ell}{\chi}; z \right),
\label{eq:ccps}
\end{equation}
where $P^{\rm L}_m(k;z)$ is the linear mass power spectrum at redshift 
$z$ (Eq.~[\ref{eq:linearps}]), and $W^{\rm h}_{i(b)}$ is the weight
function defined as 
\begin{equation}
W^{\rm h}_{i(b)}(z)
\equiv \frac{1}{\bar{n}_{i(b)}}\frac{d^2V}{d\chi d\Omega}\int\!\! dM
\frac{dn}{dM}S_{i(b)}(M)b_h(M;z). 
\label{eq:cweight}
\end{equation}
Here $d^2V/d\chi d\Omega=\chi^2$ and $b_h(M;z)$ is the bias parameter
for halos with mass $M$ and at redshift $z$, which should not be
confused with the mass bin index ${(b)}$. Again, we adopt a fitting
formula presented by \cite{Bhattacharyaetal:10} in computing
$b_h(M;z)$. When we include the
effect of the primordial non-Gaussianity on the halo bias, we employ the
method \cite{LoVerdeetal:08,Dalaletal:08}.

\begin{figure}[t]
\includegraphics[width=0.42\textwidth]{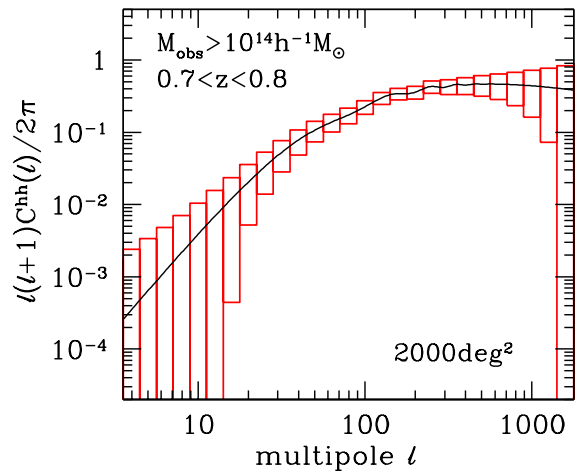}
\caption{The 
  cluster-cluster angular power spectrum for clusters with
  $M_{\rm obs}>10^{14}h^{-1}M_\odot$ and $0.7<z<0.8$. Boxes indicate
  the binned uncertainties (see Sec. \ref{subsec:ac}) for the bin size
  of 0.1~dex, assuming the survey area of 2000~deg$^2$.}  
 \label{fig:cl_hh}
\end{figure}

Fig.~\ref{fig:cl_hh} shows an example of the angular power spectrum, for
clusters in a redshift slice of $0.7<z<0.8$ and with observed masses 
$M_{\rm obs}>10^{14}h^{-1}M_\odot$. The boxes around the curve show
the expected measurement errors for our fiducial survey (see below for
details).  

\subsection{Weak Lensing Observables}

\subsubsection{Weak Lensing Basics}

Due to the geometrical nature of gravitational lensing, the lensing
shear amplitude scales with the critical surface density $\Sigma_{\rm
crit}$ for a given population of source galaxies 
\begin{equation}
\Sigma_{\rm crit}^{-1}(z) \equiv 4\pi G \chi(z)
(1+z)^{-1}\left[1-\chi(z) \sang{\frac{1}{\chi(z_s)}}\right],
\label{eq:sigmacr}
\end{equation}
where $\langle \chi(z_s)^{-1}\rangle$ is the inverse distance averaged
over the source galaxy population defined by $z_s>z_{s,{\rm min}}$
(see Eq.~[\ref{eq:ave_chis}]). 
Since we are interested in the statistical weak lensing, we here defined
the {\it mean} critical surface density for a lensing object at redshift
$z$, averaged over the redshift distribution of source galaxies as in
Eq.~(\ref{eq:idea_stacked}).

To compute the lensing efficiency, we need to specify a population of
source galaxies. We simply assume that photometric galaxies follow the
redshift distribution given by
\begin{equation}
\frac{dp}{dz_s}=\frac{z_s^2}{2z_0^3}\exp\left(-\frac{z_s}{z_0}\right), 
\label{eq:dngaldz}
\end{equation}
where $z_0$ is free parameter, and $dp/dz_s$ satisfies the normalization
condition $\int_0^\infty\!dz_s dp/dz_s=1$. With this form, the mean
redshift $z_{ m}$ is given as $z_m=\sang{z_s}=3z_0$. 
As our fiducial choice, we assume $z_0=1/3$ corresponding to the mean
redshift $z_m=1$, which roughly mimics the depth of the Subaru 
Hyper Suprime-cam (HSC) Survey.

For the method studied in this paper, we focus on a sub-population
of source galaxies for extracting the lensing signals (see also
Sec.~\ref{sec:idea} for discussions). The source galaxies are
defined from galaxies at redshifts greater than the minimum redshift
$z_{s,{\rm min}}$. Our fiducial choice is $z_{s,{\rm min}}=1.5$ such
that source galaxies are behind all clusters of interest. Thus the
source redshift dependence of the lensing efficiency is only via the
single quantity, $\sang{\chi_s^{-1}}$. We will employ the mean source
redshift of the whole galaxy distribution (i.e., {\it not} the mean
redshift of the sub-population defined by $z_s>z_{s,{\rm min}}$),
$z_m$, to describe the source redshift dependence in making parameter
forecasts.  

The lensing shearing effect on source galaxy images is caused by the
foreground structure along the line of sight. The shear amplitude for a
given galaxy in the angular direction $\bmf{\theta}$ on the sky is
closely related to the weighted mass density field along the line of
sight (e.g., see \cite{BartelmannSchneider:01}):
\begin{equation}
\vec{\gamma}(\bmf{\theta}) \leftarrow \kappa(\bmf{\theta})
\equiv \int\!\!d\chi W^{\kappa}(z)
\delta_m(\chi\bmf{\theta},\chi), 
\label{eq:kappa}
\end{equation}
where the vector notation $\vec{\gamma}$ is used to explicitly show
that the shear field is a spin-2 field, i.e., has two components at
each position on the sky, $\kappa$ is the so-called convergence field,
and $\delta_m(\chi\bmf{\theta},\chi)$ is the mass density fluctuation
field. Note that lensing does not occur for a completely homogeneous
universe with $\delta_m=0$. The function $W^{\kappa }(z)$ is the
lensing weight function defined by 
\begin{equation}
W^{\kappa}(z)\equiv \frac{\bar{\rho}_{m}(z)}{(1+z)\Sigma_{\rm crit}(z)},
\label{eq:kappa_weight}
\end{equation}
where $\bar{\rho}_m(z)$ denotes the mean mass density of the universe
at redshift $z$.

Using the Limber's approximation, we can compute the lensing power
spectrum for the source galaxy population defined above as
\begin{equation}
C^{\kappa\kappa}(\ell)=\int\!d\chi \left[W^{\kappa}(z)\right]^2 
\chi^{-2} P^{\rm NL}_m\!\left(k=\frac{\ell}{\chi};z\right), 
\label{eq:kappaps}
\end{equation}
where $P^{\rm NL}_m(k)$ denotes the nonlinear mass power spectrum.  We
use the fitting formula in \cite{Smithetal:03} for computing $P^{\rm
NL}_m(k)$ for a given cosmological model.

\subsubsection{Shear-Cluster Power Spectrum: Stacked Lensing}

As briefly described in Sec.~\ref{sec:idea}, if a catalog of galaxy
clusters in the similar redshift range is available, the contribution
of the clusters to the lensing signals of background source galaxies
can be extracted by measuring a cross-correlation between the cluster
distribution and the shapes of source galaxies -- the so-called
shear-cluster correlation or the stacked lensing.  For the limit that
a redshift slice of clusters, centered at $z_l$, is sufficiently thin,
the shear-cluster correlation is given by
\begin{equation}
\sang{\gamma_+}(\theta; z_l)
\equiv \frac{1}{\bar{n}_{\rm cl}}\sang{n_{\rm cl}(\bmf{\theta}';z_l)
\gamma_+(\bmf{\theta}-\bmf{\theta}')}=\frac{\Delta
\Sigma_m(\chi_l\theta;z_l)}{\Sigma_{\rm crit}(z_l)}, 
\end{equation}
where $\chi_l\equiv \chi(z_l)$ and $\Delta\Sigma_m(\chi_l\theta;z_l)$
is the radial profile of the projected mass density around the cluster 
center. Observationally the shear-cluster correlation can be measured
by averaging the tangential ellipticity component of source galaxy
images in each radial bins over all the pairs of background galaxies
and clusters in the sample, where the tangential ellipticity is
defined with respect to the cluster center in each cluster region
\cite{Sheldonetal:01,Sheldonetal:09,Okabeetal:10}. Note that the cluster
distribution is treated as discrete, point distribution, and a
representative point for each cluster needs to be chosen for each
cluster, the cluster center in this method \cite{TakadaBridle:07}.

There are notable advantages in the stacked lensing technique, e.g., 
compared to cosmic shear. (1) As discussed at length in
Sec.~\ref{sec:idea}, the source redshift uncertainty can be calibrated
by combining the stacked lensing signals of different cluster
redshifts. (2) The stacked lensing profile is less sensitive to
substructures within and asphericity of the individual clusters and
also to uncorrelated large-scale structure along the same line of
sight.  Since the signal probed is linear in shear, which can be both
positive and negative due to its spin-2 field nature, these
contaminating effects are averaged out via the stacking, if a
sufficient number of clusters are available and as long as the
universe is statistically isotropic and homogeneous. As a result, the
stacked lensing technique probes the {\it average} mass distribution
around the clusters, which can be considered as a spherically
symmetric profile. Therefore the stacked lensing profile is
one-dimensional, i.e., given as a function of radius from the cluster
center.  (3) The stacking boosts the signal to noise ratios of shear 
signals at very large radii, where the shear detection is limited by too
small signal for an individual cluster. The large-angle shear signal is
in the linear or weakly nonlinear regime, and hence carries relatively
clean cosmological signals.  (4) If source galaxies are not overlapped
with cluster redshifts, another serious contamination to cosmic shear, 
the intrinsic ellipticity alignment \cite{HirataSeljak:04}, does not
affect the stacked lensing signals. 

Assuming the flat-sky approximation, we can express the stacked
lensing profile in terms of the angular power spectrum
$C^{{\rm h}\kappa}_{i(b)}(\ell)$ as 
\begin{equation}
\sang{\gamma_+}_{i(b)}(\theta)= \int\!\frac{\ell d\ell}{2\pi}
C^{{\rm h}\kappa}_{i(b)}(\ell)
J_2(\ell\theta),
\label{eq:gamma_cl}
\end{equation}
where $J_2(x)$ is the second-order Bessel function. The full-sky
expression can be recently formulated in \cite{dePutterTakada:10},
where it was shown that the effect of the celestial sphere causes only
a few percent change in the shear amplitude at very large angles greater
than 10~degrees. We do not think that the inaccuracy of the flat-sky
approximation largely changes the following results, but a more
rigorous calculation will be our future work. The angular power
spectrum contains the equivalent information to the stacked lensing
profile. Hence we will use the power spectrum in conducing parameter 
forecasts for computational simplicity. For example, the covariance 
calculation is easier in Fourier space than in real space. 

The stacked shear profile arises from two contributions. The shear
signal at angular scales smaller than a typical projected virial radius
of clusters arises from the average mass distribution within one halo,
the so-called 1-halo term, while the large-angle shear arises from the
average large-scale structure surrounding the clusters, the 2-halo
term. In what follows, we model the shear-cluster correlation function
based on the halo model.

\subsubsection{Stacked Lensing: 1-Halo Term}

As described above, we need a model of the radial mass profile of halo 
in order to compute the 1-halo term at small angular scales.  In this
paper, we employ the Navarro-Frenk-White (NFW) profile
\cite{Navarroetal:97} as used in \cite{TakadaBridle:07}. The NFW
profile is defined by 
\begin{equation}
\rho(r)=\frac{\rho_s}{(r/r_s)(1+r/r_s)^2}.
\end{equation}
The density parameter $\rho_s$ is specified by imposing that the mass
enclosed within a sphere of the virial radius is equal to the virial
mass $M$ 
\begin{equation}
\rho_s=\frac{\Delta(z)\bar{\rho}_m(z)c^3}{3m_{\rm nfw}(c)}=\frac{M}{4\pi
  r_s^3m_{\rm nfw}(c)},
\end{equation}
where
\begin{equation}
m_{\rm nfw}(c)\equiv\int_0^c \frac{r}{(1+r)^2}dr=\ln(1+c)-\frac{c}{1+c}.
\end{equation}
Here $c\equiv r_{\rm vir}/r_s$ is a concentration parameter, $\Delta(z)$ is
a nonlinear overdensity that defines the virial mas based on the
spherical collapse model (we adopt a fitting formula by
\cite{NakamuraSuto:97}). Motivated by the result of $N$-body
simulations (e.g., \cite{Bullocketal:01}), we adopt the following model
for the mass and redshift dependence of the concentration parameter
\begin{equation}
c(M,z)=A_{\rm vir}\left(\frac{M}{2\times
 10^{12}h^{-1}M_\odot}\right)^{B_{\rm vir}}(1+z)^{C_{\rm vir}},
\label{eq:concent}
\end{equation}
Here $A_{\rm vir}$, $B_{\rm vir}$ and $C_{\rm vir}$ are free parameters,
and we will employ ($A_{\rm vir}$, $B_{\rm vir}$, 
$C_{\rm vir}$)=($7.85$, $-0.081$, $-0.71$) as our fiducial parameter set
according to the result in \cite{Duffyetal:08}. 

Using the method developed in \cite{Mandelbaumetal:05} (also see
\cite{TakadaBridle:07}), the Limber's approximation and the flat-sky
approximation, we can derive the power spectrum for the 1-halo term of
shear-cluster correlation between the source galaxies and the clusters
in the $i$-th redshift slice and the $b$-th mass bin. We find that the
1-halo term is described by 
\begin{eqnarray}
C^{{\rm h}\kappa, {\rm 1h}}_{i(b)}\!(\ell)&=&
\frac{1}{\bar{n}_{i(b)}}\int\!dz \frac{d^2V}{dz d\Omega}
\nonumber\\
&&\times\int\!dM \frac{dn}{dM}S_{i(b)}(M;z) \tilde{\kappa}_M(\ell;z),
\label{eq:cl_hk_1h}
\end{eqnarray}
where $\tilde{\kappa}_M(\ell)$ denotes the two-dimensional Fourier
transform of the projected NFW profile (i.e., convergence
$\kappa(\theta)$) defined as  
\begin{equation}
\tilde{\kappa}_M(\ell;z)\equiv\int2\pi\theta d\theta \kappa(\theta)
J_0(\ell\theta)=\frac{M\tilde{u}_M(k=\ell/\chi; z)}{(1+z)^{-2}\chi^2\Sigma_{\rm
crit}(z)}. 
\label{eq:kappa_halo}
\end{equation}
Here $J_0(x)$ is the zeroth order Bessel function and $\tilde{u}_M(k)$
denotes the three-dimensional Fourier transform of the normalized NFW
profile $u_M(r)=\rho(r)/M$, which has the following analytic form 
\cite{Scoccimarroetal:01}
\begin{eqnarray}
\tilde{u}_M(k)&=&\frac{1}{m_{\rm nfw}(c)}\left[\sin x\left\{{\rm Si}[x(1+c)]-{\rm
      Si}(x)\right\}\right.\nonumber\\
&&\hspace*{-2.5em}\left.+\cos x\left\{{\rm Ci}[x(1+c)]-{\rm Ci}(x)\right\}-\frac{\sin(xc)}{x(1+c)}\right],
\label{eq:nfw_uk}
\end{eqnarray}
where $x\equiv (1+z)kr_s$, and ${\rm Si}(x)$ and ${\rm Ci}(x)$ are
sine and cosine integrals.  The relation between the two-dimensional
and three-dimensional Fourier transforms of halo profile was also
discussed in Sec.~3.2 in \cite{TakadaJain:03a}.

\subsubsection{Effect of Halo Centering Offset}
\label{sec:offset}

For the derivation of the 1-halo term above, we implicitly assume that
the stacking can be done by properly selecting the halo center of each
cluster, e.g., the Fourier transform of halo profile, $\tilde{u}_M(k)$
(Eq.~[\ref{eq:nfw_uk}]), is defined with respect to the halo center. 
On individual cluster basis, a dark matter halo generally contains
many substructures, has a non-spherically symmetric mass distribution,
and has no clear boundary with its surrounding structures, all of
which reflect the collision-less nature of dark matter. Such a
complexity, as well as invisible nature of dark matter, implies that
it is not straightforward to define the halo center for each cluster.
Observationally the cluster center will be chosen from either/both of
the position of brightest cluster galaxy (BCG) or the peak of the
X-ray or SZ maps. If the significant strong/weak lensing detection
is achieved, albeit rare, the cluster center can be inferred from the
mass distribution reconstructed from the lensing. Recently we
\cite{Ogurietal:10} found that, from the detailed weak lensing
analysis of individual clusters, the BCG position is indeed close to
the center inferred from the lensing-reconstructed mass distribution,
although there are some clusters whose centers appear to be
significantly 
offseted from their BCGs. In what follows, we develop a
formulation to account for the effect of possible centering offset on
the stacked lensing.   

As derived in detail in Appendix~\ref{sec:off}, the halo centering
offset acts as a smoothing effect on the stacked lensing signal. For a 
spherically symmetric halo (in an average sense), if the halo center
is incorrectly chosen with an offset angle $\theta_{\rm off}$ from the
true center, the measured stacked lensing profile in Fourier space is
modified as 
\begin{equation}
\tilde{\kappa}_{M,{\rm off}}(\ell)=
\tilde{\kappa}_M(\ell)J_0(\ell\theta_{\rm off}). 
\end{equation}
In this paper, we are interested in the statistical average of
tangential shears around many different clusters. Naturally all
clusters do not share the same offset angle, but rather the offset
should follow some probability distribution
\begin{equation}
\tilde{\kappa}_{M,{\rm off}}(\ell)=\tilde{\kappa}_M(\ell)\int d
\theta_{\rm off} J_0(\ell \theta_{\rm off}) p_{\rm off}(\theta_{\rm off}),
\label{eq:kappa_off_prep}
\end{equation}
where $p(\theta_{\rm off})$ is the probability distribution function of
$\theta_{\rm off}$.

As implied from the mock galaxy catalog \cite{Johnstonetal:07} or
lensing measurements \cite{Ogurietal:10}, one reasonable choice of
$p_{\rm off}(\theta_{\rm off})$ is a sum of the two-dimensional
Gaussian distributions given by
\begin{equation}
p_{\rm off}(\theta_0)=\sum_i f_i\frac{\theta_0}{\sigma_{s,i}^2}\exp\left(-\frac{\theta_0^2}{2\sigma_{s,i}^2}\right).
\end{equation}
In this case, we can analytically integrate
Eq.~(\ref{eq:kappa_off_prep}) to obtain 
\begin{equation}
\tilde{\kappa}_{M,{\rm off}}(\ell)=\tilde{\kappa}_M(\ell)\sum_i 
f_i\exp\left(-\frac{1}{2}\sigma_{s,i}^2\ell^2\right),
\end{equation}
where $\sigma_s$ denotes the width of the Gaussian distribution in
radian. Again the distribution of the offset angles has been explored
by studying individual clusters based on both numerical simulations
\cite{Johnstonetal:07,HilbertWhite:10} and observations
\cite{LinMohr:04,Koesteretal:07,Shanetal:10,Ogurietal:10}. Motivated
by the results of such previous works, in this paper, we adopt a
two-component model consisting of clusters whose inferred center align 
well with the true center and clusters with their measured center
significantly offseted from the true center. Specifically, we adopt the
following expression:
\begin{equation}
\tilde{\kappa}_{M,{\rm off}}(\ell)=\tilde{\kappa}_M(\ell)\left[
f_{\rm cen}+(1-f_{\rm
  cen})\exp\left(-\frac{1}{2}\sigma_{s}^2\ell^2\right)\right], 
\label{eq:kappa_offdis}
\end{equation}
with $f_{\rm cen}$ denotes a parameter to specify the fraction of
clusters which have their inferred center to be consistent with the true
center. We employ $f_{\rm cen}=0.75+0.05\ln(M/M_{\rm piv})$ with $M_{\rm
piv}=3\times 10^{14}h^{-1}M_\odot$ and $D_A(z)\sigma_s=0.42h^{-1}{\rm Mpc} $
as our fiducial values, which are roughly consistent with the results
obtained from mock galaxy catalogs \cite{Johnstonetal:07}. 

Thus the 1-halo term of the shear-cluster power spectrum including the
halo off-centering effect can be computed by replacing
$\tilde{\kappa}_M(\ell) $ in Eq.~(\ref{eq:cl_hk_1h}) with
$\tilde{\kappa}_{M,{\rm off}}(\ell)$ in Eq.~(\ref{eq:kappa_offdis})
for a given model of the halo centering offsets.  The real-space stacked
lensing profile can be also obtained via the transform given by
Eq.~(\ref{eq:gamma_cl}).

\begin{figure}[t]
\begin{center}
\includegraphics[width=0.4\textwidth]{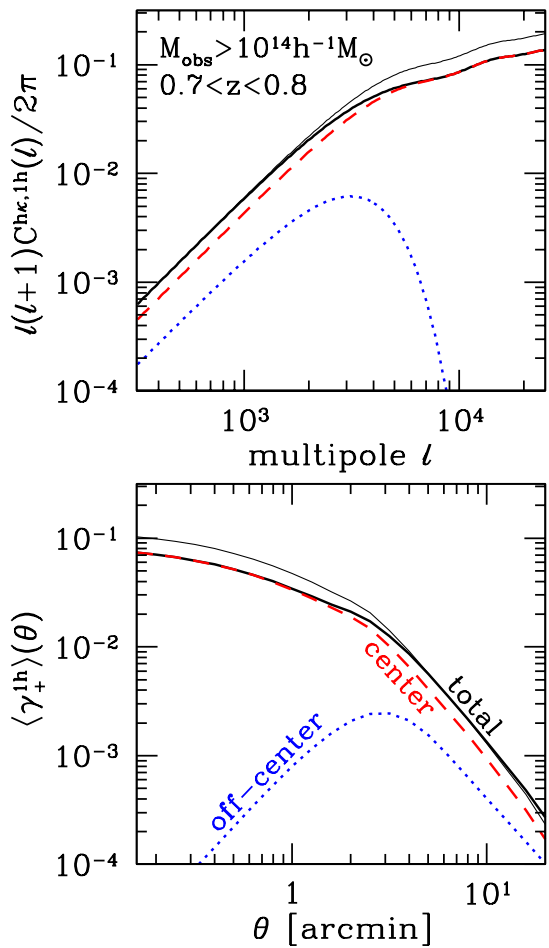}
\caption{{\it Upper panel}: The angular power spectrum of stacked
  cluster weak lensing signals from main halos (1-halo term) with the
  effect of off-centering. In this example we consider clusters with
  $M_{\rm obs}>10^{14}h^{-1}M_\odot$ located as $0.7<z<0.8$. The total
  power spectrum is shown by the thick solid curve, whereas the
  contributions from populations with and without off-centering
  effect, as expressed by Eq.~(\ref{eq:kappa_offdis}), are shown by
  the dotted and dashed curves, respectively. The thin solid curve
  shows the total power spectrum when the true centers of all the
  clusters are properly identified (i.e., $f_{\rm cen}=1$). 
  {\it Lower panel}: The corresponding stacked  shear profiles in
  real space.}
 \label{fig:stack_off}
\end{center}
\end{figure}

Fig.~\ref{fig:stack_off} shows the expected signals for the 1-halo term
of the stacked lensing profile, with and without the effect of halo
centering offset (see Fig.~4 in \cite{Johnstonetal:07} for the similar
plot). Here we consider lensing clusters with observed masses
$M_{\rm obs}>10^{14}h^{-1}M_\odot$ and in the redshift slice of
$0.7<z<0.8$. The figure explicitly shows that the halo off-centering
effect smooths out the stacked lensing signal at small angular
scales. In other words, for a fixed sample of clusters, a better
choice of the cluster center can be explored to some extent by
monitoring the shear amplitudes in order to maximize the shear
amplitudes at small radii. At least, if several choices of the cluster
center are available, e.g., from optical, X-ray and SZ data, the
stacked lensing amplitudes can be compared to obtain the best
inference of the cluster center for stacked lensing analysis. 

\subsection{Stacked Lensing: 2-Halo Term}

Now we consider the 2-halo term contribution to the stacked lensing
signal, which arises from the mass distribution surrounding lensing
clusters. Again by employing the halo model formulation (e.g., 
\cite{TakadaBridle:07}), we can derive the 2-halo term of power
spectrum for the stacked lensing between the source galaxies and the
clusters in the $i$-th redshift bin and the $b$-th mass bin as
\begin{equation}
C^{{\rm h}\kappa,{\rm 2h}}_{i(b)}(\ell) =\int\!d\chi W^{\rm h}_{i(b)}(z)
W^{\kappa}(z)\chi^{-2}P_m^{\rm L}\left(k=\frac{\ell}{\chi}; z\right),
\label{eq:cl_hk_2h}
\end{equation}
where the weight functions $W^{\rm h}(z)$ and $W^{\kappa}(z)$ are given by
Eqs.~(\ref{eq:cweight}) and (\ref{eq:kappa_weight}), respectively.
We have set $\int\!dM (dn/dM)b(M)M\tilde{u}_M(k)=1$ at angular scales
relevant for the 2-halo term; the prefactor appears in the formal
derivation of 2-halo term based on the halo model formulation (see
\cite{TakadaBridle:07}), but we used the expression above for
notational simplicity. We have checked that this simplification does
not change the following results.  

\begin{figure*}[t]
\begin{center}
\includegraphics[width=0.85\textwidth]{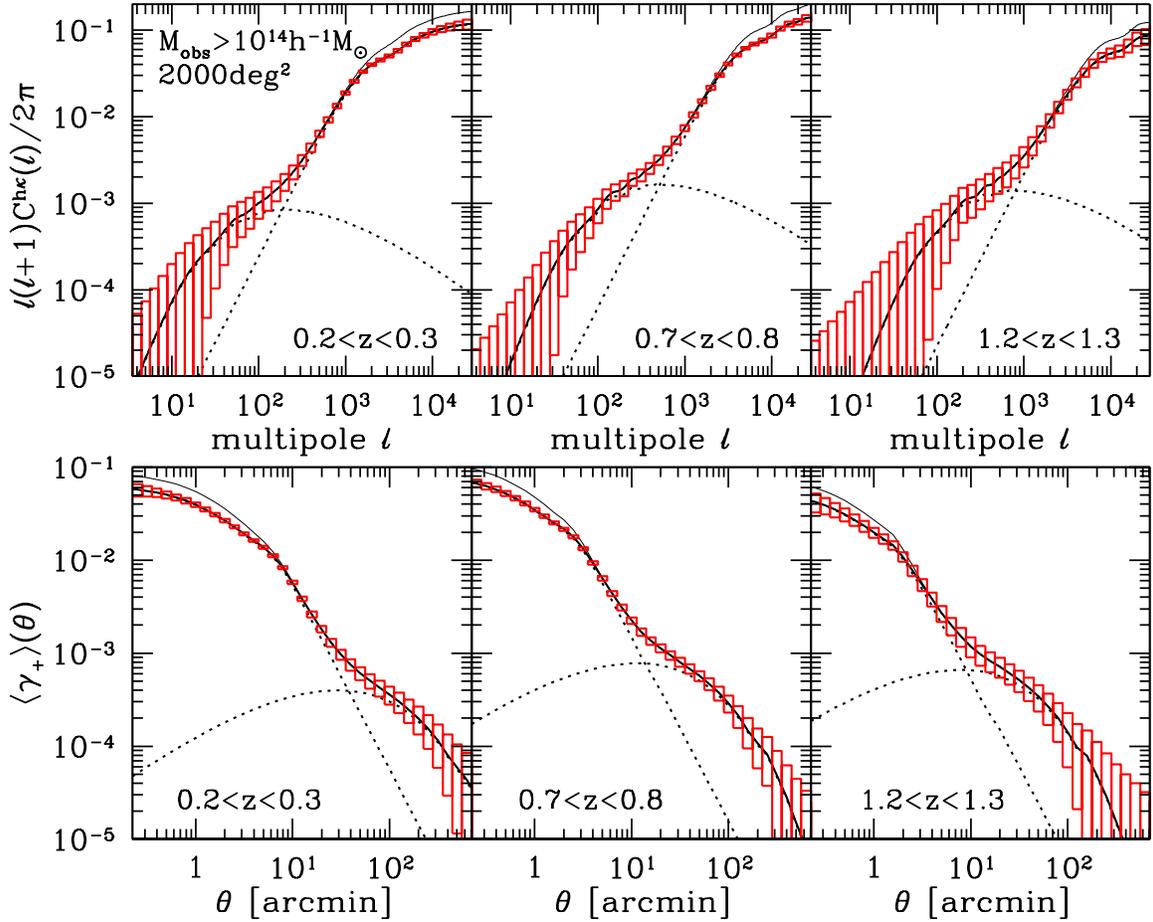}
\caption{
  The bold solid curves show 
  stacked lensing signals in Fourier ({\it upper-row panels}) 
  and real ({\it lower panels}) spaces, for three cluster redshift
  slices, $0.2<z<0.3$ ({\it left panel}), $0.7<z<0.8$ ({\it middle}),
  and $1.2<z<1.3$ ({\it right}), respectively. Note that in all the
  plots we assumed a background source galaxy sample defined by
  $z_s>1.5$. The dotted curves indicate the contributions from 1-halo
  and 2-halo terms. While the effect of off-centering is included
  assuming the two component model described by
  Eq.~(\ref{eq:kappa_offdis}), we also show the case without any
  offset by the thin solid curves for reference. The boxes around each
  curve indicate the measurement errors for the binned power spectra
  or the binned shear profiles (see Sec. \ref{subsec:ac}) for the bin
  size of 0.1~dex, assuming the survey area of 2000~deg$^2$.}  
 \label{fig:stack}
\end{center}
\end{figure*}

\subsubsection{Summary: Halo Model Calculation of Stacked Lensing}

In summary, using Eqs.~(\ref{eq:cl_hk_1h}), (\ref{eq:kappa_offdis}), 
and (\ref{eq:cl_hk_2h}), we can compute the halo model based
prediction for the angular power spectrum of the stacked lensing for a
given cosmological model: 
\begin{equation}
C^{{\rm h}\kappa}_{i(b)}(\ell)=
C^{{\rm h}\kappa,{\rm 1h}}_{i(b)}(\ell)+C^{{\rm h}\kappa,{\rm
    2h}}_{i(b)}(\ell).
\label{eq:c_hk_tot}
\end{equation}
The corresponding stacked shear profile in real space is computed
using Eq.~(\ref{eq:gamma_cl}).

Fig.~\ref{fig:stack} shows the expected stacked lensing signals in both
Fourier and real spaces, for several different redshift slices.  The
figure shows that the 1-halo term arising from the halo mass profile
dominates the signal at small angular scales $\simlt 10'$ or multipoles
greater than a few hundred, while the 2-halo term dominates at 
the smaller multipoles or $\theta\gtrsim 10'$. The boxes around each
curve show the expected $1\sigma$ measurement errors in each angular or
multiople bins for a Subaru HSC-type weak lensing survey with an area
coverage of 2000~deg$^2$, which are computed using the covariance
matrix calculations shown in the next section. Note that the error bars
are uncorrelated between different multipole bins for the power
spectrum measurement under the Gaussian error assumption, which we
will throughout this paper assume, while the error bars are highly
correlated for the real-space stacked shear measurement.

\section{Covariances of Observables and Fisher Matrices}
\label{sec:cov}

To estimate the accuracies of parameter estimation for a given survey,
we employ the Fisher matrix formalism. In this method we first need to
quantify the measurement errors of cosmological observables for given
survey parameters that are chosen to resemble planned surveys. The
covariance matrices of the observables can model the measurement
errors. Then we can propagate the measurement errors into
uncertainties of parameter estimation using the Fisher matrix
formalism.  In this section, we present the covariances between the
observables defined in Sec.~\ref{sec:signal}, and define the Fisher
matrices of each observables. 

\subsection{Number Counts}
\label{subsec:nc}

The covariance matrix of cluster number counts (Eq.~[\ref{eq:counts}])
is given by \cite{HuKravtsov03,TakadaBridle:07}
\begin{equation}
{\rm Cov}(N_{i(b)}, N_{j(b')})=N_{i(b)}\delta_{ij}^K\delta_{bb'}^{K}
+S_{i(bb')}\delta^K_{ij},
\label{eq:cov_cc}
\end{equation}
where $\delta^K_{ij}$ denotes the Kronecker delta function;
$\delta_{ij}^K=1$ if $i=j$, otherwise $\delta^K_{ij}=0$. 
The first term of Eq.~(\ref{eq:cov_cc}) denotes the shot noise arising 
due to the finite number of clusters available, and is only
non-vanishing when both the redshift bins and the mass bins are same,
$i=j$ and $b=b'$.  On the other hand, the second term of
Eq.~(\ref{eq:cov_cc}) gives the sampling variance contribution, which
arises because the number of clusters is fluctuated according to the
large-scale modes of large-scale structure within a surveyed region:
\begin{eqnarray}
S_{i(bb')}&\equiv &\Omega_s^2\bar{n}_{i(b)}\bar{n}_{i(b')}
\int\!\!d\chi W^{\rm h}_{i(b)}(z)W^{\rm h}_{i(b')}(z)\chi^{-2}\nonumber\\
&&\hspace{-5mm}\times 
\int \frac{\ell d\ell}{2\pi}
\left|\tilde{W}_{\rm s}(\ell\Theta_{\rm s})\right|^2
P_m^{\rm L}\!\left(k=\frac{\ell}{\chi};z\right).
\end{eqnarray}
Here $\tilde{W}_{\rm s}(\ell)$ being the Fourier transform of the
survey window function. We assume a circular survey geometry with
survey area $\Omega_s=\pi\Theta_{\rm s}^2$ for simplicity, with the
resulting survey window function  of 
$\tilde{W}_{\rm s}(\ell)=2J_1(\ell\Theta_{\rm s})/(\ell\Theta_{\rm s})$.
The Kronecker delta $\delta^K_{ij}$ in the second term of
Eq.~(\ref{eq:cov_cc}) imposes that the sampling variance is vanishing
for the counts of different redshift slices, i.e. $i\ne j$, assuming
that the redshift slices of clusters are sufficiently wide such that the
cluster distributions in different redshift slices are uncorrelated. 

Then we can write the Fisher matrix for the cluster number counts 
$\bmf{N}\left(\equiv N_{i(b)}\right)$ as
\begin{equation}
F_{\alpha\beta}^{N}=\sum_{I,J}\frac{\partial
  \bmf{N}_{I}}{\partial p_\alpha}[{\rm Cov}(\bmf{N},
  \bmf{N})]^{-1}_{IJ}
\frac{\partial \bmf{N}_J}{\partial p_\beta},
\label{eq:fisher_num}
\end{equation}
where the indices $I$, $J$ run over the cluster redshift and mass bins
($i$, $b$) and $p_\alpha$ denotes a set of model parameters. 

\subsection{Power Spectrum Observables}
\label{subsec:ac}

\subsubsection{Covariance Matrices and Fisher Matrix}

The angular power spectra we consider in this paper as observables are
the cluster power spectrum and the shear-cluster power spectrum (note
that we will also study how the result is changed by further adding the
cosmic shear power spectrum). Therefore the data vector can be given as
\begin{equation}
\bmf{D}\equiv \left(C_{i(bb')}^{\rm hh},C^{{\rm h}\kappa}_{i(b)}\right).
\end{equation}
We need to model the auto- and cross-covariances of these elements.

There are two contributions to the sampling variance in the power
spectrum covariance, the Gaussian and non-Gaussian error contributions,
where the later arises from nonlinear clustering of structure
formation. In this paper we ignore the non-Gaussian error covariances
for simplicity. As carefully studied in \cite{TakadaJain:09} (also see
\cite{Satoetal:09} for a simulation based study), the non-Gaussian
errors are significant at small angular scales, although the impact on 
parameter estimation is not large as long as a sufficient set of
cosmological parameters is included. For example, for the
case of cosmic shear power spectrum, the non-Gaussian errors degrade
the {\it marginalized} error of each parameter only by about
$10\%$. This also likely holds for the case of shear-cluster power
spectrum combined with the cluster experiments, although 
this assumption needs to be carefully studied.

Thus, assuming the Gaussian error covariance, the covariance matrix of
the shear-cluster power spectrum is given by
\begin{eqnarray}
{\rm Cov}\!\!\left(C^{{\rm h}\kappa}_{i(b)}, 
C^{{\rm h}\kappa}_{j(b')}\right)&=&\frac{4\pi}{\Omega_s}
\frac{\delta^{K}_{\ell\ell'}}{(2\ell+1)\Delta\ell}\nonumber\\
&&\hspace{-2em}\times
\left[\hat{C}^{{\rm hh}}_{i(bb')}\hat{C}^{\kappa\kappa}\delta^K_{ij}
+C^{{\rm h}\kappa}_{i(b)}C^{{\rm h}\kappa}_{j(b')}\right].
\label{eq:cov_s-c}
\end{eqnarray}
The spectra $\hat{C}^{\kappa\kappa}$ and $\hat{C}^{\rm hh}_{i(bb')}$
are the observed power spectra including the shot noise contamination
\begin{eqnarray}
\hat{C}^{\rm hh}_{i(bb')}&\equiv & {C}^{\rm hh}_{i(bb')}+
\frac{1}{\bar{n}_{i(b)}}\delta_{bb'}^K
\label{eq:cov_ccps}
\\
\hat{C}^{\kappa\kappa}&\equiv &
C^{\kappa\kappa}+\frac{\sigma_e^2}{2\bar{n}_{\rm gal}}.
\end{eqnarray}
where $\bar{n}_{\rm gal}$ is the mean number density of source galaxies
used for the stacked lensing analysis and $\sigma_{\rm e}$ is the rms
intrinsic ellipticity for the sum of two ellipticity components.
For our fiducial Subaru HSC-type survey, $\bar{n}_{\rm gal}\simeq
5~{\rm arcmin}^{-2}$ for source galaxies at $z_s>1.5$, and we assume
$\sigma_{\rm e}=0.35$, corresponding to $\sim 0.25$ for the rms per
component. The Kronecker delta functions $\delta^{K}_{ij}$ and
$\delta^{K}_{bb'}$ in Eqs.~(\ref{eq:cov_s-c}) and (\ref{eq:cov_ccps})
enforce that the shot noise does not contaminate the cluster power
spectra between clusters in different redshift bins and mass bins.
Note that the shear-cluster power spectrum is not contaminated 
by shot noise. 

\begin{figure}[t]
\begin{center}
\includegraphics[width=0.42\textwidth]{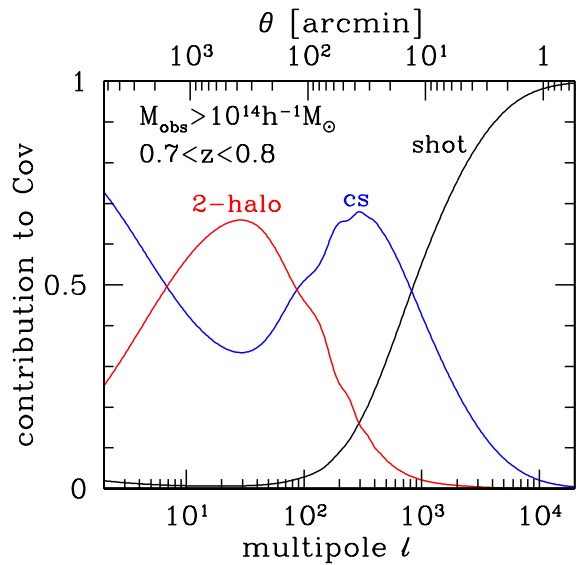}
\caption{Relative contributions to the diagonal covariance matrix
  elements of the stacked lensing signals as a function of multipole
  $\ell$, for clusters with $M_{\rm obs}>10^{14}h^{-1}M_\odot$ and
  $0.7<z<0.8$. From right to left curves, we show the fractional
  contributions from the shot noise (``shot''), the cosmic shear
  (``cs''), and the sampling variance (``2-halo'') terms. See text for
  more details.}  
 \label{fig:covcont}
\end{center}
\end{figure}

Eq.~(\ref{eq:cov_s-c}) implies that the covariance of shear-cluster
power spectrum consists of several contributions: (1) the shot noise
term, which is proportional to $\sigma_e^2/(\bar{n}_{i(b)}\bar{n}_{\rm
gal})$ and becomes important at high multipoles; (2) the cosmic shear
contribution given as $(1/\bar{n}_{i(b)})C^{\kappa\kappa}$; (3) the
sampling variance (2-halo) contribution given by $C^{{\rm hh}}_{i(bb')}
\hat{C}^{\kappa\kappa}\delta^K_{ij}+C^{{\rm h}\kappa}_{i(b)}{C}^{{\rm
h}\kappa}_{j(b')}$. Note that the contributions (1) and (2) are
non-vanishing only if $i=j$ (the same redshift bins). 
In Fig.~\ref{fig:covcont}, we examine how each terms above contribute
to the diagonal elements of covariance matrix (\ref{eq:cov_s-c}) as a
function of multipoles. As expected, the shot noise dominates at very
small angles or very high multipoles, while the contribution from the
cosmic shear is important over a range of angular scales from a few
arcminutes to a degree scale (see also
\cite{Ogurietal:10,Hoekstra:03,Dodelson:04} for the cosmic shear
contribution to cluster shear signals).  Importantly for our method,
the angular scales from a few to 10 arcminutes roughly correspond 
to the virial radii of massive clusters, and the shear signals around
the viral radii are very sensitive to the halo mass, but rather
insensitive to halo profile parameters such as the halo concentration
and the halo off-centering parameter which are more sensitive to the
shape of the shear profile (e.g., see \cite{Okabeetal:10} for such a
discussion based on the actual weak lensing measurements). Hence the
figure implies that the accuracy of halo mass estimation from the
stacked lensing measurement is mainly limited by the cosmic shear
contamination. The sampling variance (3) gives significant
contributions only at very large scales, $\ell \lesssim 10^2$ ($\theta
\gtrsim 100'$), which suggests that this term can safely be neglected
in the shear analysis of individual clusters for which tangential
shear signals can be detected out to the virial radii at most. At the
very large scales the cosmic shear term dominates again; this can be
understood from the asymptotic behaviors of these terms,
$(1/\bar{n}_{i(b)})C^{\kappa\kappa}\propto \ell^{n_s}$ for the cosmic 
shear and $(C^{{\rm h}\kappa})^2\propto (C^{\rm hh}C^{\kappa\kappa})
\propto (\ell^{n_s})^2$ for the sampling variance term in the limit of
$\ell\rightarrow 0$, where $n_s \approx 1$ is the power spectrum tilt.

Similarly, the covariance matrix of cluster power spectrum and the
cross-covariance matrix of $C^{\rm hh}$ and $C^{{\rm h}\kappa}$ are 
\begin{eqnarray}
{\rm Cov}(C^{\rm hh}_{i(bb')}, C^{\rm hh}_{j(\tilde{b}\tilde{b}')})
&=&\frac{4\pi}{\Omega_s}\frac{\delta^K_{\ell\ell'}}
{(2\ell+1)\Delta\ell}\delta^K_{ij}\nonumber\\
&&\hspace*{-12mm}\times\left[\hat{C}^{\rm hh}_{i(b\tilde{b})}
\hat{C}^{\rm hh}_{i(b'\tilde{b}')}
+\hat{C}^{\rm hh}_{i(b\tilde{b}')}\hat{C}^{\rm hh}_{i(b'\tilde{b})}\right],
\end{eqnarray}
and
\begin{eqnarray}
{\rm Cov}(C^{{\rm h}\kappa}_{i(\tilde{b})},C^{\rm hh}_{j(bb')})
&=&\frac{4\pi}{\Omega_s}\frac{\delta_{\ell\ell'}^K}{(2\ell+1)\Delta\ell
}\delta^K_{ij}\nonumber\\
&&\hspace*{-12mm}\times\left[\hat{C}^{{\rm hh}}_{i(\tilde{b}b)}
{C}^{{\rm h}\kappa}_{i(b')}+\hat{C}^{\rm hh}_{i(\tilde{b}b')}
C^{{\rm h}\kappa}_{i(b)}\right].
\end{eqnarray}

The Fisher matrices of the power spectrum observables are given as 
\begin{equation}
F_{\alpha\beta}^{\rm PS}=\sum_\ell \sum_{I,J} \frac{\partial
  \bmf{D}_I(\ell)}{\partial p_\alpha}[{\rm Cov}(\bmf{D}(\ell), 
\bmf{D}(\ell))]^{-1}_{IJ}
\frac{\partial  \bmf{D}_J(\ell)}{\partial p_\beta},
\label{eq:fisher_cor}
\end{equation}
where the indices $I$, $J$ run over the redshift and mass bins
$i$, $b$ and so on. At each multipole, if $N_z$ redshift slices,
$N_M$ halo mass bins in each redshift slice, and $N_\ell$ multipole
bins are taken, the dimension of $C_{i(bb')}^{\rm hh}$ is 
$N_z\times N_\ell\times N_M(N_M+1)/2$, and that of 
$C_{i(b)}^{{\rm h}\kappa}$ is $N_z\times N_\ell\times N_M$. The
dimensions of the covariance matrix of each power spectrum and the
cross-covariance are given by their squared or the product at each
multipole. Under the Gaussian error assumption, there is
no cross-covariance between different multipoles, and hence the
summation in Eq.~(\ref{eq:fisher_cor}) is done at each multipole.

In computing the Fisher matrix $F^{\rm PS}_{\alpha\beta}$, we include
the power spectrum information up to the maximum multipole to $\ell_{\rm
max}=10^4$, which roughly corresponds to $\theta_{\rm min}\sim 1'$.
While the choice may be conservative in the sense that significant
signals can be detected at $\ell>10^4$ (see Fig.~\ref{fig:stack}), the
stacked lensing signals near the cluster center are affected by various
systematic effects ignored in this paper, such as the non-linearity of
shear signals, the contributions from central galaxies, or more
generally the effects due to baryonic physics.  Our choice of $\ell_{\rm
max}=10^4$ is therefore intended to minimize the impact of these
systematic effects. Note that we properly include various nuisance
parameters to model variations in the stacked lensing signals that can
be quite important at $\ell_{\rm max}<10^4$, including the halo
concentration parameters and the halo off-set parameters. By
marginalizing the parameter forecasts over these nuisance parameters,
we have checked that the accuracies of cosmological parameters are not
sensitive to the modes around the maximum multipoles, because the
modes at multipoles $\ell\simgt 10^3$ are limited by the shot noise
contamination as explicitly shown below. We have also checked that the
results are not largely changed if including up to an even higher
maximum multipole such as $\ell_{\rm max}=2\times 10^4$ or 
$3\times 10^4$. 

On the other hand, the maximum angular scale we can observe should be
related with the survey area $\Omega_s$. Thus we estimate the minimum 
multipole $\ell_{\rm min}$ using the relation $\ell_{\rm min}\simeq
\pi/\sqrt{\Omega_s/\pi}$. We use $\ell_{\min}=8$ for our fiducial
survey area of 2000~deg$^2$.

The shear-cluster and cluster-cluster power spectra have fairly smooth
features in multipole space. Hence, we used the binned power spectra in
multipole space to perform the summation in Eq.~(\ref{eq:fisher_cor})
over multipoles. We have checked that a finner binning of multipoles does
not change the following results. 

For reference, the covariance matrix of the stacked lensing profile in
real space can also be computed as in \cite{Jeongetal:09} (also see
\cite{TakadaJain:09} for the similar expression on the covariance of
cosmic shear correlation functions):
\begin{eqnarray}
{\rm Cov}[\sang{\gamma_+}_{i(b)}(\theta_m), 
\sang{\gamma_+}_{j(b')}(\theta_n)]&=&
\frac{1}{\Omega_s}\int\!\frac{\ell d\ell}{2\pi}\nonumber\\
&&\hspace{-16em}\times \hat{J}_2(\ell\theta_m)
\hat{J}_2(\ell\theta_n)
\left[\hat{C}^{\rm hh}_{i(bb')}(\ell)
\hat{C}^{\kappa\kappa}(\ell)\delta_{ij}^K
+C^{{\rm h}\kappa}_{i(b)}(\ell)
C^{{\rm h}\kappa}_{j(b')}(\ell)\right].\nonumber\\
\label{eq:cov_s-c_real}
\end{eqnarray}
Here we consider the covariance between angles averaged within an
annulus between $\theta_{k,{\rm min}}$ and $\theta_{k,{\rm max}}$ and
so on.  The Bessel function averaged over such a bin is defined by
\begin{equation}
\hat{J}_2(\ell\theta_m)=\frac{2}{\theta_{m,{\rm max}}^2-\theta_{m,{\rm
    min}}^2}\int_{\theta_{m,{\rm min}}}^{\theta_{m,{\rm max}}}
\theta d\theta J_2(\ell\theta).
\end{equation}
At small scale ($\theta\ll 1$) the covariance matrix should be
dominated by the shot noise. In this limit the covariance matrix
can be simplified as
\begin{equation}
{\rm Cov}\simeq \frac{\sigma_e^2}
{2\pi(\theta_{m,{\rm max}}^2-\theta_{m,{\rm min}}^2)\Omega_s 
\bar{n}_{\rm gal}\bar{n}_{i(b)}}
\delta_{mn}^K\delta_{ij}^K\delta_{bb'}^K.
\end{equation}

Eq.~(\ref{eq:cov_s-c_real}) shows that, even for the Gaussian error
case, the stacked shear profile of different angles are correlated with
each other in contrast with the power spectrum covariance. The
correlations are indeed significant for neighboring bins (e.g., see
\cite{Satoetal:10}). Thus to correctly extract cosmological information
from the real-space stacked shear profile, we need to properly include
the covariances between the measured shear profiles of different angles,
which is not straightforward compared to the Fourier-space based
analysis \cite{Satoetal:10}. 

\subsubsection{Signal-to-Noise Ratios of Stacked Lensing}

\begin{figure*}[t]
\begin{center}
\includegraphics[width=0.75\textwidth]{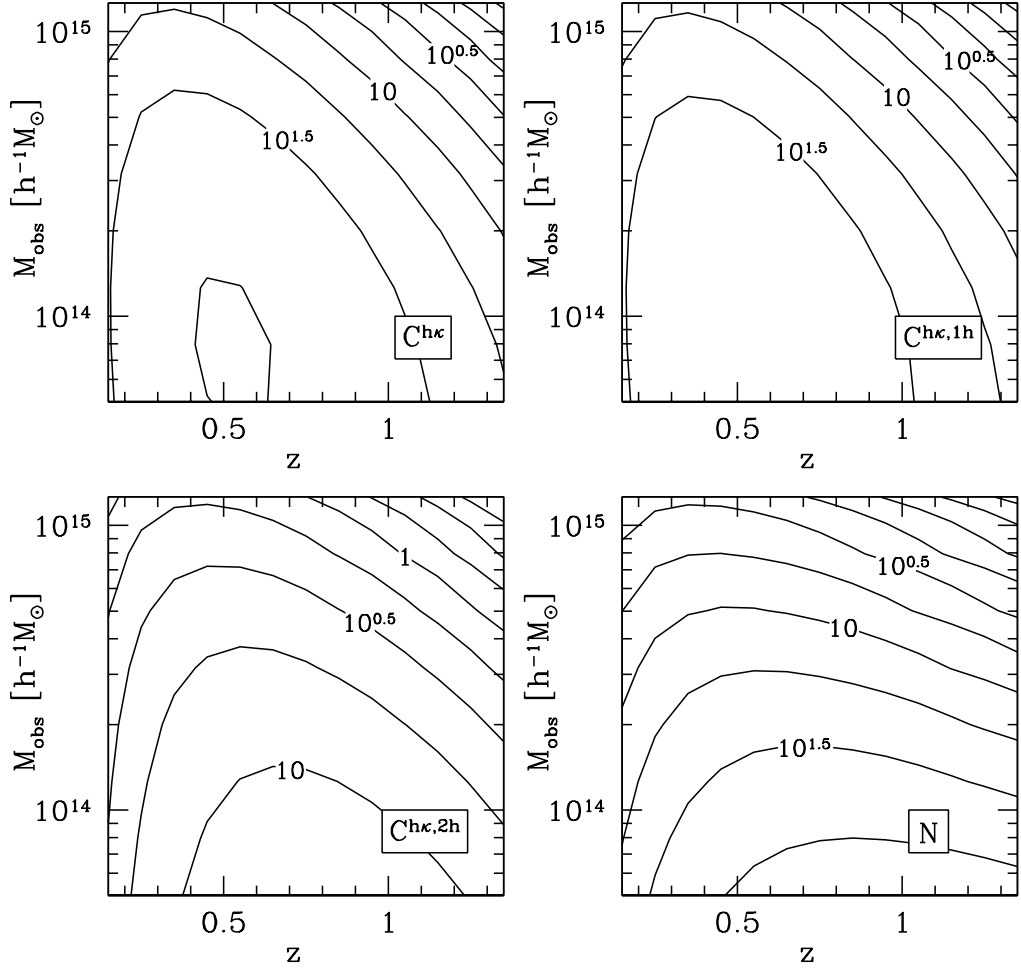}
\caption{The total signal-to-noise ratios for the stacked lensing
  signals as well as number counts of clusters, assuming the
  survey area of 2000~deg$^2$ and the bin sizes of $\Delta z=0.1$ and 
  $\Delta(\log M_{\rm obs})=0.2$. The contours are are spaced by
  $\Delta\log(S/N)=0.25$. We show the signal-to-noise ratios for the total
  stacked lensing signal $C^{{\rm h}\kappa}$ (Eq.~[\ref{eq:sn_shear}])
  ({\it upper-left panel}) and the contributions to it from 1-halo ({\it
  upper-right}) and 2-halo ({\it lower-left}) terms. The signal-to-noise
  ratios for the cluster number counts are also shown for reference ({\it
  lower-right}). Note that the signal-to-noise ratios scale with the
  survey area $\Omega_s$ (roughly) as $\propto \sqrt{\Omega_s}$.}
\label{fig:sn}
\end{center}
\end{figure*}

Using the power spectrum covariance we derived above, we here study the
{\it total} signal-to-noise ($S/N$) ratio of the stacked shear
measurement, integrated over angular scales considered, as a function
of cluster redshifts and masses. The total $S/N$ gives an indicator on
the power of a given observable, measured from survey of interest, for
constraining cosmological parameters (e.g., \cite{TakadaJain:09}).  The
total $S/N$ for measuring the stacked lensing signals of clusters in the
$i$-th redshift bin and the $b$-th mass bin can be simply expressed in
Fourier space as
\begin{equation}
\left(\frac{S}{N}\right)^2=\sum_\ell C^{{\rm h}\kappa}_{i(b)}(\ell)
[{\rm Cov}(C^{{\rm h}\kappa},
  C^{{\rm h}\kappa})]^{-1} C^{{\rm h}\kappa}_{i(b)}(\ell).
\label{eq:sn_shear}
\end{equation}
Note that the total $S/N$ in real space is similarly defined by using
the covariance matrix (\ref{eq:cov_s-c_real}), which has the same
amplitudes as the $S/N$ above, because the power spectrum and the
stacked lensing profile contain the equivalent information content at
two point level.

Fig.~\ref{fig:sn} shows contours of the expected $S/N$ for a Subaru
HSC-type survey of 2000~deg$^2$ area coverage, as a function of cluster
masses and redshifts, assuming their bin widths of $\Delta z=0.1$ and
$\Delta(\log M_{\rm obs})=0.2$, respectively. Again notice that we
assume the single population of source galaxies in this plot, source
galaxies located at $z_s>1.5$.  The figure shows that the stacked
lensing signals are detected at high significance ($S/N\gtrsim 10$)
over a wide range of cluster masses and redshifts. Most of the
signal-to-noise ratio comes mainly from the small-angle lensing
signals, where the 1-halo term (Eq.~[\ref{eq:cl_hk_1h}]) is dominated. 
The 2-halo term or equivalently the large-angle signals can also be
detected at marginal significance  ($S/N\gtrsim 1$) in most mass and
redshift bins. For comparison, we also show the signal-to-noise ratio
for the number counts, which is similarly defined using
Eqs.~(\ref{eq:counts}) and (\ref{eq:cov_cc}). The figure shows that a
wide-field survey allows a significant detection of the number counts
of clusters over a wide range of masses and redshifts. 

The $S/N$ estimate above allows us to argue how well we can measure
the mean mass of clusters in each bin from the stacked lensing signals,
which is the key in achieving the self-calibration of cluster
mass-observable relation as described in Sec.~\ref{sec:idea}.  
The halo mass is sensitive to the overall amplitude of the stacked
shear profile, with the shear amplitude at angular scales around the
project virial radius roughly scaling with the halo mass as
$\gamma_{+}\propto M$ (e.g., \cite{Okabeetal:10}).  
Hence the mass calibration accuracy is roughly estimated as
\begin{equation}
\frac{\Delta M}{M}\sim
\left(\frac{S}{N}\right)^{-1}
\left(\frac{d\ln \gamma_{+}}{d\ln M}\right)^{-1}.
\end{equation}
Fig.~\ref{fig:sn} implies that the mean mass of clusters can be
measured to a few percent accuracy for halos with masses of a few 
$10^{14}h^{-1}M_\odot$ and over redshifts $0.3\simlt z\simlt 1$. 
Since the shear signals also depend on other parameters such as
cosmological parameters, the concentration parameters and the
off-centering parameters, the genuine power of the mean halo mass
estimation has to be realized taking into account marginalization over 
other parameters, as will be carefully studied below.

\subsection{Fisher Matrix: Combining the Cluster Experiment and 
Stacked Lensing}

In summary, the Fisher matrix for the joint experiment combining the
cluster experiment (the number counts and the cluster-cluster power
spectrum) and the stacked lensing information (the shear-cluster power
spectrum) can be given as
\begin{equation}
F^{\rm CC+WL}_{\alpha\beta}=F^{\rm N}_{\alpha\beta}+F^{\rm PS}_{\alpha\beta}, 
\label{eq:fisher_tot}
\end{equation}
where $\bmf{F}^{\rm N}$ and $\bmf{F}^{\rm PS}$ are given by
Eqs.~(\ref{eq:fisher_num}) and (\ref{eq:fisher_cor}), respectively. In
Eq.~(\ref{eq:fisher_tot}) we assumed that the number counts and the
power spectrum observables are uncorrelated, where the former and the
latter probe one-point and two-point correlation information on the
underlying mass distribution.  As carefully studied in
\cite{TakadaBridle:07}, the cross-covariance between the number counts
and the power spectra arises from the three-point correlation of the
mass distribution, and an inclusion of the cross-covariance does not
largely degrade the accuracies of cosmological parameter estimation as
long as a sufficiently large number of model parameters are
included. Hence, our approach above is a good approximation as the
first step, although a more careful study including the full
covariance contribution will be needed.

\section{Results}
\label{sec:forecase}

\subsection{Parameters, Fiducial Model and Priors}

\subsubsection{Fiducial Model and Nuisance Parameters}

We now estimate accuracies of the cosmological parameter estimation
given the measurement accuracies of the observables, using the Fisher
matrix formalism. This formalism assesses how well given observables
can constrain cosmological parameters around a fiducial cosmological
model. The parameter forecasts we obtain depend on the fiducial model
and are also sensitive to the choice of free parameters. 

\begin{table*}[!t]
\begin{tabular}{lll}\hline\hline
Cosmological paras. & $\Omega_{\rm DE}$, $\Omega_{M}h^2$, 
$\Omega_b h^2$, $n_s$, $\alpha_s$, $w_0$, $w_a$, $\delta_\zeta$, 
$f_{\rm NL}$
&\\ \hline\hline
Source redshift& $z_{m}(\equiv 3z_0)$ & Eq.~(\ref{eq:dngaldz})\\
Concentration parameter: $c(M,z)$ & $A_{\rm vir}$, $B_{\rm vir}$, 
$C_{\rm vir}$ & Eq.~(\ref{eq:concent}) \\
Halo off-centering paras: $f_{\rm cen}$, $\sigma_s$ &
$f_{\rm cen}(M,z)=f_{{\rm cen},0}+p_{{\rm cen}, M}\ln (M/M_{\rm piv})+
p_{{\rm cen},z}\ln(1+z)$ &  
Eq.~(\ref{eq:kappa_offdis})\\
& $\sigma_{s}(M,z)=\sigma_{s,0}+p_{\sigma_s, M}\ln (M/M_{\rm piv})+
p_{\sigma_s,z}\ln(1+z)$ &  
\\
Mass-obs. relation: $\ln M_{\rm bias}, \sigma_{\ln M}$
&$\ln M_{\rm bias}=\ln M_{{\rm b},0}+\sum_{i=1}^3 q_{{\rm b},i}
\left[\ln (M/M_{\rm piv})\right]^i+\sum_{i=1}^{3} s_{{\rm b},i}z^i 
$ & Eq.~(\ref{eq:mobs})\\
&
$\sigma_{\ln M}=\sigma_{\ln M,0}+\sum_{i=1}^3 q_{\sigma_{\ln M},i}
\left[\ln (M/M_{\rm piv})\right]^i+\sum_{i=1}^{3} s_{\sigma_{\ln M},i}z^i$ 
&
\\
\hline\hline
\end{tabular}
\caption{Parameters included in our Fisher matrix analysis. We
 include 9 cosmological parameters as well as 24 nuisance parameters to
 quantify the impact of possible residual systematic errors on the
 cosmological parameter estimation. In total we include 33 parameters. 
 The fiducial values of each parameters are given in the text.} 
\label{tab:paras}
\end{table*}

We include all the key parameters that may affect the observables
within the cold dark matter and dark energy cosmological framework. As
for our fiducial cosmological model, which is based on the WMAP
seven-year results \cite{Komatsu:09}, we include the following 9
cosmological parameters: the density parameters of matter, baryon and
dark energy are $\Omega_Mh^2(=0.134)$, $\Omega_bh^2(=0.0226)$, and 
$\Omega_{\rm DE}(=0.734)$ (note that we assume a flat universe); 
dark energy equation of state parameters are $w_0(=-1)$ and $w_a(=0)$;
the primordial power spectrum parameters are the spectrum tilt
$n_s(=0.963)$, the running spectral index $\alpha_s(=0)$, and the
normalization parameter of primordial curvature perturbation,
$\delta_\zeta(=4.89\times 10^{-5})$, where the values in the
parentheses denote the fiducial parameter values.  As an extension of
dark energy cosmological model we also include the primordial
non-Gaussianity which is modeled by the local-type parameter  
$f_{\rm NL}(=0)$. To be more specific we adopt the model
\cite{LoVerdeetal:08,Dalaletal:08} in computing the halo mass function
and the halo bias.  

In addition, we include a number of nuisance parameters to model
systematic errors involved in the observables. For the mass-observable
relation, which is parametrized by $\ln M_{\rm bias}$ and $\sigma_{\ln
M}$, we allow more complicated dependences on redshift and halo mass by
adopting the following form:
\begin{eqnarray}
&&\hspace{-2em}\ln M_{\rm bias}(M, z)=\ln M_{\rm b, 0}
\nonumber\\
&&\hspace{-0.5em}
+\sum_{i=1}^3q_{{\rm b}, i}\left[\ln(M/M_{\rm piv})\right]^i
+\sum_{i=1}^3s_{{\rm b}, i}z^i,
\label{eq:nuis_mbias}\\
&&\hspace{-2em}\sigma_{\ln M}(M, z)=
\sigma_{\ln M, 0}
\nonumber\\
&&\hspace{-0.5em}
+\sum_{i=1}^3q_{\sigma_{\ln M}, i}\left[\ln(M/M_{\rm piv})\right]^i
+\sum_{i=1}^3s_{\sigma_{\ln M}, i}z^i,
\label{eq:nuis2}
\end{eqnarray}
where we choose $M_{\rm piv}=3\times10^{14}h^{-1}M_\odot$, and $\ln
M_{\rm b, 0}$, $q_i$, $s_i$, $\dots$ are free parameters.  The form
above has been adopted by previous work (e.g., \cite{Cunhaetal:10}). The
fiducial values for these parameters are $\sigma_{\ln M,0}=0.3$, and
the other parameters such as $\ln M_{\rm b, 0}$, $q_{{\rm b},i}$ and
$s_{{\rm b},i}$ are set to zero.

The source redshift uncertainty affecting the stacked lensing signals is
parametrized by the mean redshift parameter $z_m$ ($=1$) in the source
redshift distribution (see Eq.~[\ref{eq:dngaldz}]). The systematic
errors in the small-scale stacked shear profile are the halo
concentration parameter parametrized by three parameters, $A_{\rm
vir}$, $B_{\rm vir}$ and $C_{\rm vir}$ (see Eq.~[\ref{eq:concent}] below
for their fiducial values), and the halo off-centering effect given by
the two parameters $f_{\rm cen}$ and $\sigma_s$ 
(Eq.~[\ref{eq:kappa_offdis}]). Similarly, to include possible mass and
redshift dependences for the offset parameters, which are poorly known
theoretically, we employ a simple form given by
\begin{eqnarray}
&&\hspace{-1em}f_{\rm cen}(M, z)=f_{\rm cen, 0}
+p_{{\rm cen}, M}\ln(M/M_{\rm piv})\nonumber\\
&&\hspace{8em}+p_{{\rm cen}, z}\ln (1+z),\\
&&\hspace{-1em}\sigma_{s}(M, z)=\sigma_{s, 0}
+p_{{\sigma_s}, M}\ln(M/M_{\rm piv})\nonumber\\
&&\hspace{8em}+p_{{\sigma_s}, z}\ln (1+z).
\label{eq:nui1}
\end{eqnarray}
Note that the nuisance parameters for $\sigma_s$ are in units of
$h^{-1}$Mpc. The fiducial values are $f_{\rm cen}=0.75+0.05\ln(M/M_{\rm
piv})$, i.e., $f_{\rm cen,0}=0.75$, $p_{{\rm cen},M}=0.05$ and $p_{{\rm
cen},z}=0$; $\sigma_{s,0}=0.42 h^{-1}{\rm Mpc}/D_A(z)$,
$p_{\sigma_s,M}=0$ and $p_{\sigma_s,z}=0$.  

Thus we include 33 parameters in the Fisher matrix analysis; 9
cosmological parameters and 24 nuisance parameters (see
Table~\ref{tab:paras} for a brief summary).  Unless otherwise 
specified, we do not add any prior on the nuisance parameters.

\begin{table*}[!t]
\begin{tabular}{@{}lcccccccccc}
  \hline\hline
Survey & $\Omega_s$ & $z_m$ & $\bar{n}_{\rm
  gal, tot}$ & $\bar{n}_{\rm gal}$ & $z_{s,{\rm min}}$  
& $M_{\rm obs}$-range & $N_M$ & $z$-range & $N_z$ & $\ell$-range \\ 
& [deg$^2$] & & [arcmin$^{-2}$] & [arcmin$^{-2}$] & & 
[$h^{-1}M_\odot$] & & & & \\
\hline  
HSC (2000deg$^2$) & 2000 & 1.0 & 30 & 5.2 & 1.5 & $10^{14}<M_{\rm obs}$ & 5 
   & $0.1<z<1.4$ & 13 &  $8<\ell<10^4$ \\
HSC (1500deg$^2$) & 1500 & 1.0 & 30 & 5.2 & 1.5 & $10^{14}<M_{\rm obs}$ & 5 
   & $0.1<z<1.4$ & 13 &  $8<\ell<10^4$ \\
DES  & 5000  & 0.7 & 10 & 1.5 & 1.1 & $10^{14}<M_{\rm obs}$ & 5 
   & $0.1<z<1.0$ & 9  &  $5<\ell<10^4$ \\
LSST & 20000 & 1.2 & 50 & 14 & 1.5 & $10^{14}<M_{\rm obs}$ & 5 
   & $0.1<z<1.4$ & 13 &  $2<\ell<10^4$ \\
\hline\hline
\end{tabular}
\caption{The summary of survey parameters adopted in this paper. The
  parameters $N_M$ and $N_z$ denote the numbers of cluster
 mass and redshift bins, respectively. Note that the survey parameter
 set of HSC (2000deg$^2$) is same as the fiducial parameter set used
 in previous sections. The parameters $\bar{n}_{\rm gal, tot}$ and
  $\bar{n}_{\rm gal}$ denote the number densities of source galaxies at
  all redshifts and at redshift $z_s>z_{s,{\rm min}}$, respectively,
  where the latter is used for the weak lensing analysis in this paper.}
\label{tab:survey}
\end{table*}

We also need to specify survey parameters. Thus far we have employed
the survey parameters that resemble the planned Subaru HSC survey (see
Eq.~[\ref{eq:dngaldz}]). For a broader application of the method
proposed in this paper we also consider other survey parameters which
roughly resemble other imaging surveys being planned, DES and LSST.
Table~\ref{tab:survey} summarizes the assumed survey parameters. We
adopt the method in \cite{TakadaJain:09} (see around Eq.~[20] of the
paper) to estimate the mean redshift and the available number density of
galaxies for each survey.

For all surveys, we employ the same halo mass threshold 
$M_{\rm obs}>10^{14}h^{-1}M_\odot$, which makes it easier to compare the
results between different surveys. We consider 5 mass bins for each
redshift slice, with the bin width of $\Delta(\log M_{\rm obs})=0.2$
(for the highest mass bin, we extend the upper limit of the mass to
infinity).  The maximum cluster redshift is set to $z_{\rm max}=1.4$,
which is approximately the maximum cluster redshift detectable via
optical imaging surveys including $y$-band data, except for DES for
which we adopt $z_{\rm max}=1$ given the shallower depth. The cluster
redshift bin width is $\Delta z=0.1$ for all the surveys. For the
source galaxy population needed for the stacked lensing we use
galaxies at $z_s>1.5$ for HSC and LSST, and at $z_s>1.1$ for DES,
respectively. Assuming the depth of each survey we can compute the
available number density of source galaxies, which is needed to
quantify the intrinsic ellipticity noise. We simply assume
$\sigma_e=0.35$ for the rms intrinsic ellipticities (sum of two
components) for all the surveys. 

\begin{table*}
\begin{tabular}{lccccccccccc}\hline\hline
Method & $\sigma(\Omega_{\rm DE})$ & $\sigma(w_0)$ & $\sigma(w_a)$ 
& FoM & $z_{\rm piv}$ & $\sigma(f_{\rm NL})$ & $\sigma(\ln \Omega_Mh^2)$
&$\sigma(\ln \Omega_{\rm b}h^2)$ & $\sigma(n_s)$ & $\sigma(\alpha_s)$ & 
$\sigma(\ln\delta_\zeta)$ \\ \hline 
Planck 
& --- & --- & --- & --- & --- & --- & 0.0102 & 0.0086 & 0.0226 & 0.0068 & 0.0174 \\
+$\bmf{N}$
& 0.099 & 0.62 & 2.17 &   1.4 & 0.32 & ---   & 0.0101 & 0.0085 & 0.0218 & 0.0066 & 0.0169 \\
+$\bmf{N}+\bmf{C}^{\rm hh}$ 
& 0.035 & 0.37 & 1.13 &   7.5 & 0.46 &  15.6 & 0.0098 & 0.0079 & 0.0201 & 0.0061 & 0.0158 \\
+$\bmf{C}^{{\rm h}\kappa}$
& 0.056 & 0.44 & 1.12 &   5.1 & 0.56 &  86.9 & 0.0099 & 0.0080 & 0.0205 & 0.0063 & 0.0161 \\ \hline
+$\bmf{N}+\bmf{C}^{\rm hh}+\bmf{C}^{{\rm h}\kappa}$
& 0.023 & 0.22 & 0.60 &  28.6 & 0.56 &  14.5 & 0.0059 & 0.0068 & 0.0189 & 0.0058 & 0.0143 \\ \hline
+$\bmf{N}+\bmf{C}^{\rm hh}+\bmf{C}^{{\rm h}\kappa}+\bmf{C}^{\kappa\kappa}$ 
& 0.022 & 0.22 & 0.59 &  32.4 & 0.57 &  14.3 & 0.0053 & 0.0065 & 0.0124 & 0.0038 & 0.0104 \\
+$\bmf{N}+\bmf{C}^{\rm hh}+\bmf{C}^{{\rm h}\kappa}$+BAO
& 0.017 & 0.18 & 0.48 &  52.7 & 0.58 &  14.2 & 0.0041 & 0.0066 & 0.0188 & 0.0058 & 0.0141 \\
+$\bmf{N}+\bmf{C}^{\rm hh}+\bmf{C}^{{\rm h}\kappa}+\bmf{C}^{\kappa\kappa}$+BAO
& 0.017 & 0.18 & 0.47 &  54.8 & 0.58 &  14.0 & 0.0040 & 0.0063 & 0.0123 & 0.0038 & 0.0103 \\
\hline\hline
\end{tabular}
\caption{Expected marginalized errors (68\% C.L.) on each cosmological
 parameter for different combinations of cosmological probes: the
 column labeled as ``Planck'' shows the errors expected from the CMB
 information alone; the following columns are the results expected by
 further adding the cluster number counts ($\bmf{N}$), the
 cluster-cluster power spectra ($\bmf{C}^{\rm hh}$), the shear-cluster
 spectra ($\bmf{C}^{{\rm h}\kappa}$), the cosmic shear spectra
 ($\bmf{C}^{\kappa\kappa}$), and the geometrical constraints expected
 from the BOSS BAO information (see text for details). The errors
 include marginalization over other parameters including nuisance
 parameters (see Tab.~\ref{tab:paras}). However note that we did not
 employ any priors on the nuisance parameters. The columns labeled as
 ``FoM'' and  ``$z_{\rm piv}$'' denote the dark energy figure of merit
 and the pivot redshift, respectively  (see text for details). We
 assume the survey parameters for HSC (2000~deg$^2$) in
 Tab.~\ref{tab:survey}, for the cluster counts and the power spectrum
 measurements ($\bmf{C}^{\rm hh}, \bmf{C}^{{\rm h}\kappa}$). 
 Note that $z_{\rm piv}$ denotes the pivot redshift at which the dark
 energy equation of state is best constrained. The element labeled as
 ``---'' denotes that the parameter is not well constrained by the
 observables. } 
\label{tab:hsc_forecast} 
\end{table*}

\subsubsection{Priors}

As stated above, unless explicitly expressed, we do not use any priors
on the 24 nuisance parameters. Therefore, we can address how combining
the cluster experiments and the stacked lensing allows to self-calibrate
these nuisance parameters. 

However, using probes of large-scale structure alone is not powerful
enough to constrain all the cosmological parameters
simultaneously. Rather, combining the large-scale structure probes with
constraints from Cosmic Microwave Background (CMB) temperature and
polarization anisotropies significantly helps to lift parameter
degeneracies.  In this paper we use the CMB priors expected from the
Planck satellite mission. When computing the Fisher matrix for the CMB
we employ 9 parameters in total, 8 from our fiducial cosmological
parameter set (we do not include any constraints on $f_{\rm NL}$ from
CMB in order to examine the pure power of clusters and weak lensing in
constraining $f_{\rm NL}$) plus the Thomson scattering depth to the
last scattering surface ($\tau=0.089$). We used the publicly-available
CAMB code \cite{Lewisetal:00}, based on the CMBFAST
\cite{SeljakZaldarriaga:96}, to compute the angular power spectra of
temperature anisotropy, $E$-mode polarization and their
cross-correlation. Note that we ignored $B$-mode spectra arising from
the primordial gravitational wave. The details of our CMB Fisher
matrix calculation can be found from \cite{TakadaBridle:07}. The
Fisher matrix for the joint experiments combining the CMB information
with the method studied in this paper is simply given by adding the
CMB Fisher matrix to the matrix in Eq.~(\ref{eq:fisher_tot}), i.e., 
$\bmf{F}=\bmf{F}^{\rm CMB}+\bmf{F}^{\rm N}+\bmf{F}^{\rm PS}$. 
The Thomson scattering depth $\tau$ is marginalized over before the
CMB Fisher matrix is added to other constraints.

By the time when the imaging surveys above come online, stringent
cosmological constraints will be available from the ongoing Baryon
Oscillation Spectroscopic Survey (BOSS)
\footnote{http://cosmology.lbl.gov/BOSS/}, which gives a geometrical
probe of cosmological distances via the baryon acoustic oscillation
(BAO) experiment \cite{Eisensteinetal:05}. Using the method in
\cite{SeoEisenstein:03}, we compute the Fisher matrix expected from the
BOSS experiment. To be more precise, we will add, to the Fisher
matrix, the expected geometrical information on the Hubble expansion
rates and the angular diameter distances up to the redshifts, $z\simeq
0.65$, which are relevant for $\Omega_{M}h^2$, $\Omega_{\rm DE}$,
$w_0$, $w_a$ in our parameters. 

Finally we will also study how the cosmological constraints can be
improved by adding another power spectrum available from the same
imaging survey data, the cosmic shear power spectrum $C^{\kappa\kappa}$ 
for the same source galaxies. Since the cluster-cluster, shear-cluster
and cosmic shear power spectra at large angles depend on the halo bias
$b_h$ in different ways, roughly 
$ C^{\rm hh}\propto b_h^2P_m$, $C^{{\rm h}\kappa}\propto b_hP_m$ and
$C^{\kappa\kappa}\propto P_m$, combining all the spectra may help to
break the degeneracies between the halo bias and other parameters
further. However, note that the self-calibration of nuisance
parameters can be achieved only if focusing on the single source
galaxy population for lensing. Thus, we include the cosmic shear power 
spectrum without tomography. For the cosmic shear $C^{\kappa\kappa}$,
we include the spectrum only up to $\ell_{\rm max}=10^3$ as the cosmic
shear at small scales is subject to various uncertainties. We take
proper account of the cross-covariance of $C^{\kappa\kappa}$ with
$C^{{\rm h}\kappa}$ and $C^{\rm hh}$ when adding constraints from
$C^{\kappa\kappa}$. 

\subsection{Parameter Forecasts}

\begin{figure*}[t]
\begin{center}
\includegraphics[width=0.75\textwidth]{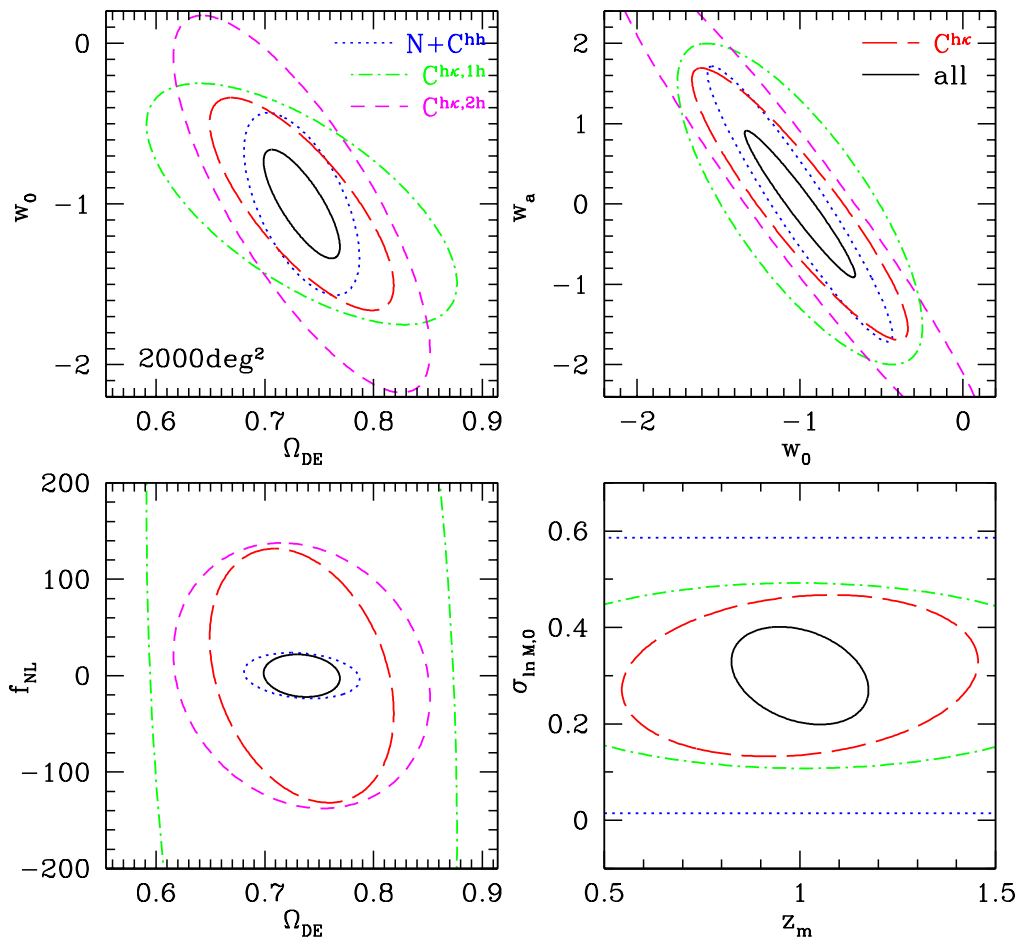}
\caption{Projected 68\% confidence limit ($\Delta \chi^2=2.3$)
  constraints in various parameter spaces; $\Omega_{\rm DE}$-$w_0$
  plane ({\it upper-left panel}),  $w_0$-$w_a$ plane ({\it upper-right}), 
  $\Omega_{\rm DE}$-$f_{\rm NL}$ plane ({\it lower-left}), and
  $z_m$-$\sigma_{\ln M,0}$ plane ({\it lower-right}). The survey area
  is assumed to 2000~deg$^2$. The different curves indicate
  constraints from different measurements; the number counts plus the  
  cluster-cluster correlation function ({\it dotted}), the 1-halo
  ({\it dot-dashed}) and 2-halo ({\it short dashed}) terms of stacked
  lensing signals and their sum ({\it long dashed}), and the total
  constraints from the combination of all measurements ({\it
  solid}). The Planck prior is added in all cases.}  
 \label{fig:ell}
\end{center}
\end{figure*}

\begin{figure}[t]
\begin{center}
\includegraphics[width=0.42\textwidth]{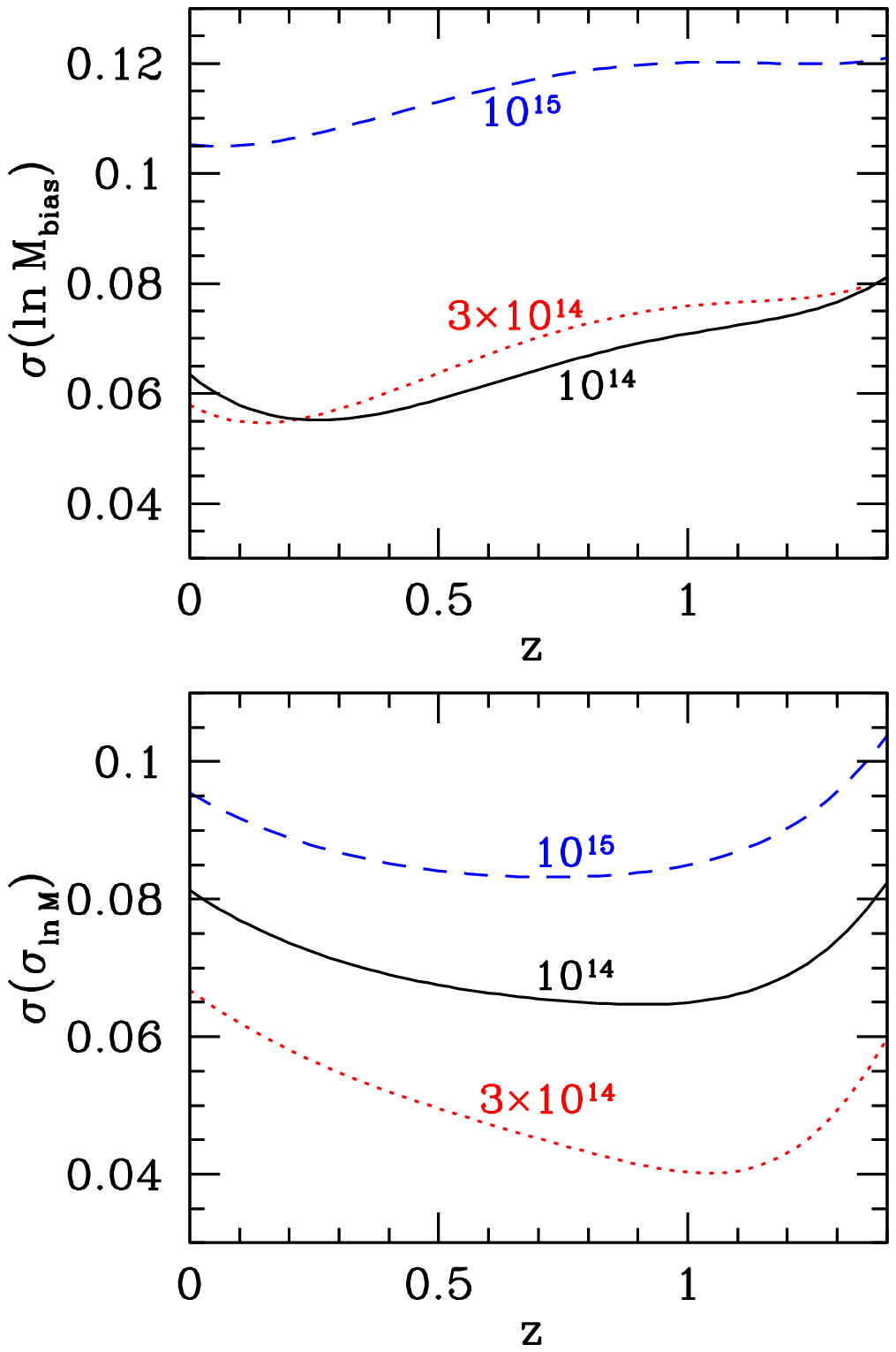}
\caption{Marginalized errors on $\ln M_{\rm bias}$ ({\it upper panel}) 
  and $\sigma_{\ln M}$ ({\it lower}), the parameters in the
  mass-observable relation (see Eq.~[\ref{eq:mobs}]), as a function of
  cluster redshifts, constrained from the combination of all the
  observables (except for the cosmic shear and BAO). The errors are
  shown for three different cluster masses, $M=10^{14}h^{-1}M_\odot$
  ({\it solid curve}), $3\times 10^{14}h^{-1}M_\odot$ ({\it dotted}),
  and $10^{15}h^{-1}M_\odot$ ({\it dashed}).} 
 \label{fig:merr}
\end{center}
\end{figure}

Table~\ref{tab:hsc_forecast} summarizes the 68\% error on each
cosmological parameter, marginalized over other parameter uncertainties
including nuisance parameters in Table~\ref{tab:paras}. Here we have
assumed survey parameters for the HSC survey of 2000~deg$^2$ coverage
given in Table~\ref{tab:survey}. The first row shows the constraints
expected from the Planck data alone, while the other rows show the
constraints by further adding the large-scale structure probes. It is
shown that the parameter constraints can be continuously improved by
adding the different observables, the cluster counts, the
cluster-cluster power spectrum, the stacked lensing, the cosmic shear
power spectrum, and the BAO experiment. We should again stress that we
did not add any priors on nuisance parameters in
Table~\ref{tab:paras}, and therefore the parameter constraints are as
a result of the self-calibration of those systematic effects attained
by combining the different probes.  The previous works
\cite{MajumdarMohr:04,LimaHu:04,LimaHu:05,Oguri:09,Cunhaetal:10} have
studied the combination of cluster number counts and cluster-cluster
correlation functions (so-called ``self-calibrated cluster counts''),
which is essentially represented by the column denoted by
``+$\bmf{N}+\bmf{C}^{\rm hh}$''. We find that an additional
calibration of cluster masses from the stacked weak lensing signals
significantly improves constraints on each of the dark energy
parameters, by a factor of 2 or so, even when various uncertainties
associated with stacked lensing signals, such as the source redshift
uncertainty, off-centering, and concentration parameters, are
marginalized over. The resulting constraints on dark energy equation
of state is in fact quite comparable to the constraints expected from
tomographic cosmic shear without any marginalization over systematic
errors \cite{TakadaJain:09}. 

\begin{figure}[t]
\begin{center}
\includegraphics[width=0.4\textwidth]{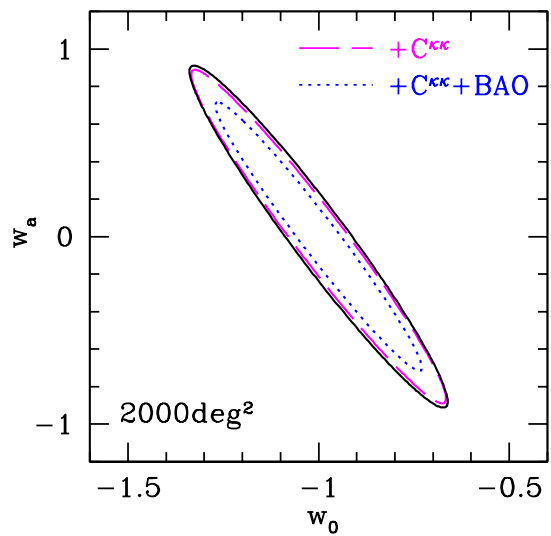}
\caption{The improvements of the marginalized constraints in the
  $w_0$-$w_a$ plane by adding the cosmic shear power spectrum
  ($C^{\kappa\kappa}$) for the same source galaxy population as that
  used in the stacked lensing analysis, and the BAO geometrical
  information expected from the SDSS-III BOSS survey. }
  \label{fig:ell_add}
\end{center}
\end{figure}

A useful parameter that quantifies the power of a given survey for
constraining dark energy equation of state is so-called the Figure
of Merit (FoM) defined in \cite{DETF}:
\begin{equation}
{\rm FoM}=\frac{1}{\sigma(w_a)\sigma(w_p)}=\frac{1}{\sqrt{{\rm
    det}({\rm Cov}[w_0,w_a])}},
\end{equation}
where $w_p$ is dark energy equation of state at the ``pivot'' redshift,
at which the dark energy equation of state is best constrained
\cite{DETF}. The FoM is inversely proportional to the area of the error
ellipse in the $w_0$-$w_a$ plane. Thus, the larger the FoM is,
constraints on dark energy models are tighter. 
Table~\ref{tab:hsc_forecast} shows that adding the stacked lensing
signals improves the FoM by a factor of $\sim 4$ compared with
the cluster experiment alone (number counts plus the cluster-cluster
power spectrum) in our fiducial setting.  This result appears to be
consistent with a recent similar analysis by \cite{RozoWuSchmidt:10},
although no systematic error associated with the stacked lensing
analysis has been considered in their Fisher matrix analysis.

It would be also informative to study how a different combination of the
observables can break the parameter degeneracies.  Fig.~\ref{fig:ell}
shows projected 68\% error ellipses in various two-parameter sub-spaces
of the parameters. The dotted and long-dashed curves show the error
ellipses expected from the cluster experiments (the number counts plus
the cluster-cluster power spectrum) and the stacked lensing,
respectively. For comparison the dot-dashed and short-dashed curves
show the results from each of the small-angle (1-halo term) and
large-angle (2-halo term) signals of the stacked lensing,
respectively. The innermost solid curve is the ellipse by combining
all the measurements (except for the BAO and the cosmic shear power
spectrum), showing that the ellipse significantly shrinks compared to
any of the ellipses of one particular observable. This improvement of
parameter constraints is due to the self-calibration of nuisance
parameters, and the lower-right panels explicitly shows one example:
the mean source redshift is directly estimated from the observables
(self-calibrated), without any prior, to an accuracy of
$\sigma(z_m)\sim 0.1$ or so.
The accuracy corresponds to $\sigma(\gamma_+)\sim 0.01-0.02$ for the
cluster redshift $z\sim 0.5-1$ (see also Fig.~\ref{fig:zdep}).
Finally our results indicate that
primordial non-Gaussianity can be constrained to 
$\sigma(f_{\rm NL})\sim 15$, which is mainly from the cluster observables 
(see also \cite{Oguri:09,Cunhaetal:10}).

To see how well cluster masses are calibrated by the combination of
stacked weak lensing and the cluster experiments, in Fig~\ref{fig:merr}
we show the marginalized 68\% error on the parameters in the
mass-observable relation, $\ln M_{\rm bias}$ and $\sigma_{\ln M}$, as
a function of cluster masses and redshifts, based on
Eqs.~(\ref{eq:nuis_mbias}) and (\ref{eq:nuis2}). We find that the halo
mass bias $\ln M_{\rm bias}$ is constrained to $\sigma(\ln M_{\rm
  bias})\sim 0.07$ for a wide range of cluster masses and redshifts,
indicating that cluster masses are determined at $\sim 7$\% accuracy,
even the other parameters are marginalized over. The mass calibrations
of more massive clusters are less accurate, simply because of the
smaller number of clusters available for the stacked lensing analysis
(see also Fig.~\ref{fig:sn}). 

Fig.~\ref{fig:ell_add} shows how the dark energy constraints, $w_0$ and
$w_a$, can be further improved by adding the cosmic shear power spectrum
of the same source galaxy population ($z_s>1.5$) and the BOSS BAO
information. The figure shows that the BOSS BAO information is useful
to improve the dark energy constraints. More quantitatively, the
marginalized errors on $w_0$ and $w_a$ are improved, from
$\sigma(w_0)=0.22$ and $\sigma(w_a)=0.60$, to $\sigma(w_0)=0.18$ and
$\sigma(w_a)=0.47$ by adding both the cosmic shear power spectrum and
the BOSS BAO information. The FoM value is improved by a factor of
1.9, from 28.6 to 54.8 (see also Table~\ref{tab:hsc_forecast}). For
comparison, the BOSS BAO information alone (with the Planck prior) has
the FoM value of 19.2. Thus the BAO information appears to be highly
complementary to the constraints from the imaging survey.  

The stacked weak lensing has been thought to provide a powerful method
to constrain the concentration-mass relation (e.g.,
\cite{Mandelbaumetal:08}). However, in our Fisher matrix analysis, the
marginalized errors on the concentration parameters, $\sigma(A_{\rm
vir})=3.9$, $\sigma(B_{\rm vir})=0.09$, and $\sigma(C_{\rm vir})=0.3$,
are not so impressive.  The reason is that these parameters are strongly
degenerate with each other, and also with off-centering parameters
($f_{\rm cen}$ and $\sigma_s$). Our result therefore may suggest
a critical difficulty in measuring the concentration-mass relation from
the stacked weak lensing technique. Rather the information can be
efficiently extracted by combining the weak and strong lensing
information on individual cluster basis (e.g.,
\cite{Broadhurstetal:05,Broadhurstetal:09,Ogurietal:09}). 

\begin{figure}[t]
\begin{center}
\includegraphics[width=0.42\textwidth]{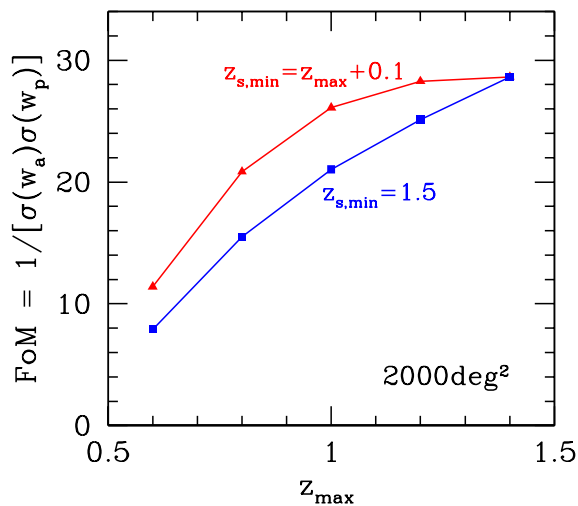}
\caption{The Figure of Merit (FoM) from the combination of the stacked
  lensing, the number counts, and the cluster-cluster correlation
  function, as a function of the maximum cluster redshift $z_{\rm
  max}$. The filled squares indicate the case that the minimum source
  redshift is fixed to $z_{s,{\rm min}}=1.5$, whereas the case that
  the minimum source redshift is set to $z_{s,{\rm min}}=z_{\rm
  max}+0.1$, is shown by the filled triangles.}  
 \label{fig:fom}
\end{center}
\end{figure}

To understand the sensitivity of cluster redshifts to the dark energy
constraints, we study how the dark energy constraints change with the
maximum cluster redshift. Fig.~\ref{fig:fom} shows the FoM as a
function of the maximum cluster redshift $z_{\rm max}$ used for the
analysis, for the same bin width of $\Delta z=0.1$. We find  
that FoM changes rapidly with the maximum cluster redshifts
particularly below 1, indicating the importance of deep cluster surveys
to include clusters out to $z\gtrsim 1$. We do gain by adding clusters
at $z>1$ if we fix the minimum source redshift $z_{s,{\rm min}}$ to
$z_{s,{\rm min}}=1.5$. However, the improvement in the FoM at $z>1$
becomes rather small if we set $z_{s,{\rm min}}=z_{\rm max}+0.1$,
because the additional information from high redshift 
clusters can be to some extent compensated by the change of the number
density of background source galaxies. 

\subsection{Effect of Priors}

\begin{figure*}[t]
\begin{center}
\includegraphics[width=0.75\textwidth]{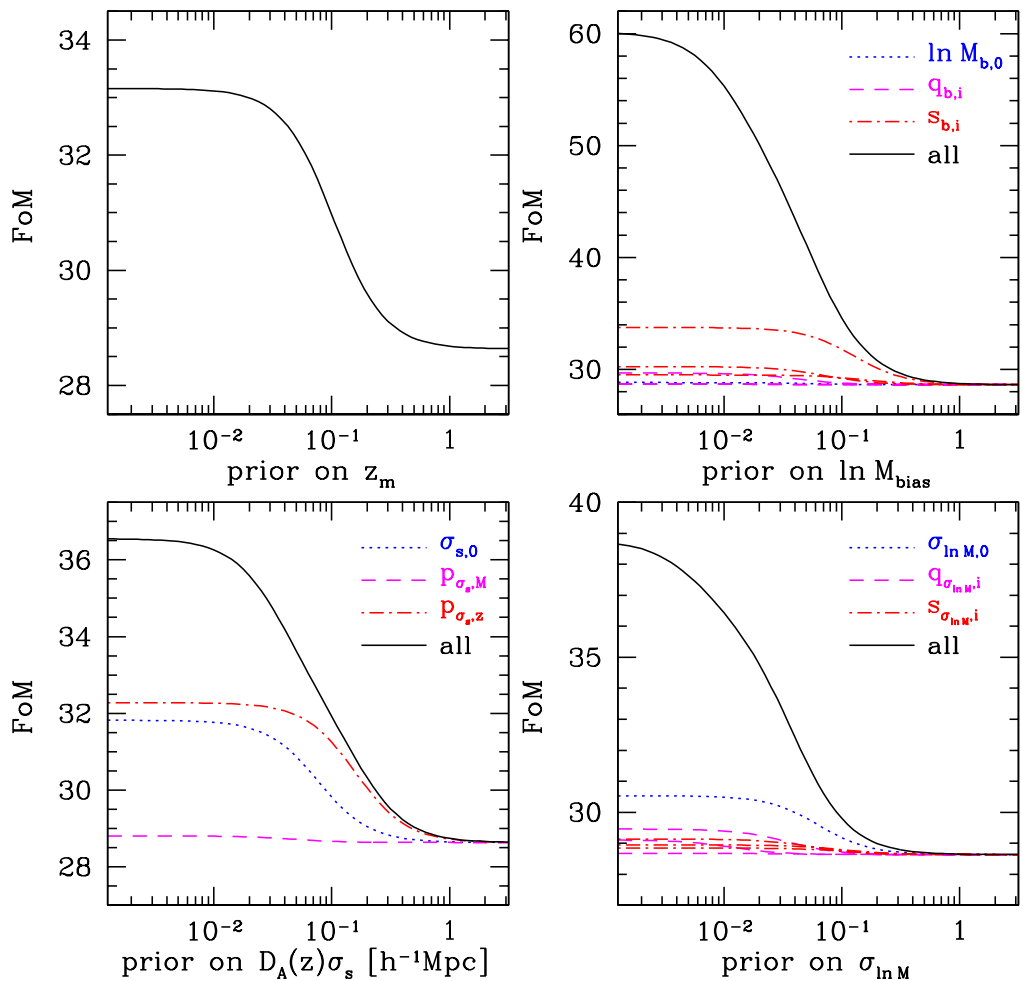}
\caption{The Figure of Merit (FoM) as a function of priors on various
 nuisance parameters to model the systematic errors: the mean source
 redshift $z_m$ ({\it upper-left panel}), the cluster off-centering 
 parameter $\sigma_s$ ({\it lower-left}; in units of $h^{-1}$Mpc), the
 mass bias parameter of the cluster-mass observable relation $\ln
 M_{\rm bias}$ ({\it upper-right}), and its mass scatter 
 $\sigma_{\ln M}$ ({\it lower-right}). For the parameters $\sigma_s$,
 $\ln M_{\rm bias}$ and $\sigma_{\ln M}$, we add the priors on
 parameters in Table~\ref{tab:paras}: the overall amplitude parameter,
 the halo mass dependence, the redshift dependence or all the
 parameters.}   
 \label{fig:prior}
\end{center}
\end{figure*}

Thus far we adopted no prior on the nuisance parameters. However, in
practice, we expect that there are external constraints on some of
these parameters. For instance, photometric redshifts, which are
required to define the source galaxy sample in the first place,
naturally provide some prior information on $z_m$. The fraction and
offset distributions of off-centering galaxies may be inferred from
mock galaxy catalogs in $N$-body simulations \cite{Johnstonetal:07} or 
directly from a detailed measurement of individual clusters
\cite{Ogurietal:10}.  For cluster observables we may be able to
further refine the mass-observable relations by using detailed studies
of individual clusters or combining different information from
optical, lensing, X-ray and SZ measurements. Here we explore
how such priors help to improve the parameter constraints. 

The result is shown in Fig.~\ref{fig:prior}. As specific examples, we
add priors on the mean redshift of the source galaxy distribution,
$z_m$, on the standard deviation of the cluster offsets,
${\sigma}_s$, on the halo mass bias of the cluster mass-observable
relation, $\ln M_{\rm bias}$, and on the mass scatter $\sigma_{\ln M}$. 
We find that the prior $\sigma(z_m)\lesssim 0.1$ indeed
improves constraints on dark energy, although the improvement of the
FoM is only up to about 10\%. This again implies that the
self-calibration of the source redshift is achieved by combining the
different observables, well enough not to degrade our cosmological
results. We, however, comment that a marginally accurate photo-$z$
information  is needed for selecting {\it source} galaxies for our
stacked lensing analysis. An imperfect photo-$z$ information causes a
contamination from foreground or cluster member galaxies to the source
galaxy population, which dilutes the stacked lensing signals. We will
come back to this issue later. On the other hands,
Fig.~\ref{fig:prior} indicates that priors on ${\sigma}_s$ help
improve the FoM. The result, in turn, confirms the previous claims
that the off-centering is one of the biggest uncertainties inherent to
the stacked weak lensing analysis \cite{Johnstonetal:07}. 

The priors on the mass-observable relations have a greater impact on the
FoM, because the dark energy constraints are mainly from cluster
observables in the method studied in this paper. A few percent level
prior on $\ln M_{\rm bias}$ or $\sigma_{\ln M}$, if available, can
improve the FoM by a factor of 2 and 1.3, respectively. The result
suggests the importance of detailed studies of individual clusters for
accurately calibrating the mass-observable relation, as has been done
in \cite{Zhangetal:08,Okabeetal:10b}.

\subsection{Predictions for Various Surveys}

Next we tune survey parameters to several future wide-field optical
imaging surveys, in which the stacked cluster lensing analysis studied
in this paper is feasible, to study/compare their powers as a dark
energy probe. Specifically, we consider the following surveys: HSC,
DES and LSST. 

\begin{table*}[!t]
\begin{tabular}{@{}cccccccccccccccccc}
  \hline\hline
& \multicolumn{5}{c}{$\bmf{N}$+$\bmf{C}^{\rm hh}$} &
& \multicolumn{5}{c}{$\bmf{C}^{{\rm h}\kappa}$} &
& \multicolumn{5}{c}{$\bmf{N}$+$\bmf{C}^{\rm hh}$+$\bmf{C}^{{\rm h}\kappa}$} \\ 
\cline{2-6} \cline{8-12} \cline {14-18}
Survey  
& $\sigma(\Omega_{\rm DE})$ & $\sigma(w_0)$ & $\sigma(w_a)$ 
& FoM & $\sigma(f_{\rm NL})$ &  
& $\sigma(\Omega_{\rm DE})$ & $\sigma(w_0)$ & $\sigma(w_a)$ 
& FoM &$\sigma(f_{\rm NL})$ &  
& $\sigma(\Omega_{\rm DE})$ & $\sigma(w_0)$ & $\sigma(w_a)$ 
& FoM & $\sigma(f_{\rm NL})$ \\ \hline
HSC (2000deg$^2$) 
& 0.035 & 0.37 & 1.13 &   7.5 &  15.6 &  
& 0.056 & 0.44 & 1.12 &   5.1 &  86.9 &
& 0.023 & 0.22 & 0.60 &  28.6 &  14.5 \\
HSC (1500deg$^2$) 
& 0.040 & 0.42 & 1.26 &   6.0 &  18.0 &
& 0.062 & 0.48 & 1.24 &   4.1 &  99.9 &
& 0.026 & 0.26 & 0.69 &  21.9 &  16.8 \\
DES  
& 0.026 & 0.27 & 0.82 &  14.0 &  10.9 &
& 0.053 & 0.52 & 1.39 &   4.7 &  66.3 &
& 0.020 & 0.22 & 0.62 &  30.7 &  10.3 \\
LSST 
& 0.011 & 0.12 & 0.38 &  59.4 &   2.3 &
& 0.016 & 0.12 & 0.29 &  54.0 &  13.0 &
& 0.007 & 0.06 & 0.16 & 322.2 &   2.2 \\
\hline\hline
\end{tabular}
\caption{Forecasts of cosmological parameters estimated from the
  Fisher matrix analysis. We compare three upcoming  surveys, HSC, DES
  and LSST. All errors are 68\% errors, after marginalizing over the
  other parameters, expected from the combination of the cluster
  number counts ($\bmf{N}$), the cluster power spectra 
  ($\bmf{C}^{\rm hh}$), and the stacked weak lensing signals  
  ($\bmf{C}^{{\rm h}\kappa}$). No prior is added on nuisance
  parameters. The Planck prior is added to all cases.} 
\label{tab:cons}
\end{table*}

\begin{figure*}[t]
\begin{center}
\includegraphics[width=0.75\textwidth]{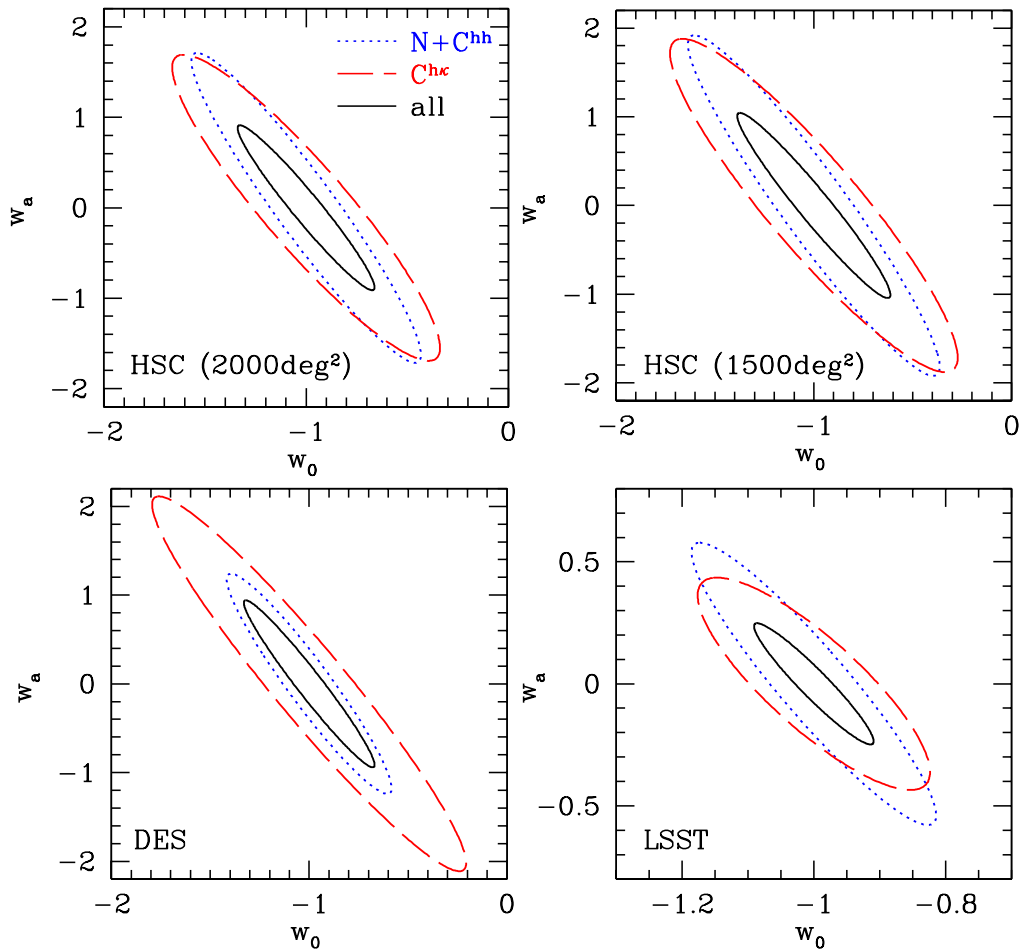}
\caption{Marginalized constraints in the $w_0$-$w_a$ plane for three
  different future surveys, HSC ({\it upper-left and upper-right
  panels}), DES ({\it lower-left}), and LSST ({\it lower-right}). The
  different curves are as in Fig.~\ref{fig:ell}, i.e., 
  the dotted curves denote constraints from the cluster number counts
  and the cluster-cluster correlation functions, the dashed curves are
  constraints from the stacked weak lensing (both 1- and 2-halo terms),
  and the solid curves show the combined constraints. While the Planck
  prior is added to all the ellipses, no prior is added to nuisance
  parameters.}  \label{fig:ell_survey}
\end{center}
\end{figure*}

The marginalized errors on dark energy parameters and $f_{\rm NL}$
are summarized in Table~\ref{tab:cons}, and the projected error
ellipses in the $w_0$-$w_a$ plane are presented in
Fig.~\ref{fig:ell_survey}. The difference of FoM between HSC of 1500
and 2000~deg$^2$, and LSST surveys can be understood by their
differences in the survey area. The FoM roughly scales with the
survey area as ${\rm FoM}\propto \Omega_{\rm s}$, and the difference
between the FoM values of HSC and LSST can approximately be explained
by this relation. On the other hand, the FoM from DES is rather small
despite its large area coverage of 5000~deg$^2$. This is because DES
is shallower than HSC and LSST, thereby yielding a narrower window of
the redshift sensitivity to dark energy parameters due to the smaller
number density of source galaxies and the lower maximum cluster
redshift (see Table~\ref{tab:survey}). In all the surveys, the FoM is
improved by a factor of $2-4$ by adding the stacked weak lensing
signals (see also Fig.~\ref{fig:ell_survey}), suggesting the powerful
complementarity of stacked weak lensing to cluster cosmology.
Table~\ref{tab:extra} summarize improvements of constraints in each
survey by adding the cosmic shear measurement for the same source
sample and the BOSS BAO information.

\begin{table*}[!t]
\begin{tabular}{@{}cccccccccccccccccc}
  \hline\hline
& \multicolumn{5}{c}{+$\bmf{C}^{\kappa\kappa}$} &
& \multicolumn{5}{c}{+BAO} &
& \multicolumn{5}{c}{+$\bmf{C}^{\kappa\kappa}$+BAO} \\ 
\cline{2-6} \cline{8-12} \cline {14-18}
Survey  
& $\sigma(\Omega_{\rm DE})$ & $\sigma(w_0)$ & $\sigma(w_a)$ 
& FoM & $\sigma(f_{\rm NL})$ & 
& $\sigma(\Omega_{\rm DE})$ & $\sigma(w_0)$ & $\sigma(w_a)$ 
& FoM & $\sigma(f_{\rm NL})$ & 
& $\sigma(\Omega_{\rm DE})$ & $\sigma(w_0)$ & $\sigma(w_a)$ 
& FoM & $\sigma(f_{\rm NL})$ \\ \hline
HSC (2000deg$^2$) 
& 0.022 & 0.22 & 0.59 &  32.4 &  14.3 &
& 0.017 & 0.18 & 0.48 &  52.7 &  14.2 &
& 0.017 & 0.18 & 0.47 &  54.8 &  14.0 \\
HSC (1500deg$^2$) 
& 0.025 & 0.25 & 0.67 &  24.8 &  16.5 &
& 0.018 & 0.20 & 0.52 &  45.0 &  16.3 &
& 0.018 & 0.19 & 0.51 &  46.7 &  16.1 \\
DES  
& 0.019 & 0.21 & 0.60 &  36.3 &  10.1 &
& 0.016 & 0.18 & 0.48 &  53.4 &  10.0 &
& 0.015 & 0.17 & 0.48 &  56.4 &   9.9 \\
LSST 
& 0.006 & 0.06 & 0.16 & 348.1 &   2.2 &
& 0.006 & 0.06 & 0.16 & 338.9 &   2.2 &
& 0.006 & 0.06 & 0.16 & 361.2 &   2.2 \\
\hline\hline
\end{tabular}
\caption{The improvements of cosmological constraints by adding extra
  constraints from the cosmic shear measurements ($\bmf{C}^{\kappa\kappa}$)
  using the same source galaxy sample as used in the stacked weak lensing
  analysis and from the baryon acoustic oscillation observations
  expected from the SDSS-III BOSS (BAO). }  
\label{tab:extra}
\end{table*}

\section{Discussions on other systematics}
\label{sec:discuss}

While our analysis properly takes account of a number of key
systematic effects, such as the mass-observable relation, the mean
source redshift for weak lensing analysis, and the offset of cluster
centers, there are still a number of potential systematics that could
affect our results. In this section, we briefly discuss these effects.

Perhaps one of the most important effects ignored in this paper is the
error on cluster redshifts. In the future imaging surveys as considered
above, cluster redshifts are thought to be measured using the
photometric redshift technique. \cite{LimaHu:07} studied how the
photo-$z$ errors of clusters affect the cluster number counts, and then
argued that photometric redshifts have to be accurate at $\lesssim 1\%$
level in order not to degrade dark energy constraints significantly. In
fact this accuracy is almost reachable in future surveys, as we can use
several tens of member galaxies to determine cluster redshifts
accurately (e.g., \cite{Linetal:06}). We should comment that, if a
redshift survey of Luminous Red Galaxies (LRGs) is available from the
imaging survey region, such as the SDSS survey, the LRG catalog can be
used to construct a secure catalog of clusters with spectroscopic
redshifts. The LRGs are very likely to be one of the brightest cluster
galaxies, and are relatively easy to measure their spectroscopic
redshifts as they are bright. Thus the LRG catalog is enormously
helpful in constructing a homogeneous catalog of clusters, by
searching for member galaxies surrounding each LRG with deeper imaging
survey. This is indeed the case for HSC, as the target region of HSC
is overlapped with the survey region of the SDSS-III BOSS.

An imperfect photometric redshift estimate from limited color
information may result in a leakage of foreground galaxies to the
source galaxy sample. This contamination not only dilutes the shear
signals particularly near the cluster centers, but also complicates
the self-calibration of the mean source redshift because such
contamination cannot be described by a single parameter. 
This is a potentially important effect which should be studied
carefully using mock catalogs or simulations.

Here we briefly estimate the possible contamination of foreground
galaxies, based on the result of \cite{Nishizawaetal:10} which
developed a clipping method to reduce the effect of photo-$z$ outliers
in weak lensing analysis. Assuming a Subaru HSC-type survey with the
$g'r'i'z'y'$ filter set, the estimated contamination rate for the
fiducial source galaxy population defined by $z_s>1.5$ is $\sim
0.6$, if no clipping is applied. However the contamination rate is
reduced to $\sim 0.2$ if 70\% of galaxies are clipped. The resulting
source galaxy number density of $\bar{n}_{\rm gal}\sim
4$~arcmin$^{-2}$ ($z_s>1.5$) after clipping is in fact similar to
the value adopted in the paper, $\bar{n}_{\rm gal}\sim 5.2$~arcmin$^{-2}$,
because of our conservative choice of the source galaxy distributions.
Although the estimated contamination rate is still large, we can
further reduce the contamination by adding near-infrared images or by 
decreasing the minimum redshift. For instance, if we add $JHK$-band
images that can be delivered from the VISTA Kilo-degree Infrared
Galaxy (VIKING) survey, the contamination rate becomes $\lesssim 3\%$,
which should be small enough not to have a significant impact on our
result. We can achieve similar contamination rate of $\lesssim 3\%$ by
decreasing the minimum source redshift from $z_{s,{\rm min}}=1.5$ to
$\sim 1.1$, which has been shown to have a small impact in terms of
the dark energy FoM (see Fig.~\ref{fig:fom}). A more comprehensive
analysis of the effect of photo-$z$ outliers on this technique is a
future work. 

One may consider that the photo-$z$ outliers to cluster redshifts can
be to some extent monitored by using an angular cross-correlation
method of the source galaxy population with objects with spectroscopic
redshifts or secure photo-$z$ estimates, such as cluster themselves
or spectroscopic LRG samples (e.g., \cite{Newman:08,Nishizawaetal:10}
for a study of the cross-correlation method).  
However, the cross-correlation method to eliminate the photo-$z$
outliers may be limited by the magnification bias, as discussed below.
First, lensing due to the foreground mass distribution causes the area
of a given patch on the sky to increase, and thus diluting the number
density of background galaxies. Second, galaxies fainter than the
limiting magnitude can be magnified to cross the threshold. Therefore,
if the magnitude cut is used in selecting source galaxies in addition
to their colors, the magnification bias may cause the observed
densities of background galaxies to be increased or decreased,
depending on the slope of the galaxy number counts. This magnification
bias causes an apparent correlation between the densities of
background galaxies and the cluster distribution (e.g.,
\cite{Broadhurstetal:05}), and hence degrades the power of the 
cross-correlation method. Here we give a rough estimate on the
effect. The apparent cross-correlation between clusters of redshift
slice $z_l$ and the source galaxy distribution is given as 
$\sang{n_{\rm gal}(z_s)n_{\rm cl}(z_l)}\propto (2-5s)
(\Delta z/H(z_l))W^\kappa(z_l)b_h(M)\xi_m(z_l)$ (see
Eq.~\ref{eq:kappa}), where $\xi_m$ is the mass correlation function 
of the cluster redshift $z_l$ and $s$ is the slope of the galaxy
number count as a function of the galaxy magnitude $m$, i.e., 
$s\equiv d\ln n_{\rm gal}/d\ln m$. On the other hand, a cross-correlation  
between (the outliers of) source galaxies and the cluster distribution
is given as $\sang{n_{\rm gal}^{\rm photo-z}(z_s)n_{\rm cl}(z_l)}\propto f_{\rm
outlier}(z_l)\Delta z b_h(M)b_g\xi_m(z_l)$, where $f_{\rm outlier}(z_l)$
is the fraction of photo-$z$ outliers leaking into the cluster redshift
$z_l$.  Hence the ratio of these cross-correlations can be found to be
$O(0.1)/f_{\rm outlier}$ for $z_l\sim 0.5$, assuming $b_g\sim 1$ and the
magnitude slope $(2-5s)\sim 1$ as implied from observations
\cite{Broadhurstetal:05}. Thus the cross-correlation method can
eliminate photo-$z$ outliers if the outliers contribute to the source
galaxy catalog by more than 10\% in a given cluster redshift slice.
This rough estimate suggests that the cross-correlation method may not
be so useful to perfectly eliminate the photo-$z$ outliers. 
Even if this is the case, the shear and the magnification bias arise
from the same mass distribution around the clusters, and therefore
this magnification bias effect on the source galaxy population may be
iteratively calibrated by combining the measurements of the stacked
shear profile and the cross-correlation function. More optimistically,
we may be able to improve the mass calibration with staked lensing by
adding information from the magnification bias 
(e.g., \cite{VanWarbekeetal:10,RozoSchmidt:10}). 

We note that the intrinsic alignments of galaxy ellipticities, which
are a significant source of systematic errors for cosmic shear
measurements (e.g., \cite{HirataSeljak:04}), are not important for our
analysis. This is because the stacked lensing is linear in shear, and
does not include any correlation between the shapes of different
galaxies. 

\begin{figure}[t]
\begin{center}
\includegraphics[width=0.4\textwidth]{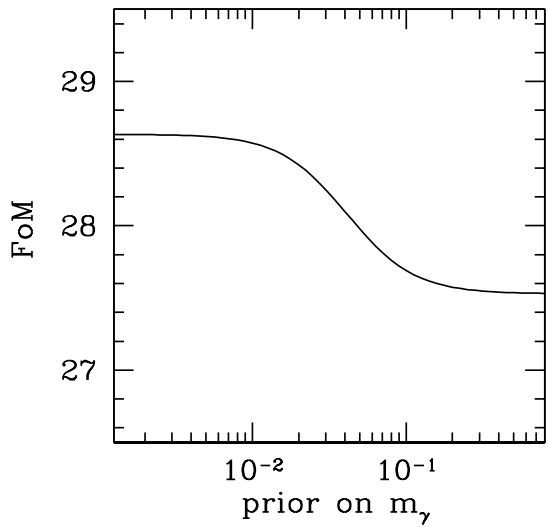}
\caption{The FoM as a function of the prior on the multiplicative shear
  error $m_\gamma$, which is a newly included nuisance parameter in
  the analysis. See text for more details.}
 \label{fig:prior_m}
\end{center}
\end{figure}

The shape measurement of faint galaxies involves several uncertainties
as extensively studied in the literature (e.g., \cite{Bridleetal:10}).
For the cosmic shear measured from upcoming wide-field surveys, the
uncertainties such as the additive bias and multiplicative errors in
the shear estimates need to be calibrated to better than 1\% or
$0.1\%$ in shear in order not to significantly degrade cosmological
parameter estimations \cite{Hutereretal:06}. Since our method focuses
on the single population of source galaxies, all clusters share the 
same systematic errors in the shear calibrations, even if they exist, 
in such a way that they are uncorrelated with the distribution of
foreground clusters. In this sense, this effect is similar to the
source redshift uncertainty, especially for the multiplicative shear
error. Thus we expect that the shape measurement errors can similarly
be self-calibrated to the comparable accuracy to that of the source
redshift uncertainty by using our method. 
To check this point, we repeat the Fisher matrix analysis for our
fiducial Subaru HSC-type survey including the multiplicative shear error
$m_\gamma\equiv\gamma^{\rm measure}/\gamma^{\rm true}-1$ as
additional nuisance parameter (i.e., 34 parameters in total). Without
any prior on $m_\gamma$, we find the FoM to be 27.5, which is only
slightly smaller than the original FoM of 28.6. This indicates that
the multiplicative shear error does not significantly degrade our
cosmological result. Furthermore, the marginalized error on the
multiplicative shear error is $\sigma(m_\gamma)=0.04$, which is the
level already achieved by the current shear measurement techniques
\cite{Massey:07}. Fig.~\ref{fig:prior_m} shows the FoM as a function
of the prior on $m_\gamma$.

In this paper we assume the Gaussian statistics in computing the
covariance of stacked weak lensing signal. The assumption is valid 
for the 1-halo term because one of the points is always fixed to the
cluster center. On the other hand, the 2-halo term dominates only at
very large scales, $\ell \lesssim 10^2$ (see Fig.~\ref{fig:stack}), 
where the effect of the non-Gaussian error is small. Thus we expect
the effect of non-Gaussian errors is small for our results. We also
ignored the cross-covariance between the cluster number counts and
the cluster power spectra (the 2-halo term and cluster-cluster angular
power spectra). \cite{TakadaBridle:07} has shown that the
cross-correlation coefficients peak around $\ell \sim 10^3$, whereas
in our calculations the information from the power spectra comes
mainly from $\ell \sim 10^2$ (see Figs.~\ref{fig:stack} and
\ref{fig:cl_hh}). Therefore, the effect of the cross-covariance should
also be small in our case.

Our results are based on the flat-sky and Limber's approximations. 
While the full calculation without relying on these approximations is
not difficult \cite{dePutterTakada:10}, their effects on our results,
i.e, forecast constraints on cosmological parameters, are expected to
be small, because we are interested in sensitivities of the signals,
rather than signals themselves, on cosmological parameters. An
exception is constraints on the primordial non-Gaussianity, 
$f_{\rm NL}$, as the constraints come mostly from very large scales
($\ell \sim \mathcal{O}(1)$) where these approximations obviously
break down. Therefore, our results on $f_{\rm NL}$ should be taken
with caution.

Also important are the effects of theoretical uncertainties, such as
the inaccuracy of the halo mass function, the halo bias
(including the assembly bias and the stochasticity),
and the radial profile. Previous work \cite{CunhaEvrard:10,Wuetal:10} has
found that theoretical uncertainties of the mass function and the halo
bias can indeed be important. Such uncertainties might also be
self-calibrated to some extent from the combination of observables
studied in this paper, but the detailed exploration is beyond the
scope of this paper. 

There are several additional systematics in cluster weak lensing
studies, including the effects of central galaxies, non-weak shear and 
magnification bias. Some of these effects on stacked cluster weak
lensing have been investigated by \cite{RozoWuSchmidt:10}. Again, our
conservative choice of $\ell_{\rm max}=10^4$ is meant to reduce these
effects, but in practical analysis more careful treatments will be
needed to minimize such effects, as done recently by 
\cite{Mandelbaumetal:10}. 

\section{Conclusion}
\label{sec:conc}

In this paper, we have estimated accuracies on cosmological parameters
derivable from a joint experiment of the cluster observables (number
counts and cluster-cluster correlation functions) and stacked weak 
lensing signals of distant galaxy shapes produced by the clusters,
both of which can be drawn from a wide-field optical imaging
survey. A striking advantage of this method is the derived
cosmological constraints become insensitive to various systematic
errors such as the source redshift uncertainty and the cluster
mass-observable relation. Our key assumption to overcome these
systematic uncertainties is to focus on a single population of
background source galaxies to extract lensing signals around cluster
at different redshifts. By cross-correlating the source galaxy shapes
with the distribution of clusters with known redshifts (or at least
secure photo-$z$ estimates), we can statistically separate out the
contribution at a given redshift slice from the total shear signals
of the source galaxies. Put another way, this method enables a
tomographic reconstruction of lensing structures. Moreover, the
systematic errors of the cluster observables and the stacked lensing
are closely related with each other, which enables an efficient
self-calibration of these systematics (see Fig.~\ref{fig:idea} for the
conceptual flow-chart). Specifically, the stacked lensing provides
direct measurements of the mean mass of clusters, and therefore  
allows to calibrate the cluster mass-observable relation, whereas 
we can calibrate the source redshift uncertainty via the relative
strengths of the stacked lensing signals at different cluster
redshifts, because the stacked lensing signals at different cluster
redshifts share the same source redshift uncertainty ({\it only if}
the single population of background source galaxies is used, as
assumed in this paper). As a result, cosmological parameters including
dark energy parameters are well constrained yet least affected by
these significant systematic effects (see
Tables~\ref{tab:hsc_forecast} and \ref{tab:cons} and
Figs.~\ref{fig:ell}, \ref{fig:fom} and \ref{fig:prior}). 

To estimate the cosmological power of the method above, we have
developed a formulation to compute all the observables based on the
halo model, and also derived the power spectra of the relevant
correlation functions. In particular we have shown that the halo
off-centering effect on the stacked lensing signals is expressed by a
simple analytic form in Fourier space (see Appendix~\ref{sec:off}). 
Moreover, the Fourier-space analysis greatly simplifies the Fisher
matrix calculations, because the power spectrum covariances have the
simpler forms than in real space. However, note that, even if the
real-space observables such as the angular correlation functions are 
used as done in the previous measurements, the cosmological power
should be equivalent to what was shown in this paper, because the
two-point correlations and the power spectra contain the equivalent
information at two-point level. We believe that the formulation
developed in this paper is useful in further extending the method,
e.g., including additional power spectrum information such as the SZ
power  spectrum available from the upcoming wide-field SZ surveys with
an overlap with optical imaging surveys, such as the HSC and the
ACT or the DES and the SPT. The additional information can further
improve the cosmological constraints, although new systematic errors
introduced by new observations should carefully be studied. This is
our future project, and will be presented elsewhere. 

The results shown in this paper can be compared with the previous work
studying the self-calibration technique for cluster observables, where
the number counts and the cluster clustering information are combined to
self-calibrate the mass-observable relation (e.g., \cite{MajumdarMohr:04}). 
We have found that adding the stacked lensing signals significantly improves
the marginalized errors on each cosmological parameter by up to a factor
of $\sim 2$, especially the dark energy FoM by a factor of $\sim 4$,
which is achieved without assuming any priors on a number of nuisance
parameters to model various systematic errors, including the
mass-observable relation, the mean source galaxy redshift, offsets of
cluster centers, and concentration parameters. The expected constraints 
on dark energy are quite comparable to those obtained from tomographic
cosmic shear measurements obtained without including any systematic
errors. We have shown that an accurate self-calibration of these
systematic errors is indeed attained, with the mean source redshift
calibrated to $\sigma(z_m)\sim 0.1$ and the mean cluster mass in each
bin to $\sigma(\ln M_{\rm bias})\sim 0.05-0.1$ (see Fig.~\ref{fig:merr}).
We have also confirmed that clustering of massive clusters is powerful
in constraining primordial non-Gaussianity. The robust constraint of
$\sigma(f_{\rm NL})\lesssim 10$ can be obtained if the joint
experiment is applied to upcoming imaging surveys.   

In summary, we have demonstrated that weak lensing information and
cluster observables, both available from the same imaging survey data,
are complementary to each other. The cosmological power is greatly
improved and an efficient self-calibration of various systematic
uncertainties can be attained when the two measurements are
combined. This is particularly promising given that there are several
planned wide-field optical imaging surveys overlapping with SZ
surveys: Subaru HSC survey and ACT, the DES and the SPT, and also the
all-sky Planck CMB survey which will be available in a few years.

\begin{acknowledgments}
We thank A.~Nishizawa for sharing results of his work with us, and 
G.~Bernstein, D.~Huterer, K.~Ichiki, B.~Jain, E.~Komatsu, E.~Krause,
R.~Mandelbaum, and D.~Spergel for useful discussions and
comments. This work is supported in part by JSPS Core-to-Core Program
``International Research Network for Dark Energy'' and World Premier
International Research Center Initiative (WPI Initiative), MEXT, Japan.  
\end{acknowledgments}

\appendix

\begin{figure*}[t]
\begin{center}
\includegraphics[width=0.85\textwidth]{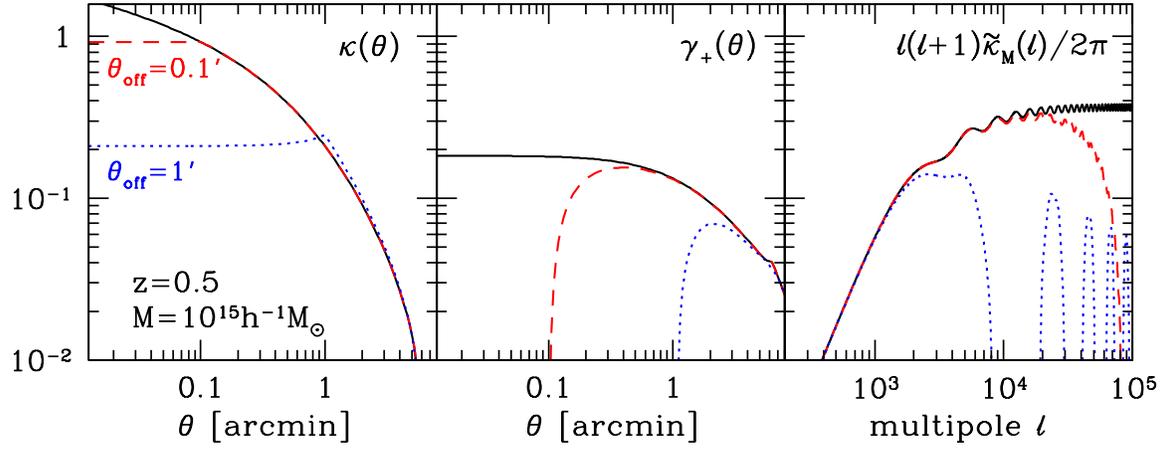}
\caption{Effect of off-centering on radial profiles of a single
 cluster with $M=10^{15}h^{-1}M_\odot$ and $z=0.5$, computed using the
 Fourier space description derived in Eq.~(\ref{eq:kappa_off}).
 From left  to right panels, we show the convergence profile,
 the tangential shear profile, and the Fourier transform of the
 convergence. Solid lines show the case with no offset 
 ($\theta_{\rm off}=0$), whereas dashed and dotted lines indicate
 profiles with the offset $\theta_{\rm off}=0.1'$ and $1'$,
 respectively.} 
\label{fig:kappa_off}
\end{center}
\end{figure*}

\begin{table}[!t]
\begin{tabular}{cll}
\hline\hline
Symbol & Definition & Eq. \\ \hline
$P^{\rm L}_m(k)$ & Linear mass power spectrum & (\ref{eq:linearps})\\
$P^{\rm NL}_m(k)$ & Nonlinear mass power spectrum & (\ref{eq:kappaps})\\
$M$& Halo mass &  (\ref{eq:mobs})\\
$M_{\rm obs}$& Cluster mass observable  & (\ref{eq:mobs}) \\
$\sigma_{\ln M}$& Variance in mass-observable rel.   &  (\ref{eq:mobs})\\
$M_{\rm bias}$& Halo mass bias   &  (\ref{eq:mobs})\\
$z_{i,{\rm min}}$, $z_{i,{\rm max}}$ 
& The $i$-th redshift bin of clusters & (\ref{eq:numdens})\\
$M_{ b,{\rm min}}$, $M_{ b,{\rm min}}$ & The $b$-th mass bin of clusters & (\ref{eq:numdens})\\
$n_{i(b)}$, $N_{i(b)}$ & Cluster number counts & (\ref{eq:numdens}), (\ref{eq:counts})\\ 
$dn/dM$ & Halo mass function  & (\ref{eq:numdens})\\
$S_{i(b)}(M)$ & Selection function of clusters & (\ref{eq:select_cluster})\\
$C^{\rm hh}_{i(bb')}(\ell)$ & Cluster power spectrum & (\ref{eq:ccps})\\
$W^{\rm h}_{i(b)}(M)$ & Weight function of clusters & (\ref{eq:cweight})\\
$b_h(M)$ & Halo bias & (\ref{eq:cweight})\\
$W^{\kappa}(z)$ & Lensing weight function & (\ref{eq:kappa_weight})\\
$C^{\kappa\kappa}(\ell)$ & Shear power spectrum & (\ref{eq:kappaps})\\
$\sang{\gamma_+}_{i(b)}(\theta)$ & Tangential shear profile & (\ref{eq:gamma_cl}) \\
$C^{{\rm h}\kappa}_{i(b)}(\ell)$ & Shear-cluster power spectrum & (\ref{eq:c_hk_tot}) \\
$C^{{\rm h}\kappa, {\rm 1h}}_{i(b)}(\ell)$ &1-halo term of $C^{{\rm h}\kappa}(\ell)$ &
     (\ref{eq:cl_hk_1h}) \\
$C^{{\rm h}\kappa, {\rm 2h}}_{i(b)}(\ell)$ &2-halo term of $C^{{\rm h}\kappa}(\ell)$ &
     (\ref{eq:cl_hk_2h}) \\
$\tilde{u}_M(k)$ & Halo mass profile & (\ref{eq:nfw_uk})\\
$\tilde{\kappa}_M(\ell)$ & Halo convergence profile & (\ref{eq:kappa_halo})\\
$\tilde{\kappa}_{M,{\rm off}}(\ell)$ & With halo centering offset  & (\ref{eq:kappa_offdis})\\
$c(M,z)$ & Halo concentration & (\ref{eq:concent})\\
\hline\hline
\end{tabular}
\caption{Symbols used in the paper.}
\label{tab:notation}
\end{table}

\section{Derivation of Off-centered Lensing Signals}
\label{sec:off}

We consider an azimuthally averaged convergence profile, whose true
center is located at $\bmf{\theta}_0$, at distance $\theta$ from
the coordinate origin.  It is expressed as
\begin{eqnarray}
\kappa_{\rm off}(\theta)&=&\int \frac{d\phi'}{2\pi}
\kappa(\bmf{\theta}'-\bmf{\theta}_0)\nonumber\\
&=&\int d^2\bmf{\theta}'\,\kappa(\bmf{\theta}'-\bmf{\theta}_0)W_\theta(\bmf{\theta}'),
\end{eqnarray}
where $W_\theta(\bmf{\theta}')$ is defined by
\begin{equation}
W_\theta(\bmf{\theta}')\equiv\frac{1}{2\pi\left|\bmf{\theta}'\right|}\delta\left(\left|\bmf{\theta}'\right|-\theta\right).
\end{equation}
The Fourier transform in angular space is given by
\begin{eqnarray}
\tilde{\kappa}_{M,{\rm off}}(\ell)&=&\int
d^2\bmf{\theta}
e^{-i\mbox{\boldmath{$\ell$}}\bmf{\theta}}\int
d^2\bmf{\theta}'\kappa(\bmf{\theta}'-\bmf{\theta}_0)W_\theta(\bmf{\theta}'),\nonumber\\
&=&\int
d^2\bmf{\theta}'J_0(\ell\theta')\int\frac{d^2\mbox{\boldmath{$\ell$}}'}{(2\pi)^2}\tilde{\kappa}_M(\ell')e^{i\mbox{\boldmath{$\ell$}}'(\bmf{\theta}'-\bmf{\theta}_0)}\nonumber\\
&=&\int \frac{d^2\mbox{\boldmath{$\ell$}}'}{2\pi} \tilde{\kappa}_M(\ell') 
  e^{-i\mbox{\boldmath{$\ell$}}'\bmf{\theta}_0}\int\theta'd\theta'
  J_0(\ell\theta')J_0(\ell'\theta')\nonumber\\
&=&\int d\ell' \tilde{\kappa}_M(\ell')
J_0(\ell'\theta_{\rm off})\delta(\ell-\ell')\nonumber\\
&=&\tilde{\kappa}_M(\ell) J_0(\ell\theta_{\rm off}),
\label{eq:kappa_off}
\end{eqnarray}
where $\theta_{\rm off}=|\bmf{\theta}_0|$, and used the integral
representation of the Bessel function
\begin{equation}
J_0(x)=\int_0^{2\pi}\frac{d\phi}{2\pi}e^{ix\cos\phi},
\end{equation}
and the closure equation
\begin{equation}
\int_0^\infty x dx J_0(ax)J_0(bx)=\frac{1}{a}\delta(a-b).
\end{equation}
Therefore the effect of off-centering is simply expressed as the
multiplication of the zeroth order Bessel function in Fourier
space. Note that Eq.~(\ref{eq:kappa_off}) is general expression
applicable to any convergence profile $\kappa$. For reference,
adopting the flat sky approximation convergence profiles 
$\kappa(\theta)$ and tangential shear profile $\gamma_+(\theta)$
with the effect of the offset are given by 
\begin{equation}
\kappa(\theta) = \int \frac{\ell d\ell}{2\pi}\tilde{\kappa}_{M,{\rm off}}(\ell)
J_0(\ell\theta),
\end{equation}
\begin{equation}
\gamma_+(\theta) = \int \frac{\ell d\ell}{2\pi}\tilde{\kappa}_{M,{\rm off}}(\ell)
J_2(\ell\theta).
\end{equation}

In Fig.~\ref{fig:kappa_off}, we show the effect of off-centering on
convergence and tangential shear profiles. Basically the effect is
such that it flattens the convergence profile, and eliminates the
tangential shear signals, below the offset size.

\section{Summary of Symbols}

In Table~\ref{tab:notation} we give a summary of symbols commonly used
in the paper.  

\bibliography{refs_mt}

\begin{thebibliography}{99}
\expandafter\ifx\csname natexlab\endcsname\relax\def\natexlab#1{#1}\fi
\expandafter\ifx\csname bibnamefont\endcsname\relax
  \def\bibnamefont#1{#1}\fi
\expandafter\ifx\csname bibfnamefont\endcsname\relax
  \def\bibfnamefont#1{#1}\fi
\expandafter\ifx\csname citenamefont\endcsname\relax
  \def\citenamefont#1{#1}\fi
\expandafter\ifx\csname url\endcsname\relax
  \def\url#1{\texttt{#1}}\fi
\expandafter\ifx\csname urlprefix\endcsname\relax\def\urlprefix{URL }\fi
\providecommand{\bibinfo}[2]{#2}
\providecommand{\eprint}[2][]{\url{#2}}

\bibitem[{\citenamefont{{Albrecht} et~al.}(2006)\citenamefont{{Albrecht},
  {Bernstein}, {Cahn}, {Freedman}, {Hewitt}, {Hu}, {Huth}, {Kamionkowski},
  {Kolb}, {Knox} et~al.}}]{DETF}
\bibinfo{author}{\bibfnamefont{A.}~\bibnamefont{{Albrecht}}},
  \bibinfo{author}{\bibfnamefont{G.}~\bibnamefont{{Bernstein}}},
  \bibinfo{author}{\bibfnamefont{R.}~\bibnamefont{{Cahn}}},
  \bibinfo{author}{\bibfnamefont{W.~L.} \bibnamefont{{Freedman}}},
  \bibinfo{author}{\bibfnamefont{J.}~\bibnamefont{{Hewitt}}},
  \bibinfo{author}{\bibfnamefont{W.}~\bibnamefont{{Hu}}},
  \bibinfo{author}{\bibfnamefont{J.}~\bibnamefont{{Huth}}},
  \bibinfo{author}{\bibfnamefont{M.}~\bibnamefont{{Kamionkowski}}},
  \bibinfo{author}{\bibfnamefont{E.~W.} \bibnamefont{{Kolb}}},
  \bibinfo{author}{\bibfnamefont{L.}~\bibnamefont{{Knox}}},
  \bibnamefont{et~al.}, \bibinfo{journal}{ArXiv Astrophysics e-prints}
  (\bibinfo{year}{2006}), \eprint{arXiv:astro-ph/0609591}.

\bibitem[{\citenamefont{{Vikhlinin} et~al.}(2009)\citenamefont{{Vikhlinin},
  {Kravtsov}, {Burenin}, {Ebeling}, {Forman}, {Hornstrup}, {Jones}, {Murray},
  {Nagai}, {Quintana} et~al.}}]{Vikhlininetal:09}
\bibinfo{author}{\bibfnamefont{A.}~\bibnamefont{{Vikhlinin}}},
  \bibinfo{author}{\bibfnamefont{A.~V.} \bibnamefont{{Kravtsov}}},
  \bibinfo{author}{\bibfnamefont{R.~A.} \bibnamefont{{Burenin}}},
  \bibinfo{author}{\bibfnamefont{H.}~\bibnamefont{{Ebeling}}},
  \bibinfo{author}{\bibfnamefont{W.~R.} \bibnamefont{{Forman}}},
  \bibinfo{author}{\bibfnamefont{A.}~\bibnamefont{{Hornstrup}}},
  \bibinfo{author}{\bibfnamefont{C.}~\bibnamefont{{Jones}}},
  \bibinfo{author}{\bibfnamefont{S.~S.} \bibnamefont{{Murray}}},
  \bibinfo{author}{\bibfnamefont{D.}~\bibnamefont{{Nagai}}},
  \bibinfo{author}{\bibfnamefont{H.}~\bibnamefont{{Quintana}}},
  \bibnamefont{et~al.}, \bibinfo{journal}{\apj} \textbf{\bibinfo{volume}{692}},
  \bibinfo{pages}{1060} (\bibinfo{year}{2009}), \eprint{0812.2720}.

\bibitem[{\citenamefont{{Mantz}
  et~al.}(2010{\natexlab{a}})\citenamefont{{Mantz}, {Allen}, {Rapetti}, and
  {Ebeling}}}]{Mantzetal:10a}
\bibinfo{author}{\bibfnamefont{A.}~\bibnamefont{{Mantz}}},
  \bibinfo{author}{\bibfnamefont{S.~W.} \bibnamefont{{Allen}}},
  \bibinfo{author}{\bibfnamefont{D.}~\bibnamefont{{Rapetti}}},
  \bibnamefont{and}
  \bibinfo{author}{\bibfnamefont{H.}~\bibnamefont{{Ebeling}}},
  \bibinfo{journal}{\mnras} \textbf{\bibinfo{volume}{406}},
  \bibinfo{pages}{1759} (\bibinfo{year}{2010}{\natexlab{a}}),
  \eprint{0909.3098}.

\bibitem[{\citenamefont{{Rozo} et~al.}(2010{\natexlab{a}})\citenamefont{{Rozo},
  {Wechsler}, {Rykoff}, {Annis}, {Becker}, {Evrard}, {Frieman}, {Hansen},
  {Hao}, {Johnston} et~al.}}]{Rozoetal:10}
\bibinfo{author}{\bibfnamefont{E.}~\bibnamefont{{Rozo}}},
  \bibinfo{author}{\bibfnamefont{R.~H.} \bibnamefont{{Wechsler}}},
  \bibinfo{author}{\bibfnamefont{E.~S.} \bibnamefont{{Rykoff}}},
  \bibinfo{author}{\bibfnamefont{J.~T.} \bibnamefont{{Annis}}},
  \bibinfo{author}{\bibfnamefont{M.~R.} \bibnamefont{{Becker}}},
  \bibinfo{author}{\bibfnamefont{A.~E.} \bibnamefont{{Evrard}}},
  \bibinfo{author}{\bibfnamefont{J.~A.} \bibnamefont{{Frieman}}},
  \bibinfo{author}{\bibfnamefont{S.~M.} \bibnamefont{{Hansen}}},
  \bibinfo{author}{\bibfnamefont{J.}~\bibnamefont{{Hao}}},
  \bibinfo{author}{\bibfnamefont{D.~E.} \bibnamefont{{Johnston}}},
  \bibnamefont{et~al.}, \bibinfo{journal}{\apj} \textbf{\bibinfo{volume}{708}},
  \bibinfo{pages}{645} (\bibinfo{year}{2010}{\natexlab{a}}),
  \eprint{0902.3702}.

\bibitem[{\citenamefont{{Schmidt} et~al.}(2009)\citenamefont{{Schmidt},
  {Vikhlinin}, and {Hu}}}]{Schmidtetal:09}
\bibinfo{author}{\bibfnamefont{F.}~\bibnamefont{{Schmidt}}},
  \bibinfo{author}{\bibfnamefont{A.}~\bibnamefont{{Vikhlinin}}},
  \bibnamefont{and} \bibinfo{author}{\bibfnamefont{W.}~\bibnamefont{{Hu}}},
  \bibinfo{journal}{\prd} \textbf{\bibinfo{volume}{80}},
  \bibinfo{pages}{083505} (\bibinfo{year}{2009}), \eprint{0908.2457}.

\bibitem[{\citenamefont{{Reyes} et~al.}(2010)\citenamefont{{Reyes},
  {Mandelbaum}, {Seljak}, {Baldauf}, {Gunn}, {Lombriser}, and
  {Smith}}}]{Reyesetal:10}
\bibinfo{author}{\bibfnamefont{R.}~\bibnamefont{{Reyes}}},
  \bibinfo{author}{\bibfnamefont{R.}~\bibnamefont{{Mandelbaum}}},
  \bibinfo{author}{\bibfnamefont{U.}~\bibnamefont{{Seljak}}},
  \bibinfo{author}{\bibfnamefont{T.}~\bibnamefont{{Baldauf}}},
  \bibinfo{author}{\bibfnamefont{J.~E.} \bibnamefont{{Gunn}}},
  \bibinfo{author}{\bibfnamefont{L.}~\bibnamefont{{Lombriser}}},
  \bibnamefont{and} \bibinfo{author}{\bibfnamefont{R.~E.}
  \bibnamefont{{Smith}}}, \bibinfo{journal}{\nat}
  \textbf{\bibinfo{volume}{464}}, \bibinfo{pages}{256} (\bibinfo{year}{2010}),
  \eprint{1003.2185}.

\bibitem[{\citenamefont{{Rapetti} et~al.}(2010)\citenamefont{{Rapetti},
  {Allen}, {Mantz}, and {Ebeling}}}]{Rapettietal:10}
\bibinfo{author}{\bibfnamefont{D.}~\bibnamefont{{Rapetti}}},
  \bibinfo{author}{\bibfnamefont{S.~W.} \bibnamefont{{Allen}}},
  \bibinfo{author}{\bibfnamefont{A.}~\bibnamefont{{Mantz}}}, \bibnamefont{and}
  \bibinfo{author}{\bibfnamefont{H.}~\bibnamefont{{Ebeling}}},
  \bibinfo{journal}{\mnras} \textbf{\bibinfo{volume}{406}},
  \bibinfo{pages}{1796} (\bibinfo{year}{2010}), \eprint{0911.1787}.

\bibitem[{\citenamefont{{Dalal} et~al.}(2008)\citenamefont{{Dalal}, {Dor{\'e}},
  {Huterer}, and {Shirokov}}}]{Dalaletal:08}
\bibinfo{author}{\bibfnamefont{N.}~\bibnamefont{{Dalal}}},
  \bibinfo{author}{\bibfnamefont{O.}~\bibnamefont{{Dor{\'e}}}},
  \bibinfo{author}{\bibfnamefont{D.}~\bibnamefont{{Huterer}}},
  \bibnamefont{and}
  \bibinfo{author}{\bibfnamefont{A.}~\bibnamefont{{Shirokov}}},
  \bibinfo{journal}{\prd} \textbf{\bibinfo{volume}{77}},
  \bibinfo{pages}{123514} (\bibinfo{year}{2008}), \eprint{arXiv:0710.4560}.

\bibitem[{\citenamefont{{Oguri}}(2009)}]{Oguri:09}
\bibinfo{author}{\bibfnamefont{M.}~\bibnamefont{{Oguri}}},
  \bibinfo{journal}{Physical Review Letters} \textbf{\bibinfo{volume}{102}},
  \bibinfo{pages}{211301} (\bibinfo{year}{2009}), \eprint{0905.0920}.

\bibitem[{\citenamefont{{Roncarelli} et~al.}(2010)\citenamefont{{Roncarelli},
  {Moscardini}, {Branchini}, {Dolag}, {Grossi}, {Iannuzzi}, and
  {Matarrese}}}]{Roncarellietal:10}
\bibinfo{author}{\bibfnamefont{M.}~\bibnamefont{{Roncarelli}}},
  \bibinfo{author}{\bibfnamefont{L.}~\bibnamefont{{Moscardini}}},
  \bibinfo{author}{\bibfnamefont{E.}~\bibnamefont{{Branchini}}},
  \bibinfo{author}{\bibfnamefont{K.}~\bibnamefont{{Dolag}}},
  \bibinfo{author}{\bibfnamefont{M.}~\bibnamefont{{Grossi}}},
  \bibinfo{author}{\bibfnamefont{F.}~\bibnamefont{{Iannuzzi}}},
  \bibnamefont{and}
  \bibinfo{author}{\bibfnamefont{S.}~\bibnamefont{{Matarrese}}},
  \bibinfo{journal}{\mnras} \textbf{\bibinfo{volume}{402}},
  \bibinfo{pages}{923} (\bibinfo{year}{2010}), \eprint{0909.4714}.

\bibitem[{\citenamefont{{Cunha} et~al.}(2010)\citenamefont{{Cunha}, {Huterer},
  and {Dor{\'e}}}}]{Cunhaetal:10}
\bibinfo{author}{\bibfnamefont{C.}~\bibnamefont{{Cunha}}},
  \bibinfo{author}{\bibfnamefont{D.}~\bibnamefont{{Huterer}}},
  \bibnamefont{and}
  \bibinfo{author}{\bibfnamefont{O.}~\bibnamefont{{Dor{\'e}}}},
  \bibinfo{journal}{\prd} \textbf{\bibinfo{volume}{82}},
  \bibinfo{pages}{023004} (\bibinfo{year}{2010}), \eprint{1003.2416}.

\bibitem[{\citenamefont{{Sartoris} et~al.}(2010)\citenamefont{{Sartoris},
  {Borgani}, {Fedeli}, {Matarrese}, {Moscardini}, {Rosati}, and
  {Weller}}}]{Sartorisetal:10}
\bibinfo{author}{\bibfnamefont{B.}~\bibnamefont{{Sartoris}}},
  \bibinfo{author}{\bibfnamefont{S.}~\bibnamefont{{Borgani}}},
  \bibinfo{author}{\bibfnamefont{C.}~\bibnamefont{{Fedeli}}},
  \bibinfo{author}{\bibfnamefont{S.}~\bibnamefont{{Matarrese}}},
  \bibinfo{author}{\bibfnamefont{L.}~\bibnamefont{{Moscardini}}},
  \bibinfo{author}{\bibfnamefont{P.}~\bibnamefont{{Rosati}}}, \bibnamefont{and}
  \bibinfo{author}{\bibfnamefont{J.}~\bibnamefont{{Weller}}},
  \bibinfo{journal}{\mnras} \textbf{\bibinfo{volume}{407}},
  \bibinfo{pages}{2339} (\bibinfo{year}{2010}), \eprint{1003.0841}.

\bibitem[{\citenamefont{{Reiprich} and
  {B{\"o}hringer}}(2002)}]{ReiprichBohringer:02}
\bibinfo{author}{\bibfnamefont{T.~H.} \bibnamefont{{Reiprich}}}
  \bibnamefont{and}
  \bibinfo{author}{\bibfnamefont{H.}~\bibnamefont{{B{\"o}hringer}}},
  \bibinfo{journal}{\apj} \textbf{\bibinfo{volume}{567}}, \bibinfo{pages}{716}
  (\bibinfo{year}{2002}), \eprint{arXiv:astro-ph/0111285}.

\bibitem[{\citenamefont{{Mantz} et~al.}(2008)\citenamefont{{Mantz}, {Allen},
  {Ebeling}, and {Rapetti}}}]{Mantzetal:08}
\bibinfo{author}{\bibfnamefont{A.}~\bibnamefont{{Mantz}}},
  \bibinfo{author}{\bibfnamefont{S.~W.} \bibnamefont{{Allen}}},
  \bibinfo{author}{\bibfnamefont{H.}~\bibnamefont{{Ebeling}}},
  \bibnamefont{and}
  \bibinfo{author}{\bibfnamefont{D.}~\bibnamefont{{Rapetti}}},
  \bibinfo{journal}{\mnras} \textbf{\bibinfo{volume}{387}},
  \bibinfo{pages}{1179} (\bibinfo{year}{2008}), \eprint{0709.4294}.

\bibitem[{\citenamefont{{Rykoff} et~al.}(2008)\citenamefont{{Rykoff}, {Evrard},
  {McKay}, {Becker}, {Johnston}, {Koester}, {Nord}, {Rozo}, {Sheldon}, {Stanek}
  et~al.}}]{Rykoffetal:08}
\bibinfo{author}{\bibfnamefont{E.~S.} \bibnamefont{{Rykoff}}},
  \bibinfo{author}{\bibfnamefont{A.~E.} \bibnamefont{{Evrard}}},
  \bibinfo{author}{\bibfnamefont{T.~A.} \bibnamefont{{McKay}}},
  \bibinfo{author}{\bibfnamefont{M.~R.} \bibnamefont{{Becker}}},
  \bibinfo{author}{\bibfnamefont{D.~E.} \bibnamefont{{Johnston}}},
  \bibinfo{author}{\bibfnamefont{B.~P.} \bibnamefont{{Koester}}},
  \bibinfo{author}{\bibfnamefont{B.}~\bibnamefont{{Nord}}},
  \bibinfo{author}{\bibfnamefont{E.}~\bibnamefont{{Rozo}}},
  \bibinfo{author}{\bibfnamefont{E.~S.} \bibnamefont{{Sheldon}}},
  \bibinfo{author}{\bibfnamefont{R.}~\bibnamefont{{Stanek}}},
  \bibnamefont{et~al.}, \bibinfo{journal}{\mnras}
  \textbf{\bibinfo{volume}{387}}, \bibinfo{pages}{L28} (\bibinfo{year}{2008}),
  \eprint{0802.1069}.

\bibitem[{\citenamefont{{Finoguenov} et~al.}(2001)\citenamefont{{Finoguenov},
  {Reiprich}, and {B{\"o}hringer}}}]{Finoguenoetal:01}
\bibinfo{author}{\bibfnamefont{A.}~\bibnamefont{{Finoguenov}}},
  \bibinfo{author}{\bibfnamefont{T.~H.} \bibnamefont{{Reiprich}}},
  \bibnamefont{and}
  \bibinfo{author}{\bibfnamefont{H.}~\bibnamefont{{B{\"o}hringer}}},
  \bibinfo{journal}{\aap} \textbf{\bibinfo{volume}{368}}, \bibinfo{pages}{749}
  (\bibinfo{year}{2001}), \eprint{arXiv:astro-ph/0010190}.

\bibitem[{\citenamefont{{Shimizu} et~al.}(2003)\citenamefont{{Shimizu},
  {Kitayama}, {Sasaki}, and {Suto}}}]{Shimizuetal:03}
\bibinfo{author}{\bibfnamefont{M.}~\bibnamefont{{Shimizu}}},
  \bibinfo{author}{\bibfnamefont{T.}~\bibnamefont{{Kitayama}}},
  \bibinfo{author}{\bibfnamefont{S.}~\bibnamefont{{Sasaki}}}, \bibnamefont{and}
  \bibinfo{author}{\bibfnamefont{Y.}~\bibnamefont{{Suto}}},
  \bibinfo{journal}{\apj} \textbf{\bibinfo{volume}{590}}, \bibinfo{pages}{197}
  (\bibinfo{year}{2003}), \eprint{arXiv:astro-ph/0212284}.

\bibitem[{\citenamefont{{Sanderson} et~al.}(2003)\citenamefont{{Sanderson},
  {Ponman}, {Finoguenov}, {Lloyd-Davies}, and {Markevitch}}}]{Sandersonetal:03}
\bibinfo{author}{\bibfnamefont{A.~J.~R.} \bibnamefont{{Sanderson}}},
  \bibinfo{author}{\bibfnamefont{T.~J.} \bibnamefont{{Ponman}}},
  \bibinfo{author}{\bibfnamefont{A.}~\bibnamefont{{Finoguenov}}},
  \bibinfo{author}{\bibfnamefont{E.~J.} \bibnamefont{{Lloyd-Davies}}},
  \bibnamefont{and}
  \bibinfo{author}{\bibfnamefont{M.}~\bibnamefont{{Markevitch}}},
  \bibinfo{journal}{\mnras} \textbf{\bibinfo{volume}{340}},
  \bibinfo{pages}{989} (\bibinfo{year}{2003}), \eprint{arXiv:astro-ph/0301049}.

\bibitem[{\citenamefont{{Arnaud} et~al.}(2005)\citenamefont{{Arnaud},
  {Pointecouteau}, and {Pratt}}}]{Arnaudetal:05}
\bibinfo{author}{\bibfnamefont{M.}~\bibnamefont{{Arnaud}}},
  \bibinfo{author}{\bibfnamefont{E.}~\bibnamefont{{Pointecouteau}}},
  \bibnamefont{and} \bibinfo{author}{\bibfnamefont{G.~W.}
  \bibnamefont{{Pratt}}}, \bibinfo{journal}{\aap}
  \textbf{\bibinfo{volume}{441}}, \bibinfo{pages}{893} (\bibinfo{year}{2005}),
  \eprint{arXiv:astro-ph/0502210}.

\bibitem[{\citenamefont{{Okabe}
  et~al.}(2010{\natexlab{a}})\citenamefont{{Okabe}, {Takada}, {Umetsu},
  {Futamase}, and {Smith}}}]{Okabeetal:10}
\bibinfo{author}{\bibfnamefont{N.}~\bibnamefont{{Okabe}}},
  \bibinfo{author}{\bibfnamefont{M.}~\bibnamefont{{Takada}}},
  \bibinfo{author}{\bibfnamefont{K.}~\bibnamefont{{Umetsu}}},
  \bibinfo{author}{\bibfnamefont{T.}~\bibnamefont{{Futamase}}},
  \bibnamefont{and} \bibinfo{author}{\bibfnamefont{G.~P.}
  \bibnamefont{{Smith}}}, \bibinfo{journal}{Publ.~Astron.~Soc.~Japan}
  \textbf{\bibinfo{volume}{62}}, \bibinfo{pages}{811}
  (\bibinfo{year}{2010}{\natexlab{a}}), \eprint{0903.1103}.

\bibitem[{\citenamefont{{Johnston} et~al.}(2007)\citenamefont{{Johnston},
  {Sheldon}, {Wechsler}, {Rozo}, {Koester}, {Frieman}, {McKay}, {Evrard},
  {Becker}, and {Annis}}}]{Johnstonetal:07}
\bibinfo{author}{\bibfnamefont{D.~E.} \bibnamefont{{Johnston}}},
  \bibinfo{author}{\bibfnamefont{E.~S.} \bibnamefont{{Sheldon}}},
  \bibinfo{author}{\bibfnamefont{R.~H.} \bibnamefont{{Wechsler}}},
  \bibinfo{author}{\bibfnamefont{E.}~\bibnamefont{{Rozo}}},
  \bibinfo{author}{\bibfnamefont{B.~P.} \bibnamefont{{Koester}}},
  \bibinfo{author}{\bibfnamefont{J.~A.} \bibnamefont{{Frieman}}},
  \bibinfo{author}{\bibfnamefont{T.~A.} \bibnamefont{{McKay}}},
  \bibinfo{author}{\bibfnamefont{A.~E.} \bibnamefont{{Evrard}}},
  \bibinfo{author}{\bibfnamefont{M.~R.} \bibnamefont{{Becker}}},
  \bibnamefont{and} \bibinfo{author}{\bibfnamefont{J.}~\bibnamefont{{Annis}}},
  \bibinfo{journal}{ArXiv e-prints}  (\bibinfo{year}{2007}),
  \eprint{0709.1159}.

\bibitem[{\citenamefont{{Bonamente} et~al.}(2008)\citenamefont{{Bonamente},
  {Joy}, {LaRoque}, {Carlstrom}, {Nagai}, and {Marrone}}}]{Bonamenteetal:08}
\bibinfo{author}{\bibfnamefont{M.}~\bibnamefont{{Bonamente}}},
  \bibinfo{author}{\bibfnamefont{M.}~\bibnamefont{{Joy}}},
  \bibinfo{author}{\bibfnamefont{S.~J.} \bibnamefont{{LaRoque}}},
  \bibinfo{author}{\bibfnamefont{J.~E.} \bibnamefont{{Carlstrom}}},
  \bibinfo{author}{\bibfnamefont{D.}~\bibnamefont{{Nagai}}}, \bibnamefont{and}
  \bibinfo{author}{\bibfnamefont{D.~P.} \bibnamefont{{Marrone}}},
  \bibinfo{journal}{\apj} \textbf{\bibinfo{volume}{675}}, \bibinfo{pages}{106}
  (\bibinfo{year}{2008}), \eprint{0708.0815}.

\bibitem[{\citenamefont{{Marrone} et~al.}(2009)\citenamefont{{Marrone},
  {Smith}, {Richard}, {Joy}, {Bonamente}, {Hasler}, {Hamilton-Morris}, {Kneib},
  {Culverhouse}, {Carlstrom} et~al.}}]{Marroneetal:09}
\bibinfo{author}{\bibfnamefont{D.~P.} \bibnamefont{{Marrone}}},
  \bibinfo{author}{\bibfnamefont{G.~P.} \bibnamefont{{Smith}}},
  \bibinfo{author}{\bibfnamefont{J.}~\bibnamefont{{Richard}}},
  \bibinfo{author}{\bibfnamefont{M.}~\bibnamefont{{Joy}}},
  \bibinfo{author}{\bibfnamefont{M.}~\bibnamefont{{Bonamente}}},
  \bibinfo{author}{\bibfnamefont{N.}~\bibnamefont{{Hasler}}},
  \bibinfo{author}{\bibfnamefont{V.}~\bibnamefont{{Hamilton-Morris}}},
  \bibinfo{author}{\bibfnamefont{J.}~\bibnamefont{{Kneib}}},
  \bibinfo{author}{\bibfnamefont{T.}~\bibnamefont{{Culverhouse}}},
  \bibinfo{author}{\bibfnamefont{J.~E.} \bibnamefont{{Carlstrom}}},
  \bibnamefont{et~al.}, \bibinfo{journal}{\apjl}
  \textbf{\bibinfo{volume}{701}}, \bibinfo{pages}{L114} (\bibinfo{year}{2009}),
  \eprint{0907.1687}.

\bibitem[{\citenamefont{{Kravtsov} et~al.}(2006)\citenamefont{{Kravtsov},
  {Vikhlinin}, and {Nagai}}}]{Kravtsovetal:06}
\bibinfo{author}{\bibfnamefont{A.~V.} \bibnamefont{{Kravtsov}}},
  \bibinfo{author}{\bibfnamefont{A.}~\bibnamefont{{Vikhlinin}}},
  \bibnamefont{and} \bibinfo{author}{\bibfnamefont{D.}~\bibnamefont{{Nagai}}},
  \bibinfo{journal}{\apj} \textbf{\bibinfo{volume}{650}}, \bibinfo{pages}{128}
  (\bibinfo{year}{2006}), \eprint{arXiv:astro-ph/0603205}.

\bibitem[{\citenamefont{{Zhang} et~al.}(2008)\citenamefont{{Zhang},
  {Finoguenov}, {B{\"o}hringer}, {Kneib}, {Smith}, {Kneissl}, {Okabe}, and
  {Dahle}}}]{Zhangetal:08}
\bibinfo{author}{\bibfnamefont{Y.}~\bibnamefont{{Zhang}}},
  \bibinfo{author}{\bibfnamefont{A.}~\bibnamefont{{Finoguenov}}},
  \bibinfo{author}{\bibfnamefont{H.}~\bibnamefont{{B{\"o}hringer}}},
  \bibinfo{author}{\bibfnamefont{J.}~\bibnamefont{{Kneib}}},
  \bibinfo{author}{\bibfnamefont{G.~P.} \bibnamefont{{Smith}}},
  \bibinfo{author}{\bibfnamefont{R.}~\bibnamefont{{Kneissl}}},
  \bibinfo{author}{\bibfnamefont{N.}~\bibnamefont{{Okabe}}}, \bibnamefont{and}
  \bibinfo{author}{\bibfnamefont{H.}~\bibnamefont{{Dahle}}},
  \bibinfo{journal}{\aap} \textbf{\bibinfo{volume}{482}}, \bibinfo{pages}{451}
  (\bibinfo{year}{2008}), \eprint{0802.0770}.

\bibitem[{\citenamefont{{Mantz}
  et~al.}(2010{\natexlab{b}})\citenamefont{{Mantz}, {Allen}, {Ebeling},
  {Rapetti}, and {Drlica-Wagner}}}]{Mantzetal:10b}
\bibinfo{author}{\bibfnamefont{A.}~\bibnamefont{{Mantz}}},
  \bibinfo{author}{\bibfnamefont{S.~W.} \bibnamefont{{Allen}}},
  \bibinfo{author}{\bibfnamefont{H.}~\bibnamefont{{Ebeling}}},
  \bibinfo{author}{\bibfnamefont{D.}~\bibnamefont{{Rapetti}}},
  \bibnamefont{and}
  \bibinfo{author}{\bibfnamefont{A.}~\bibnamefont{{Drlica-Wagner}}},
  \bibinfo{journal}{\mnras} \textbf{\bibinfo{volume}{406}},
  \bibinfo{pages}{1773} (\bibinfo{year}{2010}{\natexlab{b}}),
  \eprint{0909.3099}.

\bibitem[{\citenamefont{{Bartelmann} and
  {Schneider}}(2001)}]{BartelmannSchneider:01}
\bibinfo{author}{\bibfnamefont{M.}~\bibnamefont{{Bartelmann}}}
  \bibnamefont{and}
  \bibinfo{author}{\bibfnamefont{P.}~\bibnamefont{{Schneider}}},
  \bibinfo{journal}{Phys.~Rep.} \textbf{\bibinfo{volume}{340}},
  \bibinfo{pages}{291} (\bibinfo{year}{2001}), \eprint{arXiv:astro-ph/9912508}.

\bibitem[{\citenamefont{{Fischer} et~al.}(2000)\citenamefont{{Fischer},
  {McKay}, {Sheldon}, {Connolly}, {Stebbins}, {Frieman}, {Jain}, {Joffre},
  {Johnston}, {Bernstein} et~al.}}]{Fischeretal:00}
\bibinfo{author}{\bibfnamefont{P.}~\bibnamefont{{Fischer}}},
  \bibinfo{author}{\bibfnamefont{T.~A.} \bibnamefont{{McKay}}},
  \bibinfo{author}{\bibfnamefont{E.}~\bibnamefont{{Sheldon}}},
  \bibinfo{author}{\bibfnamefont{A.}~\bibnamefont{{Connolly}}},
  \bibinfo{author}{\bibfnamefont{A.}~\bibnamefont{{Stebbins}}},
  \bibinfo{author}{\bibfnamefont{J.~A.} \bibnamefont{{Frieman}}},
  \bibinfo{author}{\bibfnamefont{B.}~\bibnamefont{{Jain}}},
  \bibinfo{author}{\bibfnamefont{M.}~\bibnamefont{{Joffre}}},
  \bibinfo{author}{\bibfnamefont{D.}~\bibnamefont{{Johnston}}},
  \bibinfo{author}{\bibfnamefont{G.}~\bibnamefont{{Bernstein}}},
  \bibnamefont{et~al.}, \bibinfo{journal}{Astron.~J.}
  \textbf{\bibinfo{volume}{120}}, \bibinfo{pages}{1198} (\bibinfo{year}{2000}),
  \eprint{arXiv:astro-ph/9912119}.

\bibitem[{\citenamefont{{McKay} et~al.}(2001)\citenamefont{{McKay}, {Sheldon},
  {Racusin}, {Fischer}, {Seljak}, {Stebbins}, {Johnston}, {Frieman}, {Bahcall},
  {Brinkmann} et~al.}}]{McKayetal:01}
\bibinfo{author}{\bibfnamefont{T.~A.} \bibnamefont{{McKay}}},
  \bibinfo{author}{\bibfnamefont{E.~S.} \bibnamefont{{Sheldon}}},
  \bibinfo{author}{\bibfnamefont{J.}~\bibnamefont{{Racusin}}},
  \bibinfo{author}{\bibfnamefont{P.}~\bibnamefont{{Fischer}}},
  \bibinfo{author}{\bibfnamefont{U.}~\bibnamefont{{Seljak}}},
  \bibinfo{author}{\bibfnamefont{A.}~\bibnamefont{{Stebbins}}},
  \bibinfo{author}{\bibfnamefont{D.}~\bibnamefont{{Johnston}}},
  \bibinfo{author}{\bibfnamefont{J.~A.} \bibnamefont{{Frieman}}},
  \bibinfo{author}{\bibfnamefont{N.}~\bibnamefont{{Bahcall}}},
  \bibinfo{author}{\bibfnamefont{J.}~\bibnamefont{{Brinkmann}}},
  \bibnamefont{et~al.}, \bibinfo{journal}{ArXiv Astrophysics e-prints}
  (\bibinfo{year}{2001}), \eprint{arXiv:astro-ph/0108013}.

\bibitem[{\citenamefont{{Sheldon} et~al.}(2001)\citenamefont{{Sheldon},
  {Annis}, {B{\"o}hringer}, {Fischer}, {Frieman}, {Joffre}, {Johnston},
  {McKay}, {Miller}, {Nichol} et~al.}}]{Sheldonetal:01}
\bibinfo{author}{\bibfnamefont{E.~S.} \bibnamefont{{Sheldon}}},
  \bibinfo{author}{\bibfnamefont{J.}~\bibnamefont{{Annis}}},
  \bibinfo{author}{\bibfnamefont{H.}~\bibnamefont{{B{\"o}hringer}}},
  \bibinfo{author}{\bibfnamefont{P.}~\bibnamefont{{Fischer}}},
  \bibinfo{author}{\bibfnamefont{J.~A.} \bibnamefont{{Frieman}}},
  \bibinfo{author}{\bibfnamefont{M.}~\bibnamefont{{Joffre}}},
  \bibinfo{author}{\bibfnamefont{D.}~\bibnamefont{{Johnston}}},
  \bibinfo{author}{\bibfnamefont{T.~A.} \bibnamefont{{McKay}}},
  \bibinfo{author}{\bibfnamefont{C.}~\bibnamefont{{Miller}}},
  \bibinfo{author}{\bibfnamefont{R.~C.} \bibnamefont{{Nichol}}},
  \bibnamefont{et~al.}, \bibinfo{journal}{\apj} \textbf{\bibinfo{volume}{554}},
  \bibinfo{pages}{881} (\bibinfo{year}{2001}), \eprint{arXiv:astro-ph/0103029}.

\bibitem[{\citenamefont{{Guzik} and {Seljak}}(2001)}]{GuzikSeljak:01}
\bibinfo{author}{\bibfnamefont{J.}~\bibnamefont{{Guzik}}} \bibnamefont{and}
  \bibinfo{author}{\bibfnamefont{U.}~\bibnamefont{{Seljak}}},
  \bibinfo{journal}{\mnras} \textbf{\bibinfo{volume}{321}},
  \bibinfo{pages}{439} (\bibinfo{year}{2001}), \eprint{arXiv:astro-ph/0007067}.

\bibitem[{\citenamefont{{Mandelbaum} et~al.}(2005)\citenamefont{{Mandelbaum},
  {Tasitsiomi}, {Seljak}, {Kravtsov}, and {Wechsler}}}]{Mandelbaumetal:05}
\bibinfo{author}{\bibfnamefont{R.}~\bibnamefont{{Mandelbaum}}},
  \bibinfo{author}{\bibfnamefont{A.}~\bibnamefont{{Tasitsiomi}}},
  \bibinfo{author}{\bibfnamefont{U.}~\bibnamefont{{Seljak}}},
  \bibinfo{author}{\bibfnamefont{A.~V.} \bibnamefont{{Kravtsov}}},
  \bibnamefont{and} \bibinfo{author}{\bibfnamefont{R.~H.}
  \bibnamefont{{Wechsler}}}, \bibinfo{journal}{\mnras}
  \textbf{\bibinfo{volume}{362}}, \bibinfo{pages}{1451} (\bibinfo{year}{2005}),
  \eprint{arXiv:astro-ph/0410711}.

\bibitem[{\citenamefont{{Sheldon} et~al.}(2009)\citenamefont{{Sheldon},
  {Johnston}, {Scranton}, {Koester}, {McKay}, {Oyaizu}, {Cunha}, {Lima}, {Lin},
  {Frieman} et~al.}}]{Sheldonetal:09}
\bibinfo{author}{\bibfnamefont{E.~S.} \bibnamefont{{Sheldon}}},
  \bibinfo{author}{\bibfnamefont{D.~E.} \bibnamefont{{Johnston}}},
  \bibinfo{author}{\bibfnamefont{R.}~\bibnamefont{{Scranton}}},
  \bibinfo{author}{\bibfnamefont{B.~P.} \bibnamefont{{Koester}}},
  \bibinfo{author}{\bibfnamefont{T.~A.} \bibnamefont{{McKay}}},
  \bibinfo{author}{\bibfnamefont{H.}~\bibnamefont{{Oyaizu}}},
  \bibinfo{author}{\bibfnamefont{C.}~\bibnamefont{{Cunha}}},
  \bibinfo{author}{\bibfnamefont{M.}~\bibnamefont{{Lima}}},
  \bibinfo{author}{\bibfnamefont{H.}~\bibnamefont{{Lin}}},
  \bibinfo{author}{\bibfnamefont{J.~A.} \bibnamefont{{Frieman}}},
  \bibnamefont{et~al.}, \bibinfo{journal}{\apj} \textbf{\bibinfo{volume}{703}},
  \bibinfo{pages}{2217} (\bibinfo{year}{2009}), \eprint{0709.1153}.

\bibitem[{\citenamefont{{Huterer} et~al.}(2006)\citenamefont{{Huterer},
  {Takada}, {Bernstein}, and {Jain}}}]{Hutereretal:06}
\bibinfo{author}{\bibfnamefont{D.}~\bibnamefont{{Huterer}}},
  \bibinfo{author}{\bibfnamefont{M.}~\bibnamefont{{Takada}}},
  \bibinfo{author}{\bibfnamefont{G.}~\bibnamefont{{Bernstein}}},
  \bibnamefont{and} \bibinfo{author}{\bibfnamefont{B.}~\bibnamefont{{Jain}}},
  \bibinfo{journal}{\mnras} \textbf{\bibinfo{volume}{366}},
  \bibinfo{pages}{101} (\bibinfo{year}{2006}), \eprint{arXiv:astro-ph/0506030}.

\bibitem[{\citenamefont{{Nishizawa} et~al.}(2010)\citenamefont{{Nishizawa},
  {Takada}, {Hamana}, and {Furusawa}}}]{Nishizawaetal:10}
\bibinfo{author}{\bibfnamefont{A.~J.} \bibnamefont{{Nishizawa}}},
  \bibinfo{author}{\bibfnamefont{M.}~\bibnamefont{{Takada}}},
  \bibinfo{author}{\bibfnamefont{T.}~\bibnamefont{{Hamana}}}, \bibnamefont{and}
  \bibinfo{author}{\bibfnamefont{H.}~\bibnamefont{{Furusawa}}},
  \bibinfo{journal}{\apj} \textbf{\bibinfo{volume}{718}}, \bibinfo{pages}{1252}
  (\bibinfo{year}{2010}), \eprint{1002.2476}.

\bibitem[{\citenamefont{{Hearin} et~al.}(2010)\citenamefont{{Hearin},
  {Zentner}, {Ma}, and {Huterer}}}]{Hearinetal:10}
\bibinfo{author}{\bibfnamefont{A.~P.} \bibnamefont{{Hearin}}},
  \bibinfo{author}{\bibfnamefont{A.~R.} \bibnamefont{{Zentner}}},
  \bibinfo{author}{\bibfnamefont{Z.}~\bibnamefont{{Ma}}}, \bibnamefont{and}
  \bibinfo{author}{\bibfnamefont{D.}~\bibnamefont{{Huterer}}},
  \bibinfo{journal}{\apj} \textbf{\bibinfo{volume}{720}}, \bibinfo{pages}{1351}
  (\bibinfo{year}{2010}), \eprint{1002.3383}.

\bibitem[{\citenamefont{{Mandelbaum} et~al.}(2010)\citenamefont{{Mandelbaum},
  {Seljak}, {Baldauf}, and {Smith}}}]{Mandelbaumetal:10}
\bibinfo{author}{\bibfnamefont{R.}~\bibnamefont{{Mandelbaum}}},
  \bibinfo{author}{\bibfnamefont{U.}~\bibnamefont{{Seljak}}},
  \bibinfo{author}{\bibfnamefont{T.}~\bibnamefont{{Baldauf}}},
  \bibnamefont{and} \bibinfo{author}{\bibfnamefont{R.~E.}
  \bibnamefont{{Smith}}}, \bibinfo{journal}{\mnras}
  \textbf{\bibinfo{volume}{405}}, \bibinfo{pages}{2078} (\bibinfo{year}{2010}),
  \eprint{0911.4972}.

\bibitem[{\citenamefont{{Majumdar} and {Mohr}}(2004)}]{MajumdarMohr:04}
\bibinfo{author}{\bibfnamefont{S.}~\bibnamefont{{Majumdar}}} \bibnamefont{and}
  \bibinfo{author}{\bibfnamefont{J.~J.} \bibnamefont{{Mohr}}},
  \bibinfo{journal}{\apj} \textbf{\bibinfo{volume}{613}}, \bibinfo{pages}{41}
  (\bibinfo{year}{2004}), \eprint{arXiv:astro-ph/0305341}.

\bibitem[{\citenamefont{{Lima} and {Hu}}(2004)}]{LimaHu:04}
\bibinfo{author}{\bibfnamefont{M.}~\bibnamefont{{Lima}}} \bibnamefont{and}
  \bibinfo{author}{\bibfnamefont{W.}~\bibnamefont{{Hu}}},
  \bibinfo{journal}{\prd} \textbf{\bibinfo{volume}{70}},
  \bibinfo{pages}{043504} (\bibinfo{year}{2004}),
  \eprint{arXiv:astro-ph/0401559}.

\bibitem[{\citenamefont{{Lima} and {Hu}}(2005)}]{LimaHu:05}
\bibinfo{author}{\bibfnamefont{M.}~\bibnamefont{{Lima}}} \bibnamefont{and}
  \bibinfo{author}{\bibfnamefont{W.}~\bibnamefont{{Hu}}},
  \bibinfo{journal}{\prd} \textbf{\bibinfo{volume}{72}},
  \bibinfo{pages}{043006} (\bibinfo{year}{2005}),
  \eprint{arXiv:astro-ph/0503363}.

\bibitem[{\citenamefont{{Miyazaki} et~al.}(2006)\citenamefont{{Miyazaki},
  {Komiyama}, {Nakaya}, {Doi}, {Furusawa}, {Gillingham}, {Kamata}, {Takeshi},
  and {Nariai}}}]{Miyazakietal:06}
\bibinfo{author}{\bibfnamefont{S.}~\bibnamefont{{Miyazaki}}},
  \bibinfo{author}{\bibfnamefont{Y.}~\bibnamefont{{Komiyama}}},
  \bibinfo{author}{\bibfnamefont{H.}~\bibnamefont{{Nakaya}}},
  \bibinfo{author}{\bibfnamefont{Y.}~\bibnamefont{{Doi}}},
  \bibinfo{author}{\bibfnamefont{H.}~\bibnamefont{{Furusawa}}},
  \bibinfo{author}{\bibfnamefont{P.}~\bibnamefont{{Gillingham}}},
  \bibinfo{author}{\bibfnamefont{Y.}~\bibnamefont{{Kamata}}},
  \bibinfo{author}{\bibfnamefont{K.}~\bibnamefont{{Takeshi}}},
  \bibnamefont{and} \bibinfo{author}{\bibfnamefont{K.}~\bibnamefont{{Nariai}}},
  in \emph{\bibinfo{booktitle}{Society of Photo-Optical Instrumentation
  Engineers (SPIE) Conference Series}} (\bibinfo{year}{2006}), vol.
  \bibinfo{volume}{6269} of \emph{\bibinfo{series}{Society of Photo-Optical
  Instrumentation Engineers (SPIE) Conference Series}}.

\bibitem[{\citenamefont{{The Dark Energy Survey Collaboration}}(2005)}]{DES}
\bibinfo{author}{\bibnamefont{{The Dark Energy Survey Collaboration}}},
  \bibinfo{journal}{ArXiv Astrophysics e-prints}  (\bibinfo{year}{2005}),
  \eprint{arXiv:astro-ph/0510346}.

\bibitem[{\citenamefont{{LSST Science Collaborations}
  et~al.}(2009)\citenamefont{{LSST Science Collaborations}, {Abell}, {Allison},
  {Anderson}, {Andrew}, {Angel}, {Armus}, {Arnett}, {Asztalos}, {Axelrod}
  et~al.}}]{LSST}
\bibinfo{author}{\bibnamefont{{LSST Science Collaborations}}},
  \bibinfo{author}{\bibfnamefont{P.~A.} \bibnamefont{{Abell}}},
  \bibinfo{author}{\bibfnamefont{J.}~\bibnamefont{{Allison}}},
  \bibinfo{author}{\bibfnamefont{S.~F.} \bibnamefont{{Anderson}}},
  \bibinfo{author}{\bibfnamefont{J.~R.} \bibnamefont{{Andrew}}},
  \bibinfo{author}{\bibfnamefont{J.~R.~P.} \bibnamefont{{Angel}}},
  \bibinfo{author}{\bibfnamefont{L.}~\bibnamefont{{Armus}}},
  \bibinfo{author}{\bibfnamefont{D.}~\bibnamefont{{Arnett}}},
  \bibinfo{author}{\bibfnamefont{S.~J.} \bibnamefont{{Asztalos}}},
  \bibinfo{author}{\bibfnamefont{T.~S.} \bibnamefont{{Axelrod}}},
  \bibnamefont{et~al.}, \bibinfo{journal}{ArXiv e-prints}
  (\bibinfo{year}{2009}), \eprint{0912.0201}.

\bibitem[{\citenamefont{{Hincks} et~al.}(2009)\citenamefont{{Hincks},
  {Acquaviva}, {Ade}, {Aguirre}, {Amiri}, {Appel}, {Barrientos}, {Battistelli},
  {Bond}, {Brown} et~al.}}]{ACT:09}
\bibinfo{author}{\bibfnamefont{A.~D.} \bibnamefont{{Hincks}}},
  \bibinfo{author}{\bibfnamefont{V.}~\bibnamefont{{Acquaviva}}},
  \bibinfo{author}{\bibfnamefont{P.}~\bibnamefont{{Ade}}},
  \bibinfo{author}{\bibfnamefont{P.}~\bibnamefont{{Aguirre}}},
  \bibinfo{author}{\bibfnamefont{M.}~\bibnamefont{{Amiri}}},
  \bibinfo{author}{\bibfnamefont{J.~W.} \bibnamefont{{Appel}}},
  \bibinfo{author}{\bibfnamefont{L.~F.} \bibnamefont{{Barrientos}}},
  \bibinfo{author}{\bibfnamefont{E.~S.} \bibnamefont{{Battistelli}}},
  \bibinfo{author}{\bibfnamefont{J.~R.} \bibnamefont{{Bond}}},
  \bibinfo{author}{\bibfnamefont{B.}~\bibnamefont{{Brown}}},
  \bibnamefont{et~al.}, \bibinfo{journal}{ArXiv e-prints}
  (\bibinfo{year}{2009}), \eprint{0907.0461}.

\bibitem[{\citenamefont{{Niemack} et~al.}(2010)\citenamefont{{Niemack}, {Ade},
  {Aguirre}, {Barrientos}, {Beall}, {Bond}, {Britton}, {Cho}, {Das}, {Devlin}
  et~al.}}]{ACTPol:10}
\bibinfo{author}{\bibfnamefont{M.~D.} \bibnamefont{{Niemack}}},
  \bibinfo{author}{\bibfnamefont{P.~A.~R.} \bibnamefont{{Ade}}},
  \bibinfo{author}{\bibfnamefont{J.}~\bibnamefont{{Aguirre}}},
  \bibinfo{author}{\bibfnamefont{F.}~\bibnamefont{{Barrientos}}},
  \bibinfo{author}{\bibfnamefont{J.~A.} \bibnamefont{{Beall}}},
  \bibinfo{author}{\bibfnamefont{J.~R.} \bibnamefont{{Bond}}},
  \bibinfo{author}{\bibfnamefont{J.}~\bibnamefont{{Britton}}},
  \bibinfo{author}{\bibfnamefont{H.~M.} \bibnamefont{{Cho}}},
  \bibinfo{author}{\bibfnamefont{S.}~\bibnamefont{{Das}}},
  \bibinfo{author}{\bibfnamefont{M.~J.} \bibnamefont{{Devlin}}},
  \bibnamefont{et~al.}, in \emph{\bibinfo{booktitle}{Society of Photo-Optical
  Instrumentation Engineers (SPIE) Conference Series}} (\bibinfo{year}{2010}),
  vol. \bibinfo{volume}{7741} of \emph{\bibinfo{series}{Society of
  Photo-Optical Instrumentation Engineers (SPIE) Conference Series}},
  \eprint{1006.5049}.

\bibitem[{\citenamefont{{Vanderlinde} et~al.}(2010)\citenamefont{{Vanderlinde},
  {Crawford}, {de Haan}, {Dudley}, {Shaw}, {Ade}, {Aird}, {Benson}, {Bleem},
  {Brodwin} et~al.}}]{SPT:10}
\bibinfo{author}{\bibfnamefont{K.}~\bibnamefont{{Vanderlinde}}},
  \bibinfo{author}{\bibfnamefont{T.~M.} \bibnamefont{{Crawford}}},
  \bibinfo{author}{\bibfnamefont{T.}~\bibnamefont{{de Haan}}},
  \bibinfo{author}{\bibfnamefont{J.~P.} \bibnamefont{{Dudley}}},
  \bibinfo{author}{\bibfnamefont{L.}~\bibnamefont{{Shaw}}},
  \bibinfo{author}{\bibfnamefont{P.~A.~R.} \bibnamefont{{Ade}}},
  \bibinfo{author}{\bibfnamefont{K.~A.} \bibnamefont{{Aird}}},
  \bibinfo{author}{\bibfnamefont{B.~A.} \bibnamefont{{Benson}}},
  \bibinfo{author}{\bibfnamefont{L.~E.} \bibnamefont{{Bleem}}},
  \bibinfo{author}{\bibfnamefont{M.}~\bibnamefont{{Brodwin}}},
  \bibnamefont{et~al.}, \bibinfo{journal}{\apj} \textbf{\bibinfo{volume}{722}},
  \bibinfo{pages}{1180} (\bibinfo{year}{2010}), \eprint{1003.0003}.

\bibitem[{\citenamefont{{Komatsu} et~al.}(2010)\citenamefont{{Komatsu},
  {Smith}, {Dunkley}, {Bennett}, {Gold}, {Hinshaw}, {Jarosik}, {Larson},
  {Nolta}, {Page} et~al.}}]{Komatsuetal:10}
\bibinfo{author}{\bibfnamefont{E.}~\bibnamefont{{Komatsu}}},
  \bibinfo{author}{\bibfnamefont{K.~M.} \bibnamefont{{Smith}}},
  \bibinfo{author}{\bibfnamefont{J.}~\bibnamefont{{Dunkley}}},
  \bibinfo{author}{\bibfnamefont{C.~L.} \bibnamefont{{Bennett}}},
  \bibinfo{author}{\bibfnamefont{B.}~\bibnamefont{{Gold}}},
  \bibinfo{author}{\bibfnamefont{G.}~\bibnamefont{{Hinshaw}}},
  \bibinfo{author}{\bibfnamefont{N.}~\bibnamefont{{Jarosik}}},
  \bibinfo{author}{\bibfnamefont{D.}~\bibnamefont{{Larson}}},
  \bibinfo{author}{\bibfnamefont{M.~R.} \bibnamefont{{Nolta}}},
  \bibinfo{author}{\bibfnamefont{L.}~\bibnamefont{{Page}}},
  \bibnamefont{et~al.}, \bibinfo{journal}{ArXiv e-prints}
  (\bibinfo{year}{2010}), \eprint{1001.4538}.

\bibitem[{\citenamefont{{Hu}}(1999)}]{Hu:99}
\bibinfo{author}{\bibfnamefont{W.}~\bibnamefont{{Hu}}},
  \bibinfo{journal}{\apjl} \textbf{\bibinfo{volume}{522}}, \bibinfo{pages}{L21}
  (\bibinfo{year}{1999}), \eprint{arXiv:astro-ph/9904153}.

\bibitem[{\citenamefont{{Huterer}}(2002)}]{Huterer:02}
\bibinfo{author}{\bibfnamefont{D.}~\bibnamefont{{Huterer}}},
  \bibinfo{journal}{\prd} \textbf{\bibinfo{volume}{65}},
  \bibinfo{pages}{063001} (\bibinfo{year}{2002}),
  \eprint{arXiv:astro-ph/0106399}.

\bibitem[{\citenamefont{{Takada} and {Jain}}(2004)}]{TakadaJain:04}
\bibinfo{author}{\bibfnamefont{M.}~\bibnamefont{{Takada}}} \bibnamefont{and}
  \bibinfo{author}{\bibfnamefont{B.}~\bibnamefont{{Jain}}},
  \bibinfo{journal}{\mnras} \textbf{\bibinfo{volume}{348}},
  \bibinfo{pages}{897} (\bibinfo{year}{2004}), \eprint{arXiv:astro-ph/0310125}.

\bibitem[{\citenamefont{{Chevallier} and
  {Polarski}}(2001)}]{ChevallierPolarski:01}
\bibinfo{author}{\bibfnamefont{M.}~\bibnamefont{{Chevallier}}}
  \bibnamefont{and}
  \bibinfo{author}{\bibfnamefont{D.}~\bibnamefont{{Polarski}}},
  \bibinfo{journal}{International Journal of Modern Physics D}
  \textbf{\bibinfo{volume}{10}}, \bibinfo{pages}{213} (\bibinfo{year}{2001}),
  \eprint{arXiv:gr-qc/0009008}.

\bibitem[{\citenamefont{{Wang} and {Steinhardt}}(1998)}]{WangSteinhardt98}
\bibinfo{author}{\bibfnamefont{L.}~\bibnamefont{{Wang}}} \bibnamefont{and}
  \bibinfo{author}{\bibfnamefont{P.~J.} \bibnamefont{{Steinhardt}}},
  \bibinfo{journal}{\apj} \textbf{\bibinfo{volume}{508}}, \bibinfo{pages}{483}
  (\bibinfo{year}{1998}), \eprint{arXiv:astro-ph/9804015}.

\bibitem[{\citenamefont{{Takada} et~al.}(2006)\citenamefont{{Takada},
  {Komatsu}, and {Futamase}}}]{Takadaetal:06}
\bibinfo{author}{\bibfnamefont{M.}~\bibnamefont{{Takada}}},
  \bibinfo{author}{\bibfnamefont{E.}~\bibnamefont{{Komatsu}}},
  \bibnamefont{and}
  \bibinfo{author}{\bibfnamefont{T.}~\bibnamefont{{Futamase}}},
  \bibinfo{journal}{\prd} \textbf{\bibinfo{volume}{73}},
  \bibinfo{pages}{083520} (\bibinfo{year}{2006}),
  \eprint{arXiv:astro-ph/0512374}.

\bibitem[{\citenamefont{{Eisenstein} and {Hu}}(1998)}]{EisensteinHu:98}
\bibinfo{author}{\bibfnamefont{D.~J.} \bibnamefont{{Eisenstein}}}
  \bibnamefont{and} \bibinfo{author}{\bibfnamefont{W.}~\bibnamefont{{Hu}}},
  \bibinfo{journal}{\apj} \textbf{\bibinfo{volume}{496}}, \bibinfo{pages}{605}
  (\bibinfo{year}{1998}), \eprint{arXiv:astro-ph/9709112}.

\bibitem[{\citenamefont{{Komatsu} et~al.}(2009)\citenamefont{{Komatsu},
  {Dunkley}, {Nolta}, {Bennett}, {Gold}, {Hinshaw}, {Jarosik}, {Larson},
  {Limon}, {Page} et~al.}}]{Komatsu:09}
\bibinfo{author}{\bibfnamefont{E.}~\bibnamefont{{Komatsu}}},
  \bibinfo{author}{\bibfnamefont{J.}~\bibnamefont{{Dunkley}}},
  \bibinfo{author}{\bibfnamefont{M.~R.} \bibnamefont{{Nolta}}},
  \bibinfo{author}{\bibfnamefont{C.~L.} \bibnamefont{{Bennett}}},
  \bibinfo{author}{\bibfnamefont{B.}~\bibnamefont{{Gold}}},
  \bibinfo{author}{\bibfnamefont{G.}~\bibnamefont{{Hinshaw}}},
  \bibinfo{author}{\bibfnamefont{N.}~\bibnamefont{{Jarosik}}},
  \bibinfo{author}{\bibfnamefont{D.}~\bibnamefont{{Larson}}},
  \bibinfo{author}{\bibfnamefont{M.}~\bibnamefont{{Limon}}},
  \bibinfo{author}{\bibfnamefont{L.}~\bibnamefont{{Page}}},
  \bibnamefont{et~al.}, \bibinfo{journal}{Astropys.J.Suppl.}
  \textbf{\bibinfo{volume}{180}}, \bibinfo{pages}{330} (\bibinfo{year}{2009}),
  \eprint{arXiv:0803.0547}.

\bibitem[{\citenamefont{{Bhattacharya}
  et~al.}(2010)\citenamefont{{Bhattacharya}, {Heitmann}, {White}, {Luki{\'c}},
  {Wagner}, and {Habib}}}]{Bhattacharyaetal:10}
\bibinfo{author}{\bibfnamefont{S.}~\bibnamefont{{Bhattacharya}}},
  \bibinfo{author}{\bibfnamefont{K.}~\bibnamefont{{Heitmann}}},
  \bibinfo{author}{\bibfnamefont{M.}~\bibnamefont{{White}}},
  \bibinfo{author}{\bibfnamefont{Z.}~\bibnamefont{{Luki{\'c}}}},
  \bibinfo{author}{\bibfnamefont{C.}~\bibnamefont{{Wagner}}}, \bibnamefont{and}
  \bibinfo{author}{\bibfnamefont{S.}~\bibnamefont{{Habib}}},
  \bibinfo{journal}{ArXiv e-prints}  (\bibinfo{year}{2010}),
  \eprint{1005.2239}.

\bibitem[{\citenamefont{{Limber}}(1954)}]{Limber:54}
\bibinfo{author}{\bibfnamefont{D.~N.} \bibnamefont{{Limber}}},
  \bibinfo{journal}{\apj} \textbf{\bibinfo{volume}{119}}, \bibinfo{pages}{655}
  (\bibinfo{year}{1954}).

\bibitem[{\citenamefont{{Lo Verde} et~al.}(2008)\citenamefont{{Lo Verde},
  {Miller}, {Shandera}, and {Verde}}}]{LoVerdeetal:08}
\bibinfo{author}{\bibfnamefont{M.}~\bibnamefont{{Lo Verde}}},
  \bibinfo{author}{\bibfnamefont{A.}~\bibnamefont{{Miller}}},
  \bibinfo{author}{\bibfnamefont{S.}~\bibnamefont{{Shandera}}},
  \bibnamefont{and} \bibinfo{author}{\bibfnamefont{L.}~\bibnamefont{{Verde}}},
  \bibinfo{journal}{\jcap} \textbf{\bibinfo{volume}{4}}, \bibinfo{pages}{14}
  (\bibinfo{year}{2008}), \eprint{0711.4126}.

\bibitem[{\citenamefont{{Smith} et~al.}(2003)\citenamefont{{Smith}, {Peacock},
  {Jenkins}, {White}, {Frenk}, {Pearce}, {Thomas}, {Efstathiou}, and
  {Couchman}}}]{Smithetal:03}
\bibinfo{author}{\bibfnamefont{R.~E.} \bibnamefont{{Smith}}},
  \bibinfo{author}{\bibfnamefont{J.~A.} \bibnamefont{{Peacock}}},
  \bibinfo{author}{\bibfnamefont{A.}~\bibnamefont{{Jenkins}}},
  \bibinfo{author}{\bibfnamefont{S.~D.~M.} \bibnamefont{{White}}},
  \bibinfo{author}{\bibfnamefont{C.~S.} \bibnamefont{{Frenk}}},
  \bibinfo{author}{\bibfnamefont{F.~R.} \bibnamefont{{Pearce}}},
  \bibinfo{author}{\bibfnamefont{P.~A.} \bibnamefont{{Thomas}}},
  \bibinfo{author}{\bibfnamefont{G.}~\bibnamefont{{Efstathiou}}},
  \bibnamefont{and} \bibinfo{author}{\bibfnamefont{H.~M.~P.}
  \bibnamefont{{Couchman}}}, \bibinfo{journal}{\mnras}
  \textbf{\bibinfo{volume}{341}}, \bibinfo{pages}{1311} (\bibinfo{year}{2003}),
  \eprint{arXiv:astro-ph/0207664}.

\bibitem[{\citenamefont{{Takada} and {Bridle}}(2007)}]{TakadaBridle:07}
\bibinfo{author}{\bibfnamefont{M.}~\bibnamefont{{Takada}}} \bibnamefont{and}
  \bibinfo{author}{\bibfnamefont{S.}~\bibnamefont{{Bridle}}},
  \bibinfo{journal}{New Journal of Physics} \textbf{\bibinfo{volume}{9}},
  \bibinfo{pages}{446} (\bibinfo{year}{2007}), \eprint{arXiv:0705.0163}.

\bibitem[{\citenamefont{{Hirata} and {Seljak}}(2004)}]{HirataSeljak:04}
\bibinfo{author}{\bibfnamefont{C.~M.} \bibnamefont{{Hirata}}} \bibnamefont{and}
  \bibinfo{author}{\bibfnamefont{U.}~\bibnamefont{{Seljak}}},
  \bibinfo{journal}{\prd} \textbf{\bibinfo{volume}{70}},
  \bibinfo{pages}{063526} (\bibinfo{year}{2004}),
  \eprint{arXiv:astro-ph/0406275}.

\bibitem[{\citenamefont{{de Putter} and {Takada}}(2010)}]{dePutterTakada:10}
\bibinfo{author}{\bibfnamefont{R.}~\bibnamefont{{de Putter}}} \bibnamefont{and}
  \bibinfo{author}{\bibfnamefont{M.}~\bibnamefont{{Takada}}},
  \bibinfo{journal}{ArXiv e-prints}  (\bibinfo{year}{2010}),
  \eprint{1007.4809}.

\bibitem[{\citenamefont{{Navarro} et~al.}(1997)\citenamefont{{Navarro},
  {Frenk}, and {White}}}]{Navarroetal:97}
\bibinfo{author}{\bibfnamefont{J.~F.} \bibnamefont{{Navarro}}},
  \bibinfo{author}{\bibfnamefont{C.~S.} \bibnamefont{{Frenk}}},
  \bibnamefont{and} \bibinfo{author}{\bibfnamefont{S.~D.~M.}
  \bibnamefont{{White}}}, \bibinfo{journal}{\apj}
  \textbf{\bibinfo{volume}{490}}, \bibinfo{pages}{493} (\bibinfo{year}{1997}),
  \eprint{arXiv:astro-ph/9611107}.

\bibitem[{\citenamefont{{Nakamura} and {Suto}}(1997)}]{NakamuraSuto:97}
\bibinfo{author}{\bibfnamefont{T.~T.} \bibnamefont{{Nakamura}}}
  \bibnamefont{and} \bibinfo{author}{\bibfnamefont{Y.}~\bibnamefont{{Suto}}},
  \bibinfo{journal}{Progress of Theoretical Physics}
  \textbf{\bibinfo{volume}{97}}, \bibinfo{pages}{49} (\bibinfo{year}{1997}),
  \eprint{arXiv:astro-ph/9612074}.

\bibitem[{\citenamefont{{Bullock} et~al.}(2001)\citenamefont{{Bullock},
  {Kolatt}, {Sigad}, {Somerville}, {Kravtsov}, {Klypin}, {Primack}, and
  {Dekel}}}]{Bullocketal:01}
\bibinfo{author}{\bibfnamefont{J.~S.} \bibnamefont{{Bullock}}},
  \bibinfo{author}{\bibfnamefont{T.~S.} \bibnamefont{{Kolatt}}},
  \bibinfo{author}{\bibfnamefont{Y.}~\bibnamefont{{Sigad}}},
  \bibinfo{author}{\bibfnamefont{R.~S.} \bibnamefont{{Somerville}}},
  \bibinfo{author}{\bibfnamefont{A.~V.} \bibnamefont{{Kravtsov}}},
  \bibinfo{author}{\bibfnamefont{A.~A.} \bibnamefont{{Klypin}}},
  \bibinfo{author}{\bibfnamefont{J.~R.} \bibnamefont{{Primack}}},
  \bibnamefont{and} \bibinfo{author}{\bibfnamefont{A.}~\bibnamefont{{Dekel}}},
  \bibinfo{journal}{\mnras} \textbf{\bibinfo{volume}{321}},
  \bibinfo{pages}{559} (\bibinfo{year}{2001}), \eprint{arXiv:astro-ph/9908159}.

\bibitem[{\citenamefont{{Duffy} et~al.}(2008)\citenamefont{{Duffy}, {Schaye},
  {Kay}, and {Dalla Vecchia}}}]{Duffyetal:08}
\bibinfo{author}{\bibfnamefont{A.~R.} \bibnamefont{{Duffy}}},
  \bibinfo{author}{\bibfnamefont{J.}~\bibnamefont{{Schaye}}},
  \bibinfo{author}{\bibfnamefont{S.~T.} \bibnamefont{{Kay}}}, \bibnamefont{and}
  \bibinfo{author}{\bibfnamefont{C.}~\bibnamefont{{Dalla Vecchia}}},
  \bibinfo{journal}{\mnras} \textbf{\bibinfo{volume}{390}},
  \bibinfo{pages}{L64} (\bibinfo{year}{2008}), \eprint{0804.2486}.

\bibitem[{\citenamefont{{Scoccimarro} et~al.}(2001)\citenamefont{{Scoccimarro},
  {Sheth}, {Hui}, and {Jain}}}]{Scoccimarroetal:01}
\bibinfo{author}{\bibfnamefont{R.}~\bibnamefont{{Scoccimarro}}},
  \bibinfo{author}{\bibfnamefont{R.~K.} \bibnamefont{{Sheth}}},
  \bibinfo{author}{\bibfnamefont{L.}~\bibnamefont{{Hui}}}, \bibnamefont{and}
  \bibinfo{author}{\bibfnamefont{B.}~\bibnamefont{{Jain}}},
  \bibinfo{journal}{\apj} \textbf{\bibinfo{volume}{546}}, \bibinfo{pages}{20}
  (\bibinfo{year}{2001}), \eprint{arXiv:astro-ph/0006319}.

\bibitem[{\citenamefont{{Takada} and {Jain}}(2003)}]{TakadaJain:03a}
\bibinfo{author}{\bibfnamefont{M.}~\bibnamefont{{Takada}}} \bibnamefont{and}
  \bibinfo{author}{\bibfnamefont{B.}~\bibnamefont{{Jain}}},
  \bibinfo{journal}{\mnras} \textbf{\bibinfo{volume}{340}},
  \bibinfo{pages}{580} (\bibinfo{year}{2003}), \eprint{arXiv:astro-ph/0209167}.

\bibitem[{\citenamefont{{Oguri} et~al.}(2010)\citenamefont{{Oguri}, {Takada},
  {Okabe}, and {Smith}}}]{Ogurietal:10}
\bibinfo{author}{\bibfnamefont{M.}~\bibnamefont{{Oguri}}},
  \bibinfo{author}{\bibfnamefont{M.}~\bibnamefont{{Takada}}},
  \bibinfo{author}{\bibfnamefont{N.}~\bibnamefont{{Okabe}}}, \bibnamefont{and}
  \bibinfo{author}{\bibfnamefont{G.~P.} \bibnamefont{{Smith}}},
  \bibinfo{journal}{\mnras} \textbf{\bibinfo{volume}{405}},
  \bibinfo{pages}{2215} (\bibinfo{year}{2010}), \eprint{1004.4214}.

\bibitem[{\citenamefont{{Hilbert} and {White}}(2010)}]{HilbertWhite:10}
\bibinfo{author}{\bibfnamefont{S.}~\bibnamefont{{Hilbert}}} \bibnamefont{and}
  \bibinfo{author}{\bibfnamefont{S.~D.~M.} \bibnamefont{{White}}},
  \bibinfo{journal}{\mnras} \textbf{\bibinfo{volume}{404}},
  \bibinfo{pages}{486} (\bibinfo{year}{2010}), \eprint{0907.4371}.

\bibitem[{\citenamefont{{Lin} and {Mohr}}(2004)}]{LinMohr:04}
\bibinfo{author}{\bibfnamefont{Y.}~\bibnamefont{{Lin}}} \bibnamefont{and}
  \bibinfo{author}{\bibfnamefont{J.~J.} \bibnamefont{{Mohr}}},
  \bibinfo{journal}{\apj} \textbf{\bibinfo{volume}{617}}, \bibinfo{pages}{879}
  (\bibinfo{year}{2004}), \eprint{arXiv:astro-ph/0408557}.

\bibitem[{\citenamefont{{Koester} et~al.}(2007)\citenamefont{{Koester},
  {McKay}, {Annis}, {Wechsler}, {Evrard}, {Bleem}, {Becker}, {Johnston},
  {Sheldon}, {Nichol} et~al.}}]{Koesteretal:07}
\bibinfo{author}{\bibfnamefont{B.~P.} \bibnamefont{{Koester}}},
  \bibinfo{author}{\bibfnamefont{T.~A.} \bibnamefont{{McKay}}},
  \bibinfo{author}{\bibfnamefont{J.}~\bibnamefont{{Annis}}},
  \bibinfo{author}{\bibfnamefont{R.~H.} \bibnamefont{{Wechsler}}},
  \bibinfo{author}{\bibfnamefont{A.}~\bibnamefont{{Evrard}}},
  \bibinfo{author}{\bibfnamefont{L.}~\bibnamefont{{Bleem}}},
  \bibinfo{author}{\bibfnamefont{M.}~\bibnamefont{{Becker}}},
  \bibinfo{author}{\bibfnamefont{D.}~\bibnamefont{{Johnston}}},
  \bibinfo{author}{\bibfnamefont{E.}~\bibnamefont{{Sheldon}}},
  \bibinfo{author}{\bibfnamefont{R.}~\bibnamefont{{Nichol}}},
  \bibnamefont{et~al.}, \bibinfo{journal}{\apj} \textbf{\bibinfo{volume}{660}},
  \bibinfo{pages}{239} (\bibinfo{year}{2007}), \eprint{arXiv:astro-ph/0701265}.

\bibitem[{\citenamefont{{Shan} et~al.}(2010)\citenamefont{{Shan}, {Qin},
  {Fort}, {Tao}, {Wu}, and {Zhao}}}]{Shanetal:10}
\bibinfo{author}{\bibfnamefont{H.}~\bibnamefont{{Shan}}},
  \bibinfo{author}{\bibfnamefont{B.}~\bibnamefont{{Qin}}},
  \bibinfo{author}{\bibfnamefont{B.}~\bibnamefont{{Fort}}},
  \bibinfo{author}{\bibfnamefont{C.}~\bibnamefont{{Tao}}},
  \bibinfo{author}{\bibfnamefont{X.}~\bibnamefont{{Wu}}}, \bibnamefont{and}
  \bibinfo{author}{\bibfnamefont{H.}~\bibnamefont{{Zhao}}},
  \bibinfo{journal}{\mnras} \textbf{\bibinfo{volume}{406}},
  \bibinfo{pages}{1134} (\bibinfo{year}{2010}), \eprint{1004.1475}.

\bibitem[{\citenamefont{{Hu} and {Kravtsov}}(2003)}]{HuKravtsov03}
\bibinfo{author}{\bibfnamefont{W.}~\bibnamefont{{Hu}}} \bibnamefont{and}
  \bibinfo{author}{\bibfnamefont{A.~V.} \bibnamefont{{Kravtsov}}},
  \bibinfo{journal}{\apj} \textbf{\bibinfo{volume}{584}}, \bibinfo{pages}{702}
  (\bibinfo{year}{2003}), \eprint{arXiv:astro-ph/0203169}.

\bibitem[{\citenamefont{{Takada} and {Jain}}(2009)}]{TakadaJain:09}
\bibinfo{author}{\bibfnamefont{M.}~\bibnamefont{{Takada}}} \bibnamefont{and}
  \bibinfo{author}{\bibfnamefont{B.}~\bibnamefont{{Jain}}},
  \bibinfo{journal}{\mnras} \textbf{\bibinfo{volume}{395}},
  \bibinfo{pages}{2065} (\bibinfo{year}{2009}), \eprint{0810.4170}.

\bibitem[{\citenamefont{{Sato} et~al.}(2009)\citenamefont{{Sato}, {Hamana},
  {Takahashi}, {Takada}, {Yoshida}, {Matsubara}, and {Sugiyama}}}]{Satoetal:09}
\bibinfo{author}{\bibfnamefont{M.}~\bibnamefont{{Sato}}},
  \bibinfo{author}{\bibfnamefont{T.}~\bibnamefont{{Hamana}}},
  \bibinfo{author}{\bibfnamefont{R.}~\bibnamefont{{Takahashi}}},
  \bibinfo{author}{\bibfnamefont{M.}~\bibnamefont{{Takada}}},
  \bibinfo{author}{\bibfnamefont{N.}~\bibnamefont{{Yoshida}}},
  \bibinfo{author}{\bibfnamefont{T.}~\bibnamefont{{Matsubara}}},
  \bibnamefont{and}
  \bibinfo{author}{\bibfnamefont{N.}~\bibnamefont{{Sugiyama}}},
  \bibinfo{journal}{\apj} \textbf{\bibinfo{volume}{701}}, \bibinfo{pages}{945}
  (\bibinfo{year}{2009}), \eprint{0906.2237}.

\bibitem[{\citenamefont{{Hoekstra}}(2003)}]{Hoekstra:03}
\bibinfo{author}{\bibfnamefont{H.}~\bibnamefont{{Hoekstra}}},
  \bibinfo{journal}{\mnras} \textbf{\bibinfo{volume}{339}},
  \bibinfo{pages}{1155} (\bibinfo{year}{2003}),
  \eprint{arXiv:astro-ph/0208351}.

\bibitem[{\citenamefont{{Dodelson}}(2004)}]{Dodelson:04}
\bibinfo{author}{\bibfnamefont{S.}~\bibnamefont{{Dodelson}}},
  \bibinfo{journal}{\prd} \textbf{\bibinfo{volume}{70}},
  \bibinfo{pages}{023008} (\bibinfo{year}{2004}),
  \eprint{arXiv:astro-ph/0309277}.

\bibitem[{\citenamefont{{Jeong} et~al.}(2009)\citenamefont{{Jeong}, {Komatsu},
  and {Jain}}}]{Jeongetal:09}
\bibinfo{author}{\bibfnamefont{D.}~\bibnamefont{{Jeong}}},
  \bibinfo{author}{\bibfnamefont{E.}~\bibnamefont{{Komatsu}}},
  \bibnamefont{and} \bibinfo{author}{\bibfnamefont{B.}~\bibnamefont{{Jain}}},
  \bibinfo{journal}{\prd} \textbf{\bibinfo{volume}{80}},
  \bibinfo{pages}{123527} (\bibinfo{year}{2009}), \eprint{0910.1361}.

\bibitem[{\citenamefont{{Sato} et~al.}(2010)\citenamefont{{Sato}, {Takada},
  {Hamana}, and {Matsubara}}}]{Satoetal:10}
\bibinfo{author}{\bibfnamefont{M.}~\bibnamefont{{Sato}}},
  \bibinfo{author}{\bibfnamefont{M.}~\bibnamefont{{Takada}}},
  \bibinfo{author}{\bibfnamefont{T.}~\bibnamefont{{Hamana}}}, \bibnamefont{and}
  \bibinfo{author}{\bibfnamefont{T.}~\bibnamefont{{Matsubara}}},
  \bibinfo{journal}{ArXiv e-prints}  (\bibinfo{year}{2010}),
  \eprint{1009.2558}.

\bibitem[{\citenamefont{{Lewis} et~al.}(2000)\citenamefont{{Lewis},
  {Challinor}, and {Lasenby}}}]{Lewisetal:00}
\bibinfo{author}{\bibfnamefont{A.}~\bibnamefont{{Lewis}}},
  \bibinfo{author}{\bibfnamefont{A.}~\bibnamefont{{Challinor}}},
  \bibnamefont{and}
  \bibinfo{author}{\bibfnamefont{A.}~\bibnamefont{{Lasenby}}},
  \bibinfo{journal}{\apj} \textbf{\bibinfo{volume}{538}}, \bibinfo{pages}{473}
  (\bibinfo{year}{2000}), \eprint{arXiv:astro-ph/9911177}.

\bibitem[{\citenamefont{{Seljak} and
  {Zaldarriaga}}(1996)}]{SeljakZaldarriaga:96}
\bibinfo{author}{\bibfnamefont{U.}~\bibnamefont{{Seljak}}} \bibnamefont{and}
  \bibinfo{author}{\bibfnamefont{M.}~\bibnamefont{{Zaldarriaga}}},
  \bibinfo{journal}{\apj} \textbf{\bibinfo{volume}{469}}, \bibinfo{pages}{437}
  (\bibinfo{year}{1996}), \eprint{arXiv:astro-ph/9603033}.

\bibitem[{\citenamefont{{Eisenstein} et~al.}(2005)\citenamefont{{Eisenstein},
  {Zehavi}, {Hogg}, {Scoccimarro}, {Blanton}, {Nichol}, {Scranton}, {Seo},
  {Tegmark}, {Zheng} et~al.}}]{Eisensteinetal:05}
\bibinfo{author}{\bibfnamefont{D.~J.} \bibnamefont{{Eisenstein}}},
  \bibinfo{author}{\bibfnamefont{I.}~\bibnamefont{{Zehavi}}},
  \bibinfo{author}{\bibfnamefont{D.~W.} \bibnamefont{{Hogg}}},
  \bibinfo{author}{\bibfnamefont{R.}~\bibnamefont{{Scoccimarro}}},
  \bibinfo{author}{\bibfnamefont{M.~R.} \bibnamefont{{Blanton}}},
  \bibinfo{author}{\bibfnamefont{R.~C.} \bibnamefont{{Nichol}}},
  \bibinfo{author}{\bibfnamefont{R.}~\bibnamefont{{Scranton}}},
  \bibinfo{author}{\bibfnamefont{H.}~\bibnamefont{{Seo}}},
  \bibinfo{author}{\bibfnamefont{M.}~\bibnamefont{{Tegmark}}},
  \bibinfo{author}{\bibfnamefont{Z.}~\bibnamefont{{Zheng}}},
  \bibnamefont{et~al.}, \bibinfo{journal}{\apj} \textbf{\bibinfo{volume}{633}},
  \bibinfo{pages}{560} (\bibinfo{year}{2005}), \eprint{arXiv:astro-ph/0501171}.

\bibitem[{\citenamefont{{Seo} and {Eisenstein}}(2003)}]{SeoEisenstein:03}
\bibinfo{author}{\bibfnamefont{H.}~\bibnamefont{{Seo}}} \bibnamefont{and}
  \bibinfo{author}{\bibfnamefont{D.~J.} \bibnamefont{{Eisenstein}}},
  \bibinfo{journal}{\apj} \textbf{\bibinfo{volume}{598}}, \bibinfo{pages}{720}
  (\bibinfo{year}{2003}), \eprint{arXiv:astro-ph/0307460}.

\bibitem[{\citenamefont{{Rozo} et~al.}(2010{\natexlab{b}})\citenamefont{{Rozo},
  {Wu}, and {Schmidt}}}]{RozoWuSchmidt:10}
\bibinfo{author}{\bibfnamefont{E.}~\bibnamefont{{Rozo}}},
  \bibinfo{author}{\bibfnamefont{H.}~\bibnamefont{{Wu}}}, \bibnamefont{and}
  \bibinfo{author}{\bibfnamefont{F.}~\bibnamefont{{Schmidt}}},
  \bibinfo{journal}{ArXiv e-prints}  (\bibinfo{year}{2010}{\natexlab{b}}),
  \eprint{1009.0756}.

\bibitem[{\citenamefont{{Mandelbaum} et~al.}(2008)\citenamefont{{Mandelbaum},
  {Seljak}, and {Hirata}}}]{Mandelbaumetal:08}
\bibinfo{author}{\bibfnamefont{R.}~\bibnamefont{{Mandelbaum}}},
  \bibinfo{author}{\bibfnamefont{U.}~\bibnamefont{{Seljak}}}, \bibnamefont{and}
  \bibinfo{author}{\bibfnamefont{C.~M.} \bibnamefont{{Hirata}}},
  \bibinfo{journal}{\jcap} \textbf{\bibinfo{volume}{8}}, \bibinfo{pages}{6}
  (\bibinfo{year}{2008}), \eprint{0805.2552}.

\bibitem[{\citenamefont{{Broadhurst} et~al.}(2005)\citenamefont{{Broadhurst},
  {Takada}, {Umetsu}, {Kong}, {Arimoto}, {Chiba}, and
  {Futamase}}}]{Broadhurstetal:05}
\bibinfo{author}{\bibfnamefont{T.}~\bibnamefont{{Broadhurst}}},
  \bibinfo{author}{\bibfnamefont{M.}~\bibnamefont{{Takada}}},
  \bibinfo{author}{\bibfnamefont{K.}~\bibnamefont{{Umetsu}}},
  \bibinfo{author}{\bibfnamefont{X.}~\bibnamefont{{Kong}}},
  \bibinfo{author}{\bibfnamefont{N.}~\bibnamefont{{Arimoto}}},
  \bibinfo{author}{\bibfnamefont{M.}~\bibnamefont{{Chiba}}}, \bibnamefont{and}
  \bibinfo{author}{\bibfnamefont{T.}~\bibnamefont{{Futamase}}},
  \bibinfo{journal}{\apjl} \textbf{\bibinfo{volume}{619}},
  \bibinfo{pages}{L143} (\bibinfo{year}{2005}),
  \eprint{arXiv:astro-ph/0412192}.

\bibitem[{\citenamefont{{Broadhurst} et~al.}(2008)\citenamefont{{Broadhurst},
  {Umetsu}, {Medezinski}, {Oguri}, and {Rephaeli}}}]{Broadhurstetal:09}
\bibinfo{author}{\bibfnamefont{T.}~\bibnamefont{{Broadhurst}}},
  \bibinfo{author}{\bibfnamefont{K.}~\bibnamefont{{Umetsu}}},
  \bibinfo{author}{\bibfnamefont{E.}~\bibnamefont{{Medezinski}}},
  \bibinfo{author}{\bibfnamefont{M.}~\bibnamefont{{Oguri}}}, \bibnamefont{and}
  \bibinfo{author}{\bibfnamefont{Y.}~\bibnamefont{{Rephaeli}}},
  \bibinfo{journal}{\apjl} \textbf{\bibinfo{volume}{685}}, \bibinfo{pages}{L9}
  (\bibinfo{year}{2008}), \eprint{0805.2617}.

\bibitem[{\citenamefont{{Oguri} et~al.}(2009)\citenamefont{{Oguri}, {Hennawi},
  {Gladders}, {Dahle}, {Natarajan}, {Dalal}, {Koester}, {Sharon}, and
  {Bayliss}}}]{Ogurietal:09}
\bibinfo{author}{\bibfnamefont{M.}~\bibnamefont{{Oguri}}},
  \bibinfo{author}{\bibfnamefont{J.~F.} \bibnamefont{{Hennawi}}},
  \bibinfo{author}{\bibfnamefont{M.~D.} \bibnamefont{{Gladders}}},
  \bibinfo{author}{\bibfnamefont{H.}~\bibnamefont{{Dahle}}},
  \bibinfo{author}{\bibfnamefont{P.}~\bibnamefont{{Natarajan}}},
  \bibinfo{author}{\bibfnamefont{N.}~\bibnamefont{{Dalal}}},
  \bibinfo{author}{\bibfnamefont{B.~P.} \bibnamefont{{Koester}}},
  \bibinfo{author}{\bibfnamefont{K.}~\bibnamefont{{Sharon}}}, \bibnamefont{and}
  \bibinfo{author}{\bibfnamefont{M.}~\bibnamefont{{Bayliss}}},
  \bibinfo{journal}{\apj} \textbf{\bibinfo{volume}{699}}, \bibinfo{pages}{1038}
  (\bibinfo{year}{2009}), \eprint{0901.4372}.

\bibitem[{\citenamefont{{Okabe}
  et~al.}(2010{\natexlab{b}})\citenamefont{{Okabe}, {Zhang}, {Finoguenov},
  {Takada}, {Smith}, {Umetsu}, and {Futamase}}}]{Okabeetal:10b}
\bibinfo{author}{\bibfnamefont{N.}~\bibnamefont{{Okabe}}},
  \bibinfo{author}{\bibfnamefont{Y.}~\bibnamefont{{Zhang}}},
  \bibinfo{author}{\bibfnamefont{A.}~\bibnamefont{{Finoguenov}}},
  \bibinfo{author}{\bibfnamefont{M.}~\bibnamefont{{Takada}}},
  \bibinfo{author}{\bibfnamefont{G.~P.} \bibnamefont{{Smith}}},
  \bibinfo{author}{\bibfnamefont{K.}~\bibnamefont{{Umetsu}}}, \bibnamefont{and}
  \bibinfo{author}{\bibfnamefont{T.}~\bibnamefont{{Futamase}}},
  \bibinfo{journal}{\apj} \textbf{\bibinfo{volume}{721}}, \bibinfo{pages}{875}
  (\bibinfo{year}{2010}{\natexlab{b}}), \eprint{1007.3816}.

\bibitem[{\citenamefont{{Lima} and {Hu}}(2007)}]{LimaHu:07}
\bibinfo{author}{\bibfnamefont{M.}~\bibnamefont{{Lima}}} \bibnamefont{and}
  \bibinfo{author}{\bibfnamefont{W.}~\bibnamefont{{Hu}}},
  \bibinfo{journal}{\prd} \textbf{\bibinfo{volume}{76}},
  \bibinfo{pages}{123013} (\bibinfo{year}{2007}), \eprint{0709.2871}.

\bibitem[{\citenamefont{{Lin} et~al.}(2006)\citenamefont{{Lin}, {Lima},
  {Oyaizu}, {Cunha}, {Frieman}, {Annis}, {Koester}, {Hao}, {McKay}, and
  {Sheldon}}}]{Linetal:06}
\bibinfo{author}{\bibfnamefont{H.}~\bibnamefont{{Lin}}},
  \bibinfo{author}{\bibfnamefont{M.}~\bibnamefont{{Lima}}},
  \bibinfo{author}{\bibfnamefont{H.}~\bibnamefont{{Oyaizu}}},
  \bibinfo{author}{\bibfnamefont{C.}~\bibnamefont{{Cunha}}},
  \bibinfo{author}{\bibfnamefont{J.}~\bibnamefont{{Frieman}}},
  \bibinfo{author}{\bibfnamefont{J.}~\bibnamefont{{Annis}}},
  \bibinfo{author}{\bibfnamefont{B.}~\bibnamefont{{Koester}}},
  \bibinfo{author}{\bibfnamefont{J.}~\bibnamefont{{Hao}}},
  \bibinfo{author}{\bibfnamefont{T.}~\bibnamefont{{McKay}}}, \bibnamefont{and}
  \bibinfo{author}{\bibfnamefont{E.}~\bibnamefont{{Sheldon}}}, in
  \emph{\bibinfo{booktitle}{Bulletin of the American Astronomical Society}}
  (\bibinfo{year}{2006}), vol.~\bibinfo{volume}{38} of
  \emph{\bibinfo{series}{Bulletin of the American Astronomical Society}}, pp.
  \bibinfo{pages}{1196--+}.

\bibitem[{\citenamefont{{Newman}}(2008)}]{Newman:08}
\bibinfo{author}{\bibfnamefont{J.~A.} \bibnamefont{{Newman}}},
  \bibinfo{journal}{\apj} \textbf{\bibinfo{volume}{684}}, \bibinfo{pages}{88}
  (\bibinfo{year}{2008}), \eprint{0805.1409}.

\bibitem[{\citenamefont{{Van Waerbeke} et~al.}(2010)\citenamefont{{Van
  Waerbeke}, {Hildebrandt}, {Ford}, and {Milkeraitis}}}]{VanWarbekeetal:10}
\bibinfo{author}{\bibfnamefont{L.}~\bibnamefont{{Van Waerbeke}}},
  \bibinfo{author}{\bibfnamefont{H.}~\bibnamefont{{Hildebrandt}}},
  \bibinfo{author}{\bibfnamefont{J.}~\bibnamefont{{Ford}}}, \bibnamefont{and}
  \bibinfo{author}{\bibfnamefont{M.}~\bibnamefont{{Milkeraitis}}},
  \bibinfo{journal}{\apjl} \textbf{\bibinfo{volume}{723}}, \bibinfo{pages}{L13}
  (\bibinfo{year}{2010}), \eprint{1004.3793}.

\bibitem[{\citenamefont{{Rozo} and {Schmidt}}(2010)}]{RozoSchmidt:10}
\bibinfo{author}{\bibfnamefont{E.}~\bibnamefont{{Rozo}}} \bibnamefont{and}
  \bibinfo{author}{\bibfnamefont{F.}~\bibnamefont{{Schmidt}}},
  \bibinfo{journal}{ArXiv e-prints}  (\bibinfo{year}{2010}),
  \eprint{1009.5735}.

\bibitem[{\citenamefont{{Bridle} et~al.}(2010)\citenamefont{{Bridle}, {Balan},
  {Bethge}, {Gentile}, {Harmeling}, {Heymans}, {Hirsch}, {Hosseini}, {Jarvis},
  {Kirk} et~al.}}]{Bridleetal:10}
\bibinfo{author}{\bibfnamefont{S.}~\bibnamefont{{Bridle}}},
  \bibinfo{author}{\bibfnamefont{S.~T.} \bibnamefont{{Balan}}},
  \bibinfo{author}{\bibfnamefont{M.}~\bibnamefont{{Bethge}}},
  \bibinfo{author}{\bibfnamefont{M.}~\bibnamefont{{Gentile}}},
  \bibinfo{author}{\bibfnamefont{S.}~\bibnamefont{{Harmeling}}},
  \bibinfo{author}{\bibfnamefont{C.}~\bibnamefont{{Heymans}}},
  \bibinfo{author}{\bibfnamefont{M.}~\bibnamefont{{Hirsch}}},
  \bibinfo{author}{\bibfnamefont{R.}~\bibnamefont{{Hosseini}}},
  \bibinfo{author}{\bibfnamefont{M.}~\bibnamefont{{Jarvis}}},
  \bibinfo{author}{\bibfnamefont{D.}~\bibnamefont{{Kirk}}},
  \bibnamefont{et~al.}, \bibinfo{journal}{\mnras}
  \textbf{\bibinfo{volume}{405}}, \bibinfo{pages}{2044} (\bibinfo{year}{2010}),
  \eprint{0908.0945}.

\bibitem[{\citenamefont{{Massey} et~al.}(2007)\citenamefont{{Massey},
  {Heymans}, {Berg{\'e}}, {Bernstein}, {Bridle}, {Clowe}, {Dahle}, {Ellis},
  {Erben}, {Hetterscheidt} et~al.}}]{Massey:07}
\bibinfo{author}{\bibfnamefont{R.}~\bibnamefont{{Massey}}},
  \bibinfo{author}{\bibfnamefont{C.}~\bibnamefont{{Heymans}}},
  \bibinfo{author}{\bibfnamefont{J.}~\bibnamefont{{Berg{\'e}}}},
  \bibinfo{author}{\bibfnamefont{G.}~\bibnamefont{{Bernstein}}},
  \bibinfo{author}{\bibfnamefont{S.}~\bibnamefont{{Bridle}}},
  \bibinfo{author}{\bibfnamefont{D.}~\bibnamefont{{Clowe}}},
  \bibinfo{author}{\bibfnamefont{H.}~\bibnamefont{{Dahle}}},
  \bibinfo{author}{\bibfnamefont{R.}~\bibnamefont{{Ellis}}},
  \bibinfo{author}{\bibfnamefont{T.}~\bibnamefont{{Erben}}},
  \bibinfo{author}{\bibfnamefont{M.}~\bibnamefont{{Hetterscheidt}}},
  \bibnamefont{et~al.}, \bibinfo{journal}{\mnras}
  \textbf{\bibinfo{volume}{376}}, \bibinfo{pages}{13} (\bibinfo{year}{2007}),
  \eprint{arXiv:astro-ph/0608643}.

\bibitem[{\citenamefont{{Cunha} and {Evrard}}(2010)}]{CunhaEvrard:10}
\bibinfo{author}{\bibfnamefont{C.~E.} \bibnamefont{{Cunha}}} \bibnamefont{and}
  \bibinfo{author}{\bibfnamefont{A.~E.} \bibnamefont{{Evrard}}},
  \bibinfo{journal}{\prd} \textbf{\bibinfo{volume}{81}},
  \bibinfo{pages}{083509} (\bibinfo{year}{2010}), \eprint{0908.0526}.

\bibitem[{\citenamefont{{Wu} et~al.}(2010)\citenamefont{{Wu}, {Zentner}, and
  {Wechsler}}}]{Wuetal:10}
\bibinfo{author}{\bibfnamefont{H.}~\bibnamefont{{Wu}}},
  \bibinfo{author}{\bibfnamefont{A.~R.} \bibnamefont{{Zentner}}},
  \bibnamefont{and} \bibinfo{author}{\bibfnamefont{R.~H.}
  \bibnamefont{{Wechsler}}}, \bibinfo{journal}{\apj}
  \textbf{\bibinfo{volume}{713}}, \bibinfo{pages}{856} (\bibinfo{year}{2010}),
  \eprint{0910.3668}.

\end{thebibliography}

\end{document}